%% file: main.tex
\documentclass[opre,nonblindrev]{informs3} 

\OneAndAHalfSpacedXI 

\input{"preamble.tex"}

\usepackage{endnotes,multirow}
\let\footnote=\endnote
\let\enotesize=\normalsize
\def\notesname{Endnotes}%
\def\makeenmark{$^{\theenmark}$}
\def\enoteformat{\rightskip0pt\leftskip0pt\parindent=1.75em
  \leavevmode\llap{\theenmark.\enskip}}

\usepackage{natbib}
 \bibpunct[, ]{(}{)}{,}{a}{}{,}%
 \def\bibfont{\small}%
 \def\bibsep{\smallskipamount}%
\TheoremsNumberedThrough     
\ECRepeatTheorems

\EquationsNumberedThrough    

\usepackage[normalem]{ulem}

\begin{document}

\VOLUME{}%
\NO{}%
\MONTH{}
\YEAR{}
\FIRSTPAGE{}%
\LASTPAGE{}%
\SHORTYEAR{}
\ISSUE{} %
\LONGFIRSTPAGE{} %
\DOI{}%

\RUNAUTHOR{S. M. Pesenti, P. Millossovich, and A. Tsanakas}

\RUNTITLE{Differential Quantile-Based Sensitivity in Discontinuous Models}

\TITLE{Differential Quantile-Based Sensitivity in Discontinuous Models}

\ARTICLEAUTHORS{%
\AUTHOR{Silvana M. Pesenti}
\AFF{Department of Statistical Sciences, University of Toronto, \EMAIL{silvana.pesenti@utoronto.ca}} 
\AUTHOR{Pietro Millossovich}
\AFF{Bayes Business School (formerly Cass), City, University of London,\\ DEAMS, University of Trieste} 
\AUTHOR{Andreas Tsanakas}
\AFF{Bayes Business School (formerly Cass), City, University of London}
1. October 2024\footnote{First version: 11 October 2023.}
} 

\ABSTRACT{%
Differential sensitivity measures provide valuable tools for interpreting complex computational models, as used in applications ranging from simulation to algorithmic prediction. Taking the derivative of the model output in direction of a model parameter can reveal input-output relations and the relative importance of model parameters and input variables. Nonetheless, it is unclear how such derivatives should be taken when the model function has discontinuities and/or input variables are discrete. We present a general framework for addressing such problems, considering derivatives of quantile-based output risk measures, with respect to distortions to random input variables (risk factors), which impact the model output through step-functions. We prove that, subject to weak technical conditions, the derivatives are well-defined and we derive the corresponding formulas. We apply our results to the sensitivity analysis of compound risk models and to a numerical study of reinsurance credit risk in a multi-line insurance portfolio.
}%

\KEYWORDS{Sensitivity analysis, importance measurement, differential sensitivity measures, simulation, risk measures, credit risk.}

\maketitle

\section{Introduction}\label{sec:intro}
The interpretability of complex computational models is of fundamental importance across areas of applications, with sensitivity analysis providing tools for understanding the importance of risk factors, their interactions and their impact on a model's output \citep{Saltelli2008book, Borgonovo2016EJOR, Razavi2021future,Fissler2023EJOR}. In recent years, the field  received renewed impetus by the widespread adoption of machine learning and artificial intelligence models for prediction tasks, which are usually opaque and thus require additional work to illuminate input/output relationships. Contributions in this field range from the development of general model-agnostic model interpretation procedures \citep{ribeiro2016model, BORGONOVO_Classifiers}, to those tailored to a class of models, such as tree ensembles \citep{lundberg2018consistent} and neural networks \citep{merz2022interpreting}, or to specific applications, such as image recognition \citep{chen2019looks} and credit scoring \citep{CHEN2023}. Furthermore, the interest in model interpretation is amplified by the requirement for models' behaviour to be fair, in the sense that it does not generate discriminatory impacts on protected groups \citep{frees2021discriminating, KOZODOI20221083, lindholm2022discrimination} -- such concerns have generated further research at the interface of sensitivity analysis and algorithmic fairness \citep{benesse2022fairness, hiabu2023unifying}.

As part of sensitivity analysis, metrics are often used to assess the importance of model inputs. A broad class of such metrics is that of differential sensitivity measures, which rely on derivatives (of a statistical functional of) the model output, in the direction of a perturbation of a (random) input factor. Specifically, \cite{borgonovo2001new} introduce a local sensitivity measure by considering partial derivatives normalised by total derivatives; building on that work 
\cite{antoniano2018parameters} consider derivatives of expected loss functionals with respect to statistical parameters. Furthermore, substantial work has been carried out to reconcile and compare global sensitivity analysis based on Sobol' indices \citep{sobol2001global} with the local view given by differentiation at a particular parameter value \citep{sobol2010derivative, lamboni2013derivative, rakovec2014distributed}. 

Recent advances in sensitivity analysis pertain to perturbing quantile-based risk measures of the model output  \citep{Tanakas2016RA, Browne2017WP, Pesenti2021RA, merz2022interpreting}, which gives an alternative way of obtaining a global view of local effects. In that context, fundamental technical requirements for differential sensitivity measures include differentiability of the model function and Lipschitz continuity of the model output  in the perturbation \citep{Broadie1996MS, Hong2009OR,Hong2009MS}. These requirements are stringent, as many computational models map input factors to outputs in a discontinuous manner; examples include credit risk models \citep{Chen2008FS}, financial derivatives and insurance contracts \citep{albrecher2017reinsurance}, and tree-based predictive models  \citep{chen2016xgboost}. One should not be cavalier about differentiability  conditions, as it has been long established that lack of consideration in systems subject to  -- possibly `hidden' -- discontinuities can lead to integration failures and thus incorrect sensitivity assessments \citep{tolsma2002hidden}.

In this work, we overcome such strong conditions and derive, under rather mild assumptions, formulas for differential quantile-based sensitivity measures, in models where the input-output relationship contains step functions. This is a general setting, since many functions with a finite number of jump discontinuity points can be written via a sum of step functions. We focus on the two most common quantile-based risk measures, Value-at-Risk (VaR) and Expected Shortfall (ES), although the expressions can be generalised for the broader case of distortion risk measures and rank dependent expected utilities. We consider two types of differential sensitivity measure, marginal sensitivities and cascade sensitivities. The marginal sensitivity quantifies an input factor's sole effect on a model output's risk measure \citep{Hong2009OR, Tanakas2016RA}. In contrast, in the cascade sensitivity setting  \citep{Pesenti2021RA} a perturbation of a risk factor affects other dependent risk factors, which in turn impact the output risk measure. When the random input factors are independent, the two methods coincide. In the case of dependence, the cascade sensitivity additionally reflects the indirect effects that a risk factor may have on the output via other inputs, implicitly interpreting their statistical relationship as a functional (or causal) one, with the risk factor stressed being the driver. To prove the derived sensitivity formulas we use quantile differentiation and weak convergence of generalised functions. We find that stresses propagated via step functions naturally lead to delta functions, which in turn allow for representation as conditional expectations. Hence, our framework allows estimation of differential sensitivity measures by standard simulation-based methods \citep{Glasserman2005JCF,  Fu2009MS, Koike2022IME}.

Key to our framework is the choice of perturbation or \textit{stress} on the random input factor. In particular, the technical conditions we require pertain to the continuity of the stressing mechanism rather than the underlying random input factor. Consequently, our methods can also be applied to the calculation of differential sensitivity measures with respect to discrete random inputs for a suitably chosen stress. Sensitivity to discrete or categorical input factors is of importance in a variety of fields, such as modelling biological systems \citep{Gunawan2005BJ}, chemical processes \citep{Plyasunov2007JCP}, and insurance claims  \citep{wuthrich2023statistical}.

The manuscript is organised as follows. Section \ref{sec:diff-sensitiity} introduces the discontinuous loss model and discusses choices of stresses on a risk factor. Following that,
 expressions are derived for differential (marginal) sensitivities, with respect to the VaR and ES risk measures. The next two sections contain extensions within that framework. Section \ref{sub:sec:cascade} deals with cascade sensitivities, which reflect indirect effects via risk factors' dependence structure. Section \ref{sec:discrete} provides differential sensitivities when the considered input random variables are discrete, along with an application to compound distributions. Finally, a detailed numerical study of a reinsurance credit risk portfolio is given in Section \ref{sec: numerical example RI}. 

Additional formulas for the cascade sensitivity for the VaR are presented in Appendix \ref{app:additional-sensitivies}. Most of the proofs are delegated to Appendix \ref{app:proofs}. Finally, the electronic companion consists of the following appendices. In Appendix \ref{app::general-loss-model} the differential sensitivity formulas together with their proofs for a more general model function are recorded. Appendix \ref{elec-app-additional-proofs} contains proofs of results related to mixture stresses, Proposition \ref{prop: inv-rosenblatt}, and Theorem \ref{thm:VaR:marginal-discrete}. Appendix \ref{app:reinsurance-model} contains additional details on the reinsurance credit risk portfolio model used in Section  \ref{sec: numerical example RI}.

\section{Differential Sensitivity Measures}\label{sec:diff-sensitiity}

\subsection{Portfolio Loss Model}

We work on a probability space $(\Omega, \mathcal{A}, \P)$ and consider a discontinuous  model of the form
\begin{equation}\label{eq:loss-model}
    L := \sum_{j = 1}^m g_j(\Z) \Id_{\{X_j \le d_j\}},
\end{equation}
where:
\begin{itemize}
     \item The random vectors $\X := (X_1, \ldots, X_m)$, $\Z := (Z_1, \ldots, Z_n)$, $m, n \in \N$ are model inputs or \textit{risk factors};
    \item $L$ is the (univariate) random model output, which we typically interpret as a \textit{loss};
    \item Discontinuities emerge at those states where elements of $\X$ cross the  \textit{thresholds} $d_1, \ldots, d_m \in \R$;
    \item The functions  $g_j \colon \R^n \to\R$, $j \in \mM:=\{1, \dots, m\}$ represent the (random) jump of the model output at the points of discontinuity.
\end{itemize}
 \noindent We assume throughout that the marginal distribution functions of $X_j$, $j \in\mM$, and $Z_k$, $k \in \mN:= \{1, \ldots ,n\}$, denoted by $F_j(x) := \P(X_j \le x)$ and $F_{m+k}(z) := \P(Z_k \le z)$, respectively, are absolutely continuous and strictly increasing on their support and denote their corresponding (strictly positive, a.e. on their support) densities by $f_j$ and $f_{m+k}$ respectively. We further denote by $F(l) := \P(L \le l)$ the distribution of the loss $L$. The functions  $g_j \colon \R^n \to\R$, $j \in \mM$ are almost everywhere differentiable and $\P(g_j(\Z) = d) = 0$ for all discontinuity points $d$ of $g_j$. 

A standard example of a discontinuous loss \eqref{eq:loss-model} is a structural model of a credit risk portfolio \citep[e.g.][Ch. 11]{mcneil2015quantitative}, where $\{X_j \le d_j\}$ represents the default event of obligor $j$ and $g_j(\Z)$ the corresponding loss given default. Applications to credit risk modelling are further discussed in Example \ref{ex:CR - 1st model} and Section \ref{sec: numerical example RI}. We consider these types of discontinuity sufficient for practical modelling purposes, since the discontinuities arising more broadly in settings such as financial derivatives, reinsurance contracts, and reliability, can typically be represented through indicator functions of critical events. We note however that the model \eqref{eq:loss-model} is formulated such that $g_j$ are functions of $\Z$ only. We make this assumption throughout the paper to simplify exposition but it is not an essential limitation; our methods work also for the general case of $g_j$ depending on both $\Z$ and $\X$, i.e., for the loss $ L =  \sum_{j \in \mM} g_j(\X, \Z) \Id_{\{X_j \le d_j\}}$, a case treated in the electronic companion, Appendix \ref{app::general-loss-model}.

The risk of a loss is assessed via a risk measure $\rho\colon \mathcal{L}^1 \to \R$, where $\mathcal{L}^1$ denotes the set of integrable random variables on $(\Omega, \mathcal{A}, \P)$. The two most widely used risk measures in practice are the \emph{Value-at-Risk} (VaR) and the \emph{Expected Shortfall} (ES). The VaR at level $\alpha \in [0, 1]$ of the portfolio loss $L$ is defined as the (left-) quantile function of $L$ evaluated at $\alpha$, that is
\begin{equation*}
    \VaR_\alpha(L) := F^{-1}(\alpha)
    = \inf\{ y \in \R\, |\, F(y) \ge \alpha\}\,,
\end{equation*}
with the usual convention that $\inf\emptyset = + \infty$ (e.g., \cite{Embrechts2013MMOR}).
The Expected Shortfall at level $\alpha \in [0, 1)$ of the portfolio loss $L$ is defined by 
\begin{equation*}
    \ES_\alpha (L) := \frac{1}{1 - \alpha} \int_\alpha^1 F^{-1}(u) \, du\,.
\end{equation*}
While we focus on VaR and ES, the sensitivities can be generalised to other quantile-based functionals, such as rank dependent expected utilities or spectral risk measures \citep{acerbi2002spectral} -- in the interest of concision we do not pursue this further.
 
In Section \ref{sub:sec:marginal} we  introduce the \textit{marginal sensitivity measure}, and derive expressions in the context of the VaR/ES risk measures and the discontinuous model \eqref{eq:loss-model}. The sensitivity measure is defined via a partial derivative of a risk measure in the direction of a \emph{stressed} version of a risk factor; hence we first introduce ways of stressing risk factors.

\subsection{Stressing Risk Factors}
 Throughout the paper, we fix the index $i $ of the risk factor with respect to which sensitivity is calculated, such that stresses are applied to either $X_i$, with  $i\in \mM$, or to $Z_i$, with $i\in \mN$. We define a \textit{stress} on $X_i$ or $Z_i$ as a deformation of the risk factor given by
\begin{equation*}
    X_{i, \ep}:= \kappa_\ep(X_i) \,,
    \quad
    \text{respectively}, \quad
    Z_{i, \ep}:= \kappa_\ep(Z_i)\,,
\end{equation*} 
 where $\kappa_\ep\colon \R \to \R$ is a \textit{stress function} defined as follows.

\begin{definition}[Stress Function]\label{def:stress}
A family of functions $\kappa_\ep \colon A \to A$, $A \subseteq \R$, where $\ep  \in [0, +\infty)$, is called a \textit{stress function}, if it satisfies the following properties:
\begin{enumerate}[label  = $\roman*)$]
    \item 
    \label{asm:kappa-invert}
    For all $\ep $ in a neighbourhood of 0, the function $\kappa_\ep(x)$ is invertible in $x \in A$, denoted by $\kappa_\ep^{-1}(\cdot)$;
    
    \item 
    \label{asm:kappa-lim}
    $\displaystyle \lim_{\ep \searrow 0}\;\kappa_\ep(x) = x$, for all $x \in A$;
    
    \item  
    \label{asm:kappa-lim-invert}
    $\displaystyle\lim_{\ep \searrow 0}\;\kappa_\ep^{-1}(x) = x$, for all $x\in A$;

    \item One of the following holds: 
    \label{asm:kappa-ineq}
        \begin{enumerate}[label = $(\alph*)$]
            \item for all $\ep$ in a neighbourhood of 0 and all $x \in A$, it holds that  $\kappa_\ep(x)\ge x$; or \label{asm:kappa-ineq-ge}
            
            \item for all $\ep$ in a neighbourhood of 0 and all $x \in A$, it holds that $\kappa_\ep(x) \le x$;
            \label{asm:kappa-ineq-le}
        \end{enumerate} 
        \label{eq:kappa-ineq}

    \item 
    \label{asm:kappa-mfk}
    $\kappa_\ep(x)$ is differentiable in $\ep$ at $\ep = 0$, and we denote its derivative by
    \begin{equation*}
        \mfk(x):=\displaystyle\lim_{\ep \to 0}\frac{\kappa_\ep(x) - x}{\ep}\,,
        \qquad
        x\in A\,;
    \end{equation*}

    \item 
    \label{asm:kappa-mfk-inv}
    $\kappa_\ep^{-1}(x)$ is differentiable in $\ep$ at $\ep = 0$, and we denote its derivative by
    \begin{equation*}
        \mfkinv(x):=\displaystyle\lim_{\ep \to 0}\frac{\kappa_\ep^{-1}(x) - x}{\ep}\,,
        \qquad 
        x\in A\,.
    \end{equation*}
\end{enumerate}
\end{definition}
We further define
\begin{equation}\label{eq:c-kappa}
    c(\kappa) := 
    \begin{cases}
    +1, \qquad \text{if} \quad \kappa_\ep \text{ fulfils \ref{asm:kappa-ineq} \ref{asm:kappa-ineq-ge}}\,,
    \\
    -1, \qquad \text{if} \quad \kappa_\ep \text{ fulfils \ref{asm:kappa-ineq} \ref{asm:kappa-ineq-le}}\,.
     \end{cases}
\end{equation}

Note that the identity stress function $\kappa(x) = x$ satisfies both $iii)$ $(a)$ and $(b)$, however, it is not of interest as $\mfk(x) =\mfk^{-1}(x)= 0$ and thus the below sensitivities becomes zero.  
The requirements on the stress function are assumptions on its continuity. First, if stressing $X_i$, we typically assume that the domain of the stress function $A$ is equal to the support of $X_i$. This guarantees that $X_i$ and its stressed version $\kappa_\ep(X_i)$ have the same support. Second, properties \ref{asm:kappa-invert} to \ref{asm:kappa-lim-invert} provide that the stressed risk factor converges $\P$-a.s. to its unstressed form as $\ep \searrow 0$. Property \ref{asm:kappa-ineq} means that the stress, e.g.  $X_{i, \ep}$, either approaches $X_i$ $\P$-a.s. from above or below, thus excluding oscillatory behaviour. The last two properties imply that the stress function and its inverse are differentiable, so that the sensitivities, introduced in Sections \ref{sub:sec:marginal} and \ref{sub:sec:cascade}, exist.

Different stress functions may be used, depending on the context of the problem investigated and what type of deformation of a risk factor is interpretable within that context. Some stress functions and related quantities are summarised in Table \ref{tab:stresses}.
\begin{itemize}
    \item \textit{Additive} stresses can be used when the analyst is interested in the impact of a constant shift to the risk factor. This can be interpreted parametrically as a change in the {location} parameter of a distribution, for example the mean or median. 
    \item \textit{Proportional} stresses can be used when the analyst is interested in the impact of a scale change, such as the exposure in a particular financial instrument or loss. One can also see this as a stress on a {scale} parameter of a distribution, for example the standard deviation. An application in the context of credit risk, where the loss given default is proportionally stressed, is given in Example \ref{ex:CR - 1st model}. Proportional stresses also form the basis of Euler-type capital allocation approaches \citep{tasche1999risk}.
    \item \textit{Probability} stresses are useful when one needs to modify the probability of a given -- e.g., a component failure -- event. For example, in a credit risk model the probability of default is a key input. If default is implied by the event $\{X_i\leq d_i\}$ then one can consider a stressed version of the model that specifically increases the probability of this event. The process of using a probability stress in such a context is illustrated in Example \ref{ex:CR - 1st model}. 
    \item \textit{Mixture} stresses are useful it the context of model uncertainty. Essentially, the mixture stress reflects a perturbation of the marginal distribution $F_i$ by an alternative distribution $G$. This is a common device in sensitivity analysis and in Bayesian and robust statistics \citep{glasserman1991gradient,hampel1986robust}. We note that the mixture stress in Table \ref{tab:stresses} can be generalised to distributional stresses by choosing $F_{i,\ep}$ not as a mixture but, e.g. arising via a perturbation of densities, see e.g., \cite{Gauchy2022information} for perturbation of densities using the Fisher Information.
    \item \textit{Tail}  stresses may reflect risk management objectives. Regulatory capital requirements in finance and insurance are generally driven by the tails of probability distributions \citep{mcneil2015quantitative}. Hence, a stress that specifically seeks to alter the tail behaviour of a distribution can be suitable for a financial organisation assessing its capital requirement. We show how this approach can be applied in the numerical study of Section \ref{sec: numerical example RI}.
\end{itemize}
In Table \ref{tab:stresses}, the additive and proportional stresses with $\beta>0$ are such that property  \ref{asm:kappa-ineq} \ref{asm:kappa-ineq-ge} is satisfied and the stress stochastically increases the risk factor; this is easily modified by choosing $\beta<0$. For the mixture and tail stresses both increasing (\ref{asm:kappa-ineq} \ref{asm:kappa-ineq-ge}) and decreasing (\ref{asm:kappa-ineq} \ref{asm:kappa-ineq-le}) versions of the stresses are stated. The functions $\mfk,~\mfkinv$ are easily worked out; some additional detail for mixture stresses is given in the electronic companion, Appendix \ref{app:proofs mixture}.

\begin{table}[thbp]
  \centering
  \caption{Types of stress functions and related quantities.}
  \begin{footnotesize}
    \begin{tabular}{l c c c c}
    \toprule\toprule 
   Type of stress & $\kappa_\ep$ & $\mfk $ & $\mfk^{-1}$ & $c(\kappa)$\\[0.5em]
    \midrule 
   Additive & $x+\beta \ep$  & $\beta$ & $-\beta$ & sgn($\beta$)\\[1em]
   
   Proportional & $x (1 + \beta\ep)$ & $\beta x$ & $ -\beta x$ & sgn($\beta$) \\[1em]
   
    Probability & $F_i^{-1}\big(F_i(x) 
    + \beta \ep\big)$ & $\frac{\beta}{f_i(x)}$ & $\frac{-\beta}{f_i(x)}$ & sgn($\beta$)\\[1.5em]

    \multirow{2}{*}{Mixture}  
     & $F_{i,\ep}^{-1} \circ F_i(x)$, where &  \multirow{2}{*}{$\frac{F_i(x) - G(x)}{f_i(x)} $} &  \multirow{2}{*}{$\frac{G(x) - F_i(x)}{f_i(x)}$}  & \multirow{2}{*}{sgn$\big(F_i(x) - G(x)\big)$}
     \\[0.75em]
   &     $ F_{i,\ep}(x) := (1 - \ep)\,F_i(x) + \ep \,G(x)$ &\\[1.5em]

    \multirow{2}{*}{Tail}  & $ x + \ep \left(x - t\right) \Id_{\{ x \ge t\}}$ & $(x - t)_+$ & $ - (x - t)_+$ & 1 \\[0.5em]
   & $ x + \ep \left(x - t\right) \Id_{\{ x \le t\}} $ & $- (t - x)_+$ & $(t - x)_+$ &  $-1$ \\
       \bottomrule \bottomrule
    \end{tabular}%
      \end{footnotesize}
    \label{tab:stresses}
\end{table}%

\subsection{Marginal Sensitivity}\label{sub:sec:marginal}
For a stress $Z_{i, \ep}$ or a stress $X_{i, \ep}$, we denote the corresponding marginally stressed loss model by, respectively
\begin{align*}
    L_{\ep}(Z_i)
     &:= \sum_{j \in\mM} g_j(\Z_{-i}, Z_{i, \ep}) \,\Id_{\{ X_j \le d_j\}}\,
     \quad \text{and}
     \\
     L_{\ep}(X_i)
     &:= \sum_{\substack{j \neq i \\ j \in\mM}} g_j(\Z) \,\Id_{\{ X_j \le d_j\}} + g_i(\Z) \,\Id_{\{ X_{i, \ep} \le d_i\}}\,,
\end{align*}
where $(\Z_{-i}, Z_{i, \ep})$ is the vector $\Z$ whose $i^\text{th}$ component is replaced by $Z_{i, \ep}$. 
We call $L_\ep$, denoting either $L_\ep(Z_i)$ or $L_\ep(X_i)$, the marginally stressed loss, since only the marginal distribution of $Z_i$ or $X_i$ is altered, leaving all other input factors fixed. We denote by $F_{\ep}(\cdot)$ the distribution function and by $f_\ep(\cdot)$ the density of $L_\ep$ and by $q_{u}(\ep) := F^{-1}_\ep(u)$, $u \in [0,1]$, the quantile function of $L_{\ep}$ evaluated at $u$, for any $\ep \ge 0$. For $\ep=0$, we simply write $F:=F_0$, $f:=f_0$, and $\q := \q(0)$.
For the sensitivities to exist, we require two assumptions on the stressed loss model. 

\begin{assumption}
\label{asm: marginal VaR}
Let $0\le \alpha\le 1$. For all $\ep$ in a neighbourhood of 0 the distribution function $F_\ep$ is continuously differentiable at $F^{-1}(\alpha)$.
\end{assumption}

\begin{assumption}
\label{asm:diff-quantile}
Let $0\le \alpha\le 1$. The quantile function at level $\alpha $ of the stressed loss $L_\ep$, $\q(\ep)$, is differentiable with respect to $\ep$, that is $\frac{\partial}{\partial \ep}\q(\ep)$ exists.
\end{assumption}

\begin{definition}[Marginal Sensitivity]
The \emph{marginal sensitivity} to the risk factor $Z_i$ and $X_i$ for a risk measure $\rho$ is defined by, respectively,
\begin{equation}
\mathcal{S}_{Z_i}\,[\,\rho\,] 
:= \frac{\partial}{ \partial \ep} \,\rho \left(L_\ep(Z_i)\right)\Big|_{\ep = 0}
\quad 
\text{and}
\quad
\mathcal{S}_{X_i}\,[\,\rho\,] 
:= \frac{\partial}{ \partial \ep} \,\rho \left(L_\ep(X_i)\right)\Big|_{\ep = 0}\,,
\end{equation}
whenever the derivatives exists.
\end{definition}

\begin{theorem}[Marginal Sensitivity VaR]\label{thm:VaR:marginal}
Let Assumptions \ref{asm: marginal VaR} and \ref{asm:diff-quantile} be fulfilled for a given $\alpha \in (0,1)$.
Then, the  marginal sensitivity for $\VaR_\alpha$ to input factor $Z_i$ for a stress with stress function $\kappa_\ep$ is
\begin{equation*}
   \S_{Z_i}\,[\,\VaR_\alpha\,]
   =    \sum_{j \in \mM}\left. \E\left[\,  \mfk(Z_i)  \partial_i \,g_j(\Z) \Id_{\{X_j \le d_j\}}   ~\right|~L = \q\,\right]\,,
\end{equation*}
where $\partial_i\, g_j(\z):= \frac{\partial}{\partial z_i}g_j(\z)$ is the partial derivative in the $i^\text{th}$ component. 
The marginal sensitivity to input factor $X_i$ is given by 
\begin{equation*}
   \S_{X_i}\,[\,\VaR_\alpha\,]
   = c(\kappa) \mfkinv(d_i)\,\frac{f_i(d_i)}{f\left(\q\right)}\;\;  \E\left.\left[\left( \Id_{\left\{L \le \q +  c(\kappa)g_i(\Z)\,\right\}}
      - \Id_{\{L \le \q\}}\right) ~\right|~X_i = d_i\right]\,.
\end{equation*}
\end{theorem}

\begin{theorem}[Marginal Sensitivity ES]\label{thm:ES:marginal}
Let Assumptions \ref{asm: marginal VaR} and \ref{asm:diff-quantile} be fulfilled for a given $\alpha \in (0,1)$.
Then, the marginal sensitivity for $\ES_\alpha$ to input factor $Z_i$ for a stress with stress function $\kappa_\ep$ is
\begin{equation*}
    \S_{Z_i}\,[\,\ES_\alpha\,]
    = \sum_{j \in \mM}
     \E\left[\,\mfk(Z_i)\,
    \partial_i \,g_j(\Z)\Id_{\left\{X_j \le d_j\right\}}~|~L \ge \q \right]\,.
\end{equation*}
The marginal sensitivity to input factor $X_i$  for a stress with stress function $\kappa_\ep$ is
\begin{equation*}
   \S_{X_i}\,[\,\ES_\alpha\,]
    =
    \frac{-c(\kappa) \mfkinv(d_i)\, f_i(d_i)}{1 - \alpha}\left.\E\left[\big(L - c(\kappa) g_i(\Z)\,-  \q\big)_+ -   \big(L -  \q\big)_+ ~\right|~X_i = d_i \right]\,.
\end{equation*}
\end{theorem}
The marginal sensitivity measures to $Z_i$ for both the $\VaR$ and $\ES$ generalise the sensitivities derived in \cite{Hong2009OR, Hong2009MS} to loss functions $L$ that are not Lipschitz continuous and to general types of stresses. Related, \cite{Fu2009MS} proposes a conditional Monte-Carlo approach to estimate quantile sensitivities. Note however, that their key assumption is that the perturbed distribution function of $L_\ep$ can be written as $F_{L_\ep}(t) = \E[G(t, \ep, Y(\ep))]$, where $G$ is $\P$-a.s. continuous w.r.t. $\ep$ and $Y(\ep)$ is an arbitrary random variable. This assumption does not hold in our setting as can be seen in, e.g., Equation \eqref{eq:maginal-VaR-deriv-ep} of the Proof of Theorem \ref{thm:VaR:marginal}. Furthermore, one could derive the marginal sensitivities of $\ES_\alpha$ -- as well as those of other spectral risk measures \citep{acerbi2002spectral} -- using its representation as the integral of $\VaR_\alpha$. Interchanging the limit and the integral, however, requires that the sensitivities for $\VaR_\beta$ to exist, for all $\beta \in [\alpha, 1)$. This would imply that Assumptions \ref{asm: marginal VaR} and \ref{asm:diff-quantile} need to hold for all $\beta \in (\alpha,1)$, which is in contrast to Theorem \ref{thm:ES:marginal} that requires Assumptions \ref{asm: marginal VaR} and \ref{asm:diff-quantile} to hold for $\alpha$ only.

We now provide an expression for the marginal sensitivity of the mean. While this could be obtained as a special case of Expected Shortfall with $\alpha=0$, it is simpler to derive Corollary \ref{cor-sens-mean} as a direct consequence of Lemma \ref{lemma:dirac-marginal} in Appendix \ref{app:proofs}.

\begin{corollary}[Marginal Sensitivity Mean]\label{cor-sens-mean}
Let $\kappa_\ep$ be a stress function,  then the marginal sensitivity for the mean $(\E)$ to input factor $Z_i$ respective $X_i$ are
\begin{align*}
    \S_{Z_i}\,[\,\E\,]
    &= \sum_{j \in \mM}
     \E\left[\,\mfk(Z_i)\,
    \partial_i \,g_j(\Z)\Id_{\left\{X_j \le d_j\right\}}\right]\, 
    \quad \text{and}
    \\[0.5em]
       \S_{X_i}\,[\,\E\,]
    &=
     \mfkinv(d_i)\, f_i(d_i)\left.\E\left[ \,g_i(\Z)~\right|~X_i = d_i \right]\,.
\end{align*}    
\end{corollary}

We conclude the section with an example of how the marginal sensitivity measure can be applied in the context of a standard portfolio credit risk model, with two different analysis objectives in mind.

\begin{example}\label{ex:CR - 1st model}
    Consider a credit risk setting, where $\Z$ has the same dimension as $\X$, with $g_j(\Z)=Z_j,~j\in \mathcal M$ representing the loss given default and ${\{X_j\le d_j\}}$ the  default events with corresponding probabilities $F_j(d_j)$. Hence we have that
    $$
    L=\sum_{j \in \mathcal M} Z_j\Id_{\{X_j \le d_j\}}\,.
    $$

    A first analysis pertains to the calculation of the sensitivity of the portfolio ES with respect to the probability of the $i$-th default event. To achieve this, we need to formulate an appropriate stress function. Consider the probability stress from Table \ref{tab:stresses},
    $
        \kappa_\ep(x)=F_i^{-1}(F_i(x)-\ep),
    $
    leading to
    \begin{align*}
        X_{i,\ep} &= F_i^{-1}(F_i(X_i)-\ep) \quad \text{and}\\
        \P(X_{i,\ep} \le d_i)&=\P \left(F_i(X_i) \le F_i(d_i)+\ep\right)= F_i(d_i)+\ep.
    \end{align*}
    Hence the chosen stress function gives an additive stress on the default probability, such that the sensitivity $   \S_{X_i}\,[\,\ES_\alpha\,]$ becomes precisely the derivative of the portfolio risk in direction of the default probability of the $i$-th obligor. Using $c(\kappa)=-1$ and $\mfkinv(x)  =\frac{1}{f_i(x)}$; 
 Theorem \ref{thm:ES:marginal} yields:
 \begin{align*}
      \S_{X_i}\,[\,\ES_\alpha\,]
    &=
    \frac{1}{1 - \alpha}\left.\E\left[\big(L - (\q-Z_i)\big)_+ -   \big(L -  \q\big)_+ ~\right|~X_i = d_i \right].
 \end{align*}
The resulting sensitivity can thus be understood as the difference between two expectations, each representing the excess portfolio loss over a threshold, conditioned on the least adverse outcome of $X_i$ that gives a default of the $i$-th obligor. The difference between the two terms lies in the lower threshold used in the first term, which is reduced by the loss given default $Z_i$.

Second, we consider the sensitivity to a proportional increase  in the loss given default $Z_i$, that is, using $\kappa_\ep(z) = z(1+\ep)$. Application of  Theorem \ref{thm:ES:marginal} then gives us:
$$
    \S_{Z_i}\,[\,\ES_\alpha\,]
    =
     \E\left[\,Z_i\Id_{\left\{X_i \le d_i\right\}}~|~L \ge \q \right]\,.
$$
Note that this is precisely the Euler allocation of the risk $\ES_\alpha(L)$ to the loss $Z_i\Id_{\left\{X_i \le d_i\right\}}$ to the $i$-th obligor \citep{tasche1999risk}. 

Finally, within the same model, we turn our attention to assessment of the relative importance of common factors that drive dependence between defaults. The dependence of the critical variables $X_j$ is often modelled via factor models \citep[][Ch.~6.4,~11]{mcneil2015quantitative} and a question of interest is the relative importance of underlying factors for portfolio risk. Consider the following representation
$$
X_j :=\sum_{t= 1}^\tau \beta_{j,t}W_t+V_j,\quad j\in \mathcal{M},
$$
where $W_t,~t = 1, \ldots, \tau$, $\tau \in \N$, are the common factors, and $V_j$ are idiosyncratic error terms. We are interested in the sensitivity of the portfolio loss to the factor $W_s$. To that effect, define:
\begin{align*}
    \tilde X_{j,\ep} & := \sum_{t\neq s} \beta_{j,t}W_t+\beta_{j,s}(W_s-\ep)+V_j=X_j-\beta_{j,s}\ep\,,\\
    \tilde \kappa_{\ep}(x) &:=x-\beta_{j,s}\ep\,,\\
    \tilde L_\ep&:= \sum_{j \in \mathcal{M} } Z_j\Id_{\{\tilde\kappa_{\ep}(X_j) \le d_j\}}\,.
\end{align*}
The sensitivity of the portfolio risk to the factor $W_s$ can then be written as
\begin{align*}
    \left. \frac{\partial}{\partial \ep} \ES_\alpha(\tilde L_\ep) \right|_{\ep=0} = \sum_{j\in \mathcal M}S_{X_j}[\ES_\alpha]\,,
\end{align*}
where the sensitivities $S_{X_j}[\ES_\alpha]$ are now calculated with the stress functions     $\tilde \kappa_{\ep}$ above. Applying again Theorem \ref{thm:ES:marginal} leads to
\begin{align*}
    \left. \frac{\partial}{\partial \ep} \ES_\alpha(\tilde L_\ep) \right|_{\ep=0}=\sum_{j\in \mathcal M}\frac{f_j(d_j)\beta_{j,s}}{1 - \alpha}\left.\E\left[\big(L - ( \q-Z_j)\big)_+ -   \big(L -  \q\big)_+ ~\right|~X_i = d_i \right].
\end{align*}
Hence, intuitively, the sensitivity to the common factor $W_s$ is expressed as sum of sensitivities for each obligor, weighted by the factor loadings $\beta_{j,s}$.  
\end{example}

\section{Measuring Cascading Effects}\label{sub:sec:cascade}

The marginal sensitivity introduced in Section \ref{sub:sec:marginal} quantifies the differential impact of stressing a risk factor on the portfolio loss. Here, we provide the first generalisation/adjustment of the framework, by considering \textit{cascade sensitivity} measures, introduced in  \cite{Pesenti2021RA}. These sensitivity measures quantify not only the sensitivity to an individual input $X_i$, but also consider (joint) perturbation of  all other risk factors $X_j$, $j \neq i$, and $Z_k$, $k\in \mN$, induced by their statistical dependence on $X_i$. This is achieved by using the \emph{inverse Rosenblatt transform}, recalled next \citep{Rosenblatt1952, Ruschendorf1993}.

\begin{definition}[Inverse Rosenblatt Transform]\label{def:invRosen}
An inverse Rosenblatt transform of an $r$-dimensional random vector $\Y$, starting at $Y_i$, for fixed $i \in \{1,\dots,r\}$, is given by a function $\bPsi =  (\Psi^{(1)}, \ldots, \Psi^{(r)})^\top \colon \mathbb{R}^{r} \to \mathbb{R}^{r}$ and an $(r-1)$-dimensional random vector $\V = (V_1, \ldots, V_{r -1})$, consisting of independent standard uniform variables, independent of $Y_i$, such that
\begin{align*}
    \Y
    =  \bPsi \left( Y_i, \V\right)
    = \left(\Psi^{(1)}( Y_i, \V), \,\ldots ,\, \Psi^{(r)}( Y_i, \V)\right)\,, \quad \P\text{-a.s.}\,.
\end{align*}
In particular, $Y_k = \Psi^{(k)}( Y_i, \V)$ $\P$-a.s. for all $k \in \{1,\dots,r\}$.
\end{definition}

\begin{remark}Before proceeding with the definition of sensitivity measures in our specific model context, we provide some comments on the construction of Definition \ref{def:invRosen}. For further references on dependency models and the Rosenblatt transform in sensitivity analysis see, e.g., \cite{Mara2012variance,Mai2015analyzing,Pesenti2021RA,lamboni2021RESS}.
\begin{itemize}
    \item The key idea is that one can represent the random vector $\Y$ as a function of an independent vector starting at $Y_i$, the variable being stressed. Here the assumption of $\V$ being uniform is not material and an alternative distribution could be chosen. Assuming that $\V$ is uniform links to the \textit{standard construction} \citep{Ruschendorf1993}, where, e.g., for $i=1$, we have $  \Psi^{(1)}( Y_1, \V)=Y_1,~\Psi^{(2)}( Y_1, \V)= F_{Y_2|Y_1}^{-1}(V_1|Y_1),\dots,\Psi^{(r)}( Y_1, \V)=F_{Y_r|Y_1,\dots,Y_{r-1}}^{-1}(V_{r-1}|Y_1,\dots, Y_{r-1})$.
     This process is simplified for specific parametric models. For example, if $\Y$ is multivariate normal, then one can let $\V$ consist of independent standard normal variables and the functions $\Psi^{(k)}$ are linear; see Example 3 in \cite{Pesenti2021RA} for more details.  
    \item More generally, it is typical for dependent random vectors $\Y$ to be written as functions of independent random vectors $\boldsymbol{W}$ by some transformation $\Y=\widetilde  \bPsi (\boldsymbol{W})$.  Indicatively, this holds for elliptical (e.g., multivariate t) distributions and Archimedean copulas and in such cases the dimension of $\boldsymbol W$ is $r+1$, see \cite{mcneil2015quantitative}, Sections 6.3 and 7.4 respectively. Furthermore, none of the elements of the independent vector  $\boldsymbol{W}$ can be identified with an element $Y_i$ of $\Y$, which is to be stressed. Hence, these representations of random vectors are generally not  suitable for the purposes of this paper. Of course, if the independent components of $\boldsymbol W$ are interpretable in their own right, then they can be marginally stressed, following the reasoning of Section \ref{sub:sec:marginal}.
\end{itemize}
    \end{remark}

We now return to our setting, where $\Y=(\X, \Z)$ with dimension $r=m+n$. For simplicity, we write for the inverse Rosenblatt transform starting at $X_i=Y_i,~i\in \mM$,   such that $(\X, \Z)=\bPsi(X_i, \V)$,  $X_j = \Psi^{(j)}(X_i, \V)$, for all $j \in\mM$, and $Z_k = \Psi^{(m+k)}(X_i, \V)$ for all $k \in \mN$. 
To construct the cascade sensitivity to input $X_i$, we  replace $X_i$ by $X_{i, \ep}$, such that the stressed vector of risk factors becomes $\bPsi(X_{i, \ep}, \V)$. Thus, using the inverse Rosenblatt transform, all other risk factors are perturbed according to their dependence on $X_i$ and the portfolio loss is transformed to:
\begin{align}\label{eq:stressed-cascade-loss-X}
 L^{\bPsi}_{\ep}(X_i)
    &:= \sum_{j \in\mM} g_j\left(\{\Psi^{(m+k)}(X_{i, \ep}, \V)\}_{k\in\mN} \right) \,\Id_{\{ \Psi^{(j)}( X_{i, \ep}, \V) \le d_j\}}\,.  
\end{align}
Subsequently, to derive the cascade sensitivity measure, we apply the marginal sensitivity to the stressed portfolio loss $L^{\bPsi}_{\ep}(X_i)$. In \eqref{eq:stressed-cascade-loss-X}, $\{\Psi^{(m+k)}(X_{i, \ep}, \V)\}_{k\in\mN}$ is the part of the stressed input vector returning $\Z$, potentially impacted by the stress on $X_{i}$.

When stressing $Z_i=Y_{m+i}$, $i\in\mN$, the respective transform is $(\X, \Z)=\bPsi(Z_i, \V)$.
The process of stressing $Z_i$
is analogous to the case of $X_i$, with the transformed portfolio loss now given by  
\label{eq:stressed-cascade-loss-Z}
\begin{align}
    L^{\bPsi}_{\ep}(Z_i)
     &:= \sum_{j  \in\mM} g_j\left(\{\Psi^{(m+k)}(Z_{i, \ep}, \V)\}_{k\in\mN}\right) \,\Id_{\{ \Psi^{(j)}( Z_{i, \ep}, \V) \le d_j\}},
\end{align}
where, similarly to the case of stressing $X_i$, we have that  $\{\Psi^{(m+k)}(Z_{i}, \V)\}_{k\in\mN}$ returns a version of $\Z$ deformed by the stress on $Z_i$.

With these building blocks in place, we can now define the cascade sensitivity measure in the specific context of this paper.
\begin{definition}[Cascade Sensitivity]\label{def:cascade sens}
The \emph{cascade sensitivity} to the risk factor $Z_i$ and $X_i$ for a risk measure $\rho$ is defined by, respectively,
\begin{equation}
\mathcal{C}_{Z_i}\,[\,\rho\,] 
:= \frac{\partial}{ \partial \ep} \,\rho \big(\,L^{\bPsi}_\ep(Z_i)\, \big)\Big|_{\ep = 0}\,,
\quad  \text{and}\quad  
\mathcal{C}_{X_i}\,[\,\rho\,] 
:= \frac{\partial}{ \partial \ep} \,\rho \big(\,L^{\bPsi}_\ep(X_i)\, \big)\Big|_{\ep = 0}\,,
\end{equation}
whenever the derivatives exist.
\end{definition}

Note that if the cascade sensitivity is exists, it is independent of the choice of Rosenblatt transform, see Prop. 3.6 in \cite{Pesenti2021RA}. In order to establish existence, in this section we make the assumption that the inverse Rosenblatt transforms are differentiable and locally monotone in their first argument. This means that stressing a model input leads to perturbation of elements of $\X$ that makes them $\P$-a.s. greater (or smaller) than the original input $X_j$.

\begin{assumption}\label{asm:psi-combined}
Let $\kappa_\ep$ be a stress function and $Y,~Y_{\ep}$ be such that either  $Y:=Z_i,~Y_{i,\ep}:=Z_{i,\ep}$ or $Y:=X_i,~Y_{i,\ep}:=X_{i,\ep}$. Let $\bPsi$ be a {differentiable} inverse Rosenblatt transform starting at $Y$, such that $(\X, \Z) = \bPsi(Y, \V)$. Then, for each $j \in \mM$, one of the following holds
\begin{enumerate}[label = $(\alph*)$]
    \item \label{asm:psi-ineq-ge-Z}
    for all $\ep$ in a neighbourhood of 0, it holds $\Psi^{(j)}\left(Y_{i, \ep}, \V\right)
    \ge 
    X_j$ $\P$-a.s.; or 
    \item \label{asm:psi-ineq-le-Z}
    for all $\ep$ in a neighbourhood of 0,  it holds $\Psi^{(j)}\left(Y_{i, \ep}, \V\right)
    \le 
    X_j$ $\P$-a.s.
\end{enumerate}
In the case (a) we denote $c(\kappa; \;j)=1$ and in the case (b) $c(\kappa; \;j)=-1$.
\end{assumption}

With these assumptions in place, we can now obtain explicit formulas for the cascade sensitivity measure of Definition \ref{def:cascade sens}. In Theorems \ref{thm:ES:cascade} and \ref{thm:ES:cascade-Zi} below we deal with the case of $\ES$, while formulas for $\VaR$ are given in Appendix \ref{app:cascade}. We observe that the cascade sensitivity to both $X_i$ and $Z_i$ entails a decomposition, reflecting the indirect contribution of the vector being stressed via the other inputs $X_j,~Z_k$.

\begin{theorem}[Cascade Sensitivity ES to $X_i$]\label{thm:ES:cascade}
Let Assumptions \ref{asm: marginal VaR}, \ref{asm:diff-quantile} and \ref{asm:psi-combined} (for $Y=X_i$) be fulfilled for the stressed model $L^{\bPsi}_\ep(X_i)$ and $\alpha \in (0,1)$. Denote $\Psi_1^{(j)}(x, \bv): = \frac{\partial}{\partial x}\Psi^{(j)}(x, \bv)$. Then, the cascade sensitivity for $\ES_\alpha$ to input $X_i$ is given by
\begin{equation}\label{eq:cascade-xi-es-decomp}
    \C_{X_i}\,[\,\ES_\alpha\,]
    = 
    \sum_{j \in\mM} \C_{X_i,X_j}
    + 
    \sum_{k \in\mN} \C_{X_i,Z_k},
 \end{equation}
where, for all $k \in\mN$,
\begin{equation*}
   \C_{X_i,Z_k}
   = 
    \sum_{j \in\mM}  \left.\E\left[\mfk(X_i)\, \partial_k\, g_j(\Z) \Psi_1^{(m + k)}(X_i, \V)  \Id_{\{X_j \le d_j\}} ~\right|~L \ge \q \right]
   \,,
\end{equation*}
and for all $j \in\mM$,
\begin{small}
\begin{equation*}
   \C_{X_i,X_j}
   = 
   -  \left.\frac{c(\kappa;\, j)f_j(d_j)}{1  - \alpha} \E\left[\mfkinv(X_i) \Psi_1^{(j)}(X_i, \V)
    \left(\big(L 
    -  c(\kappa;\, j) g_j(\Z)\,-  \q\big)_+ - \big(L -  \q\big)_+\right)
    \,\right|\, X_j = d_j\right]\,.
\end{equation*}
\end{small}
\end{theorem}

\begin{theorem}[Cascade Sensitivity ES to $Z_i$]\label{thm:ES:cascade-Zi}
Let Assumptions \ref{asm: marginal VaR}, \ref{asm:diff-quantile} and \ref{asm:psi-combined} (for $Y=Z_i$)  be fulfilled for the stressed model $L^{\bPsi}_\ep(Z_i)$ and $\alpha \in (0,1)$. Then, the cascade sensitivity for $\ES_\alpha$ to input $Z_i$ is given by
\begin{align*}
    \C_{Z_i}\,[\,\ES_\alpha\,]
    = 
    \sum_{j \in\mM} \C_{Z_i,X_j}
    + 
    \sum_{k \in\mN} \C_{Z_i,Z_k},
 \end{align*}
where, for all $k \in\mN$,
\begin{equation*}
   \C_{Z_i,Z_k}
   = 
    \sum_{j \in\mM}  \E\big[\mfk(Z_i)\, \partial_k\, g_j(\Z) \Psi_1^{(m + k)}(Z_i, \V)  \Id_{\{X_j \le d_j\}} ~\big|~L \ge \q \big]
   \,,
\end{equation*}
and for $j \in\mM$,
\begin{small}
\begin{equation*}
   \C_{Z_i,X_j}
   = 
   -  \frac{c(\kappa;\, j)f_j(d_j)}{1  - \alpha} \left.\E\left[\mfkinv(Z_i) \Psi_1^{(j)}(Z_i, \V)
    \left(\big(L 
    -  c(\kappa;\, j) g_j(\Z)\,-  \q\big)_+ - \big(L -  \q\big)_+\right)
    \,\right|\, X_j = d_j\right]\,.
\end{equation*}
\end{small}
\end{theorem}

To calculate the cascade sensitivities, we need the derivative of the inverse Rosenblatt transform. This calculation is simplified by noting that the value of the cascade sensitivity is independent of the specific choice of Rosenblatt transform. Hence, when calculating, for example,  $\Psi_1^{(j)} (X_i, \V)$, we can without loss of generality use the standard construction \citep{Ruschendorf1993} $\Psi^{(j)} (X_i, \V) = F_{X_j|X_i}^{-1}(V_1|X_i)$  in Theorem \ref{thm:ES:cascade} -- analogously if $Z_i$ is being stressed (Theorem \ref{thm:ES:cascade-Zi}). As a result, it is sufficient to consider the derivatives of inverse Rosenblatt transforms corresponding to the bivariate dependence structure of, e.g., $(X_i,X_j)$. If the bivariate copula between the risk factors are known, analytical expressions for the required derivatives  may be available. We refer also to \cite{Pesenti2021RA}, where the formulas given below for the Gaussian and t copulas are derived.

For simplicity of presentation, we only provide the expressions for $\Psi_1^{(j)} (X_i, V)$, where $V$ is a suitably defined random variable such that $X_j=\Psi^{(j)} (X_i, V) $.
The formulas for $\Psi_1^{(j)} (Z_i, V)$, $\Psi_1^{(m+k)} (X_i, V)$, and $\Psi_1^{(m+k)} (Z_i, V)$, for $j \in \mM$, $k\in \mN$ follow analogously.

\begin{proposition}[Bivariate Inverse Rosenblatt Transform]
\label{prop: inv-rosenblatt}
Denote by $\Phi, ~\phi$, the distribution function and density of a standard normal variable, and by $t_\nu,~s_\nu$ the distribution function and density of a t-distributed random variable with $\nu$ degrees of freedom. 
\begin{enumerate}
    \item Assume $(X_i,X_j)$ follows a Gaussian copula with correlation parameter $r_{ij}$ and define $Y_i:=\Phi^{-1}(F_i(X_i))$ and $Y_j:=\Phi^{-1}(F_j(X_j))$. Then, 
\begin{equation*}
        \Psi_1^{(j)} (X_i, V)
        = r_{ij}\;\frac{f_i (X_i )}{\phi\left(Y_i\right)}\, \frac{ \phi\left(Y_j\right)}{f_j(X_j)}\,,
    \end{equation*}

    \item Assume $(X_i,X_j)$ follows a t copula with correlation parameter $r_{ij}$ and $\nu$ degrees of freedom and define  $Y_i:=t_\nu^{-1}(F_i(X_i))$ and $Y_j:=t_\nu^{-1}(F_j(X_j))$. Then, 
    \begin{equation*}
    \Psi_1^{(j)} (X_i, V) 
    =
    \left(r_{ij} + \frac{Y_i \, Y_j  - r_{ij} Y_i^2}{\nu + Y_i^2}\right)
    \frac{f_{i}(X_i)}{s_\nu(Y_i)}\frac{s_\nu(Y_j)}{f_{j}(X_j)} \,. 
    \end{equation*}

    \item Assume $(X_i,X_j)$ follows a Archimedean copula with generator $\psi\colon [0, + \infty] \to [0,1]$, i.e., the copula is given by
\begin{equation*}
    \Cop (u_1, u_2) = \psi\left(\psi^{-1}(u_1) + \psi^{-1}(u_2)\right) \,,\quad u_1, u_2 \in [0,1]\,,
\end{equation*}
where $\psi^{-1}$ denotes the inverse of the generator $\psi$. Then, for $i \neq j$
    \begin{equation*}
    \Psi_1^{(j)} (X_i, V)
    = 
    \frac{\dot{\psi}\left(\psi^{-1}(U_j)\right) }{\dot{\psi}\left(\psi^{-1}(U_i)\right)} 
        \left( \frac{\dot{\psi}\left(\psi^{-1}(U_i) + \psi^{-1}(U_j)\right)}{\ddot{\psi}\left(\psi^{-1}(U_i) + \psi^{-1}(U_j)\right)} \; \frac{\ddot{\psi}\left( \psi^{-1}(U_i)\right)}{\dot{\psi}\left( \psi^{-1}(U_i) \right)}
        - 1 \right)
        \frac{f_i(X_i) }{f_{j}(X_j)}\,,
    \end{equation*}
    where $U_j := F_j(X_j)$, $\dot{\psi}(x) := \frac{\partial}{\partial x} \psi(x)$, and $\ddot{\psi}(x) := \frac{\partial}{\partial x} \dot{\psi}(x)$.
\end{enumerate}
\end{proposition}

\section{Sensitivity to Discrete Random Variables}\label{sec:discrete}
In this section, we adapt the techniques developed so far, to calculate differential sensitivities to discrete risk factors. Given the different portfolio structure we consider here, we change notation to avoid confusion with previous sections. We consider the loss model 
\begin{equation}\label{eq:loss-discrete}
 T := h\left(W,\,\Y \right)   \,,
\end{equation}
where $\Y := (Y_1,\dots,Y_d)$, the function $h\colon \R^{d+1} \to \R$ is differentiable, and $W$ is a discrete random variable which sensitivity we aim to assess. Such a sensitivity calculation presents both technical and conceptual challenges. While $h$ is differentiable, the corresponding differential (or infinitesimal increment) in its first argument is hard to interpret, given the discreteness of $W$. Indeed when assessing sensitivities to discrete random variables, the realisations of $W$ are typically exogenously given, thus a stress on $W$ should manifests itself through perturbations on the probabilities. Here we propose to calculate the differential sensitivity with respect to a continuous variable, from which $W$ is obtained via a (discontinuous) transformation. In other words, we exchange the problem of discreteness with the one of non-differentiability, which we have established. 

We assume that $W$ takes values $w_1 < \cdots < w_r$ with $\P(W \le w_k) = p_k$, $k = 1, \ldots, r$, such that $0 =:p_0 < p_1< \cdots < p_r = 1$. As we propose to perturb the probabilities $p_k$ of $W$ while keeping the realisations $w_k$ and thus the support of $W$ fixed, we rewrite the loss model \eqref{eq:loss-discrete} into a form analogous to \eqref{eq:loss-model}. For this, let $V \sim \mathrm U(0,1)$ be independent of $(W,\,\Y)$ and define the uniform random variable $U$ by
\begin{equation*}
    U  := \tilde{F}_{W}(W; V)\,,
\end{equation*}
where $\tilde{F}_{W}(w;\lambda):=\P(W < w) + \lambda \P(W = w)$, $\lambda \in [0,1]$, is the generalised distributional transform of $W$ \citep{Ruschendorf2013Springer}. It then follows that $U \sim \mathrm U(0,1)$, $U$ is comonotonic to $W$, and
\begin{align*}
    W 
     = F_W^{-1}(U)
    = 
    \sum_{k = 1}^r w_k \Id_{\{p_{k-1}< U \le p_k\}}\,, \quad  \P\text{-a.s.}\,.
\end{align*}
Then, following some manipulations, the loss model admits the form:
\begin{equation*}
    T
    =
 \sum_{k = 1}^r h\left( w_k , \Y\right)\Id_{\{p_{k-1}< U \le p_k\}}
    = 
    \sum_{k = 1}^r
    \Delta_k \,h\left(W, \Y\right) \Id_{\{ U \le p_k\}}\,,
\end{equation*}
where $\Delta_k  \,h\left(W, \Y\right) := h\left( w_k , \Y\right) - h\left( w_{k+1} , \Y\right)$, for $k = 1, \ldots r-1$, and $\Delta_r h\left( W,\Y\right):= h(w_r, \Y)$.

We next stress the portfolio loss $T$ with respect to $W$ by applying a stress function to $U$. Note that by stressing $U$ instead of $W$, we perturb the distribution without altering the support of $W$. Moreover, as $U$ and $W$ are comonotonic, a stress on $U$ is by construction also a stress on $W$, which does not change the dependence to other risk factors. Hence, we write the stressed model as 
\begin{equation}\label{eq:loss-discrete-stress}
    T_{W,\ep}
    :=
    \sum_{k = 1}^r \, \Delta_k \,h\left( W,\Y\right)\Id_{\{\kappa_\ep(U) \le p_k\}}\,.
\end{equation}
Stressing the uniform variable that generates $W$ allows for a cohesive stress, given the comonotonicity of $(W,U)$. 
Next, we define the differentiable sensitivity to $W$ via a stress on $U$ by:
\begin{equation*}
    \tilde{\S}_{W}\,[\,\rho\,]
    :=
    \frac{\partial}{\partial \ep } \rho(T_{W,\ep}) \Big|_{\ep = 0}\,.
\end{equation*}
Formulas for this sensitivity are given in the following result.

\begin{theorem}[Marginal Sensitivity -- Discrete]\label{thm:VaR:marginal-discrete}
Let Assumptions \ref{asm: marginal VaR} and \ref{asm:diff-quantile} be fulfilled for the loss model \eqref{eq:loss-discrete} and for a fixed $\alpha \in (0,1)$.
Then the sensitivity for VaR to the discrete input $W$ is 
\begin{equation*}
   \tilde{\S}_{W}\,[\,\VaR_\alpha\,]
   = 
   \frac{c(\kappa)}{f(\q)} \sum_{k = 1}^r \mfk^{-1}(p_k) \E\left[ \left(
       \Id_{\{T \le \q + c(\kappa)  \Delta_k\,h(W,\Y) \}}
    -\Id_{\{T \le \q\}}
    \right)~|~ W = w_k \right]\,,
\end{equation*}
where, for simplicity of notation, $\q$ is the $\alpha$-quantile of $T$ and $f$ its density.
The sensitivity for ES to the discrete input $W$ is
\begin{equation*}
   \tilde{\S}_{W}\,[\,\ES_\alpha\,]
   = 
   -\frac{c(\kappa) }{1-\alpha}\sum_{k =1 }^r
    \mfk^{-1}(p_k) 
    \E\left[
     \Big( T -c(\kappa) \Delta_k \,h(W,\Y)  - \q\Big)_+ - (T-\q)_+
    ~|~ W =w_k
    \right]\,.
\end{equation*}

\end{theorem}

We now present an application of Theorem \ref{thm:VaR:marginal-discrete} for the ES-sensitivity calculation of the frequency and severity variables in a compound loss model. Compound distributions are canonical tools in modelling insurance claims, as well as credit and operational risk losses, and the impact of the choice of frequency distribution is well attested, see e.g., \cite{mcneil2015quantitative}. To this effect, we represent by $T=h(W,\Y)$ a compound random variable. Specifically, we set $r=d+1$ and assume that $W$ is a discrete loss frequency, taking values in $\{w_1=0,\dots,w_{d+1}=d\}$, while the $d$ elements of $\Y=(Y_1,\dots,Y_d)$ are loss severities. 
The variable $W$ has distribution $\P(W \leq k-1)=p_k,~k=1,\dots,d+1$. Furthermore, we assume that $Y_1,\dots,Y_{d}$ are i.i.d., continuously distributed with $Y_1\sim F_Y$, and independent of $W$. The portfolio loss is:
$$
T=h(W,\Y)=\sum_{\ell=1}^{W}Y_\ell\,,
$$
with the understanding that for $W=0$ we have  $T=0$. Our aim is to calculate the sensitivity of the portfolio's ES to the frequency variable, i.e. to evaluate the quantity $\tilde{\S}_{W}\,[\,\ES_\alpha\,]$, and to compare this with the impact of the vector of loss severities, $\tilde{\S}_{\Y}\,[\,\ES_\alpha\,]$, which are defined below.

The sensitivity $\tilde{\S}_{W}\,[\,\ES_\alpha\,]$ is evaluated by application of Theorem \ref{thm:VaR:marginal-discrete}. As before let $W=F_W^{-1}(U)$. To stress $U$ we need to specify a stress function $\kappa_\ep(u):(0,1) \to (0,1)$. Let 
$
\kappa_\ep(u) := \Phi\left(\Phi^{-1}(u) + \ep\right),
$
where $\Phi$ is the standard normal distribution. This choice is consistent with the well-known \textit{Wang Transform} \citep{wang2000class} in risk measure theory and satisfies the conditions of Definition \ref{def:stress}, with $c(\kappa)=1$.
Then, for $U_{\ep}:=\Phi\left(\Phi^{-1}(U) + \ep\right)$ we obtain, using Theorem \ref{thm:VaR:marginal-discrete} and
after a few manipulations not documented here, that
\begin{align*}
\tilde{\S}_{W}\,[\,\ES_\alpha\,]
  = \sum_{k=1}^{d} \big(v(p_{k})-v(p_{k+1})\big)  \,\E\Big[
     \Big( \sum_{\ell=1}^{k} Y_\ell   - \q\Big)_+ 
    \Big]\,,
\end{align*}
where $ v(p):=\frac{\phi\left(\Phi^{-1}(p)\right)}{1-\alpha}$, $p \in [0,1]$, and $\phi$ is the density of the standard normal distribution. Hence, the sensitivity becomes a linear combination of the stop-loss terms $\E[
     ( \sum_{\ell=1}^{k} Y_\ell   - \q)_+] $, with the coefficient weights driven by the distribution of the loss frequency $W$. 

We now turn our attention to stressing the loss severities $\Y$. We choose to stress all elements of $\Y$ at the same time, using a stress function consistent with the one used for $W$, i.e. the same $\kappa_\ep$. Specifically, for $U_\ell  :=F_Y(Y_\ell),~\ell =1,\dots,d$, we define the stressed portfolio
\begin{align*}
    T_{\Y,\ep}&:=\sum_{\ell=1}^WF_Y^{-1}\left(\kappa_\ep(U_\ell)\right)=\sum_{k=1}^d \Id_{\{W=k\}}\sum_{\ell=1}^kF_Y^{-1}\left(\kappa_\ep(U_\ell)\right)\,
\end{align*}
and calculate the sensitivity
$$
\tilde{\S}_{\Y}\,[\,\ES_\alpha\,]:=
\frac{\partial }{\partial \ep}\ES_{\alpha}(T_{\Y,\ep})\Big|_{\ep=0}\,.
$$
By the pointwise continuity of the mapping $\ep\mapsto T_{\Y,\ep}$ we can calculate $\tilde{\S}_{\Y}\,[\,\ES_\alpha\,]$ by standard methods \citep{Hong2009MS}, yielding:
$$
\tilde{\S}_{\Y}\,[\,\ES_\alpha\,]:=\sum_{k=1}^d \P(W=k)\sum_{\ell=1}^k \E\Big[\Id_{\{T>q_\alpha\}}\frac{v(U_\ell)}{f_Y(Y_\ell)}~|~W=k\Big]\,.
$$

\begin{example}
For the compound model discussed above, we now evaluate the sensitivities $\tilde{\S}_{W}\,[\,\ES_\alpha\,]$ and $\tilde{\S}_{\Y}\,[\,\ES_\alpha\,]$, with the following baseline assumptions. The confidence level of the risk measure is $\alpha=0.95$. The frequency $W$ follows a Negative Binomial distribution with mean $\E[W]=5$ and over-dispersion $\rm{Var}(W)/\E[W]=2.5$, truncated at the 99.9th percentile. The severities $Y_\ell$ follow Gamma distributions with shape parameter $\theta=5$, corresponding to a skewness coefficient of $0.894$. 
With these choices, we find that the sensitivities, scaled by the portfolio risk, take values  $\frac{\tilde{\S}_{W}\,[\,\ES_\alpha\,]}{\ES_\alpha(T) }= 0.414$ and $\frac{\tilde{\S}_{\Y}\,[\,\ES_\alpha\,]}{\ES_\alpha(T) }= 0.429$. This indicates that the compound sum $T$ is approximately equally sensitive to the loss frequency and severity. 

\begin{figure}[t]
    \centering
    \includegraphics[width = 0.7\textwidth]{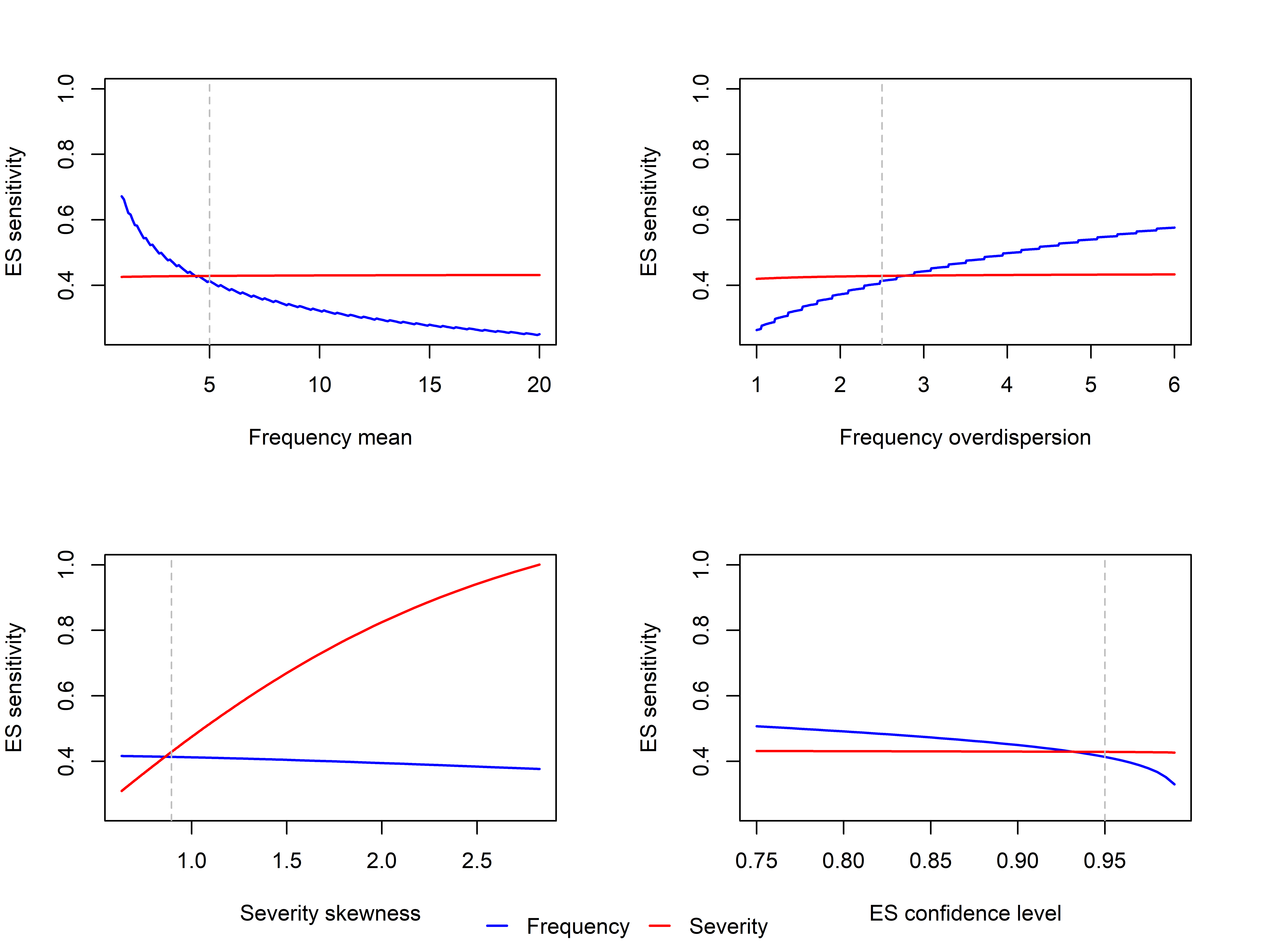}
    \caption{Changes in the scaled ES-sensitivity to the frequency (blue) and severity (red) of a compound Negative Binomial-Gamma distribution. The vertical dashed line in each plot represent baseline assumptions.}
    \label{fig:CompoundSens}
\end{figure}

In Figure \ref{fig:CompoundSens} we depict how the scaled sensitivities change after varying the baseline assumptions, one at a time, regarding frequency mean, frequency over-dispersion, the skewness of the severity distribution, and the confidence level of the ES risk measure. In each plot the baseline assumption is indicated by a vertical dashed line. We observe that, as the frequency mean increases, the importance of severities dominates, given the larger overall number of individual losses. On the other hand, when the frequency over-dispersion increases, the importance of frequency dominates, since the variance of the frequency distribution becomes the key risk driver. Furthermore, as one would expected, the sensitivity of the severities $\Y$ increases in the skewness, which reflects a riskier loss profile. Finally, as the confidence level increases, severities become more important than the frequency $W$, representing a more pronounced impact on the extreme tail of the portfolio loss.  
\end{example}

\section{Application to reinsurance credit risk modelling} \label{sec: numerical example RI}

Reinsurance credit risk modelling represents a prominent example where credit risk exposures are non-granular and inhomogeneous. Insurers buy reinsurance products in order to transfer some of the risk of (typically) higher than expected claim amounts to a third party. By taking on insurers' excess liabilities, the reinsurance market thus operates as an industry-wide risk pooling arrangement \citep{albrecher2017reinsurance}.  Credit risk then arises from the possibility that, in the event of high (industry) losses, reinsurers will not be able to make good on their obligations to insurers.  

Reinsurance credit risk has two features particularly relevant to our setting. First, dependence is of primary importance. Different reinsurers' ability to fulfil obligations is highly dependent on each other, given the systemic impact of overall (re)insurance market conditions and industry shocks.  As a result, reinsurers' default indicators should also be considered dependent on insurers' gross (i.e. before-reinsurance) losses; hence one needs to account for the event that reinsurers default precisely at those times when insurers need them most. Second, reinsurance credit risk exposures are highly inhomogeneous. Different reinsurers often reinsure different lines of business at different levels of extreme loss. Furthermore, the credit rating of reinsurers varies and insurers  typically transfer the most extreme layers of their gross losses to a small number of highly rated reinsurers -- while this is a rational strategy, it creates non-trivial concentration effects. The concern with the risk of reinsurance defaults, and particularly with their dependence, has been thoroughly reflected in actuarial modelling practice \citep{ter2008portfolio,britt2009reinsurance}.

Here we present a numerical example of differential sensitivity analysis to reinsurance defaults, working with an illustrative model of reinsurance credit risk. In equation \eqref{eq:loss-model}, we interpret the terms as follows: 
\begin{itemize}
    \item $L$ is the total reinsurance credit risk loss for an insurer.
    \item $\Z=(Z_1,\dots,Z_n)$ are the gross losses of the insurer, from its $n = 12$ lines of business (LoB).
    \item $g_j(\Z),~j\in\mM$ are the reinsurance recoveries expected from each of $m = 8$ reinsurers.
    \item $\{X_j \le d_j\}$ is the event that the $j$-th reinsurer defaults. 
\end{itemize}
The 12 LoB are marginally Lognormal distributed with the same mean and coefficient of variation (CoV) given in Table \ref{tab:LOB}, and which are consistent with the Solveny II standard formula parameters \citep{lloydsSCR}. 
In specifying the form of the $g_j$s we make the simplifying assumption that all reinsurance contracts bought consist of reinsurance layers on the gross losses $Z_1,\dots,Z_{12}$. 

  \begin{table}[htbp]
  \centering
  \caption{Name of Lines of business (LoB) and Coefficient of variation (CoV) (Source: \cite{lloydsSCR}).}
  \begin{footnotesize}
    \begin{tabular}{l c S[table-format=3.2]}
    \toprule\toprule
    LOB Name & \multicolumn{1}{l}{LoB} & \multicolumn{1}{l}{\hspace{5pt}CoV} \\
    \midrule
    Direct and Proportional Motor Vehicle Liability & $Z_1$ & 0.1 \\
    Direct and Proportional Other Motor & $Z_2$ & 0.08 \\
    Direct and Proportional Marine, Aviation and Transportation & $Z_3$ & 0.15 \\
    Direct and Proportional Fire \& Other Damage to Property & $Z_4$ & 0.08 \\
    Direct and Proportional General Liability & $Z_5$ & 0.14 \\
    Direct and Proportional Credit \& Suretyship & $Z_6$ & 0.19 \\
    Direct and Proportional Legal Expenses & $Z_7$ & 0.083 \\
    Direct and Proportional Assistance & $Z_8$ & 0.064 \\
    Direct and Proportional Miscellaneous Financial Loss & $Z_9$ & 0.13 \\
    Non-Proportional Casualty Reinsurance & $Z_{10}$ & 0.17 \\
    Non-Proportional Marine, Aviation and Transportation Reinsurance & $Z_{11}$ & 0.17 \\
    Non-Proportional Property Reinsurance & $Z_{12}$ & 0.17 \\
    \bottomrule \bottomrule
    \end{tabular}%
      \end{footnotesize}
    \label{tab:LOB}
\end{table}%

We assume that each of the first 6 reinsurers covers a layer from two LoBs, with deductibles $s_{j,k}$ and limit $t_{j,k}$. Each of reinsurers 7 and 8 covers a higher layer from 6 LoBs. Specifically, we have: 
\begin{subequations}
\begin{align*}
    g_j(\Z) 
    &= \sum_{k=2j -1}^{2j} \min\left\{\left(Z_{k}-s_{j,k}\right)_+ ,\;  t_{j,k}\right\}\,,
    \quad \text{for}\quad j = 1, \ldots, 6\,,
    \\
    g_7(\Z) 
    &= \sum_{k=1}^6 \min\left\{\left(Z_{k}-s_{7,k}\right)_+ ,\,  t_{7,k}\right\},
    \quad \text{and} \quad
     g_8(\Z) 
    = \sum_{k=7}^{12} \min\left\{\left(Z_{k}-s_{8,k}\right)_+ ,\,  t_{8,k}\right\}\,.
\end{align*}
\end{subequations}
The deductibles and limits are such that the first six reinsurers cover losses between the 55\% and 85\% quantile, whereas the last two reinsurers cover the losses between the 85\% and the 95\% quantile, i.e.,
\begin{align*}
    s_{j,k} &= F_{Z_k}^{-1}(0.55)
    \quad \text{and}\quad
    t_{j,k} = F_{Z_k}^{-1}(0.85) - s_{j,k}\,,
    \quad \text{for} \quad
    j = 1, \ldots ,6\,,
    \\    
    s_{j',k} &= F_{Z_k}^{-1}(0.85)
    \quad \text{and}\quad
    t_{j',k} = F_{Z_k}^{-1}(0.95) -s_{j,k}\,,
    \quad \text{for} \quad
     j' = 7, 8 \,.
\end{align*}
Finally, the default probabilities are set at 1.5\% for the first 6 reinsurers and 1\% for reinsurers 7 and 8. We assume that the random vector $(\X, \Z)$ is dependent with a t-copula with 4 degrees of freedom, such that the correlation matrix of $\Z$ satisfies Solvency II assumptions \citep{lloydsSCR}, while the random vector $\X$ has a homogeneous correlation matrix such that $\mathrm{Corr}(X_i,X_j)=0.05$. The joint dependence of $(\X, \Z)$ is effected via a t-distribution factor model; further details are given in the electronic companion, Appendix \ref{app:reinsurance-model}.

\begin{figure}[t]
    \centering
    \includegraphics[width =0.5\textwidth]{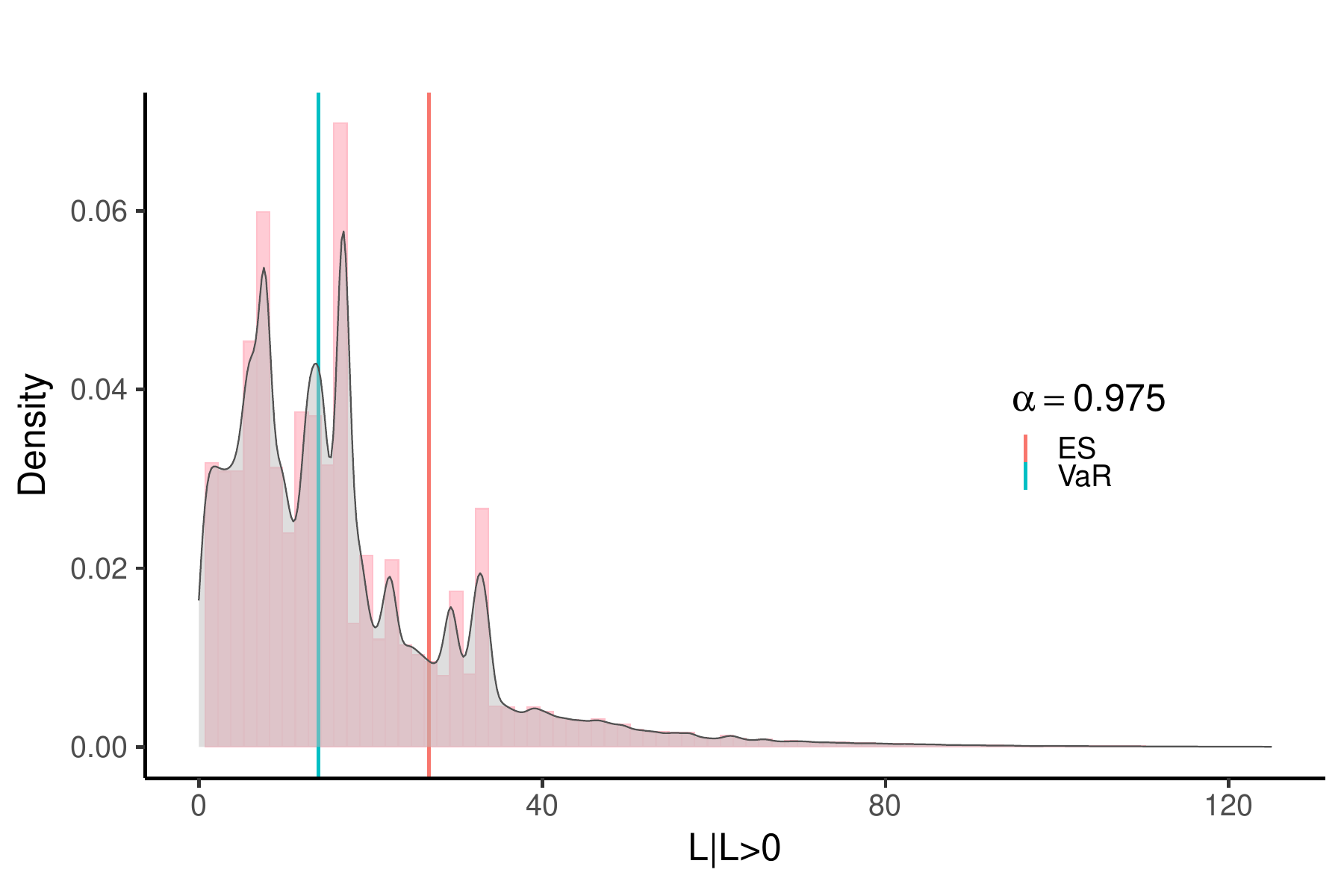}
    \caption{Histogram of the insurer's total reinsurance credit risk loss conditional on a positive loss, i.e., $L | L>0$. Vertical lines are the unconditional VaR and ES at level $\alpha = 0.975$.}
    \label{fig:hist-L}
\end{figure}

The distribution of the total credit risk loss $L$ is evaluated by Monte-Carlo simulation. Specifically, since almost all scenarios of $(\X, \Z)$ result in a credit loss of zero, i.e., a realisation $\{L = 0\}$, we generated a dataset of size 500,000 (keeping track of the total number of simulations), in which all realisations satisfy $L > 0$.  The probability that $L>0$ is 5.044\% in our dataset. Figure \ref{fig:hist-L} depicts a histogram of the insurer's total credit risk loss $L$ conditional that a loss occurred. We also report the unconditional VaR and ES at level $\alpha = 0.975$. The skewness and multimodality  of the loss distribution, driven by the portfolio's lack of homogeneity, are apparent. 

We apply stresses on each of the risk factors $X_i$ and $Z_i$. Specifically, we apply left-tail stresses (see Table \ref{tab:stresses}) on the risk factors driving defaults, i.e., $X_{i, \ep}:= X_i + \ep \left(X_i - F_{X_i}^{-1}(0.2)\right) \Id_{\{ X_i \le F_{X_i}^{-1}(0.2)\}}$, $i = 1, \ldots, 8$. These stresses increase the probability of reinsurance defaults, though in a more complex way compared to Example \ref{ex:CR - 1st model}. For each LoB, we consider a right-tailed stress $Z_{i, \ep}:= Z_i + \ep \left(Z_i - F_{Z_i}^{-1}(0.8)\right) \Id_{\{ Z_i \ge F_{Z_i}^{-1}(0.8)\}}$, $i = 1, \ldots, 12$, which increases the loss quantiles of $Z_i$, beyond its 80\% quantile.

We calculate the sensitivities with respect to the VaR and ES risk measures at level $\alpha=0.975$, according to Theorems \ref{thm:VaR:marginal} and \ref{thm:ES:marginal}.
To calculate the sensitivities to $Z_i$s, we require estimates of expectation conditioned on the event $\{L = q_\alpha\}$ and $\{L \ge q_\alpha\}$. For estimating the expectation conditional on the event of zero probability $\{L = q_\alpha\}$, we use the $\delta$-estimator \citep{Glasserman2005JCF}, though more sophisticated methods such as quasi-Monte Carlo Methods could be employed, see e.g., \cite{cambou2017SC} and \cite{basu2016SIAMNA} for convergence rates. Specifically, for $\delta >0$ with $0< \alpha - \delta , $ and $\alpha + \delta <1$, we approximate the sensitivity of VaR to $Z_i$ by
\begin{equation*}
    \widehat{\S}_{Z_i}\,[\,\VaR_\alpha\,]
   =
   \frac{1}{2\delta}\sum_{j = 1}^8 \E\Big[\,  \mfk(Z_i)  \partial_i \,g_j(\Z) \Id_{\{X_j \le d_j\}}   \Id_{\{ L \in (F^{-1}(\alpha - \delta),\, F^{-1}(\alpha + \delta))\}} \,\Big]\,.
\end{equation*}
Mathematically, we replace the conditioning event $\{L = \q\}$ by an event of probability $2\delta$, i.e. by $\{L \in (F^{-1}(\alpha - \delta),\, F^{-1}(\alpha + \delta)\}$. A value of $\delta=0.005$ was used throughout.
We use our sample of $(\X, \Z,L~|~L>0)$, which contains  500,000 simulated scenarios, and estimate the sensitivities using bootstrap with replacement and a bootstrap size of 450,000. The reported sensitivities are averaged over 100 bootstrap estimates.        

\begin{figure}[t]
    \centering
    \includegraphics[width =0.45\textwidth]{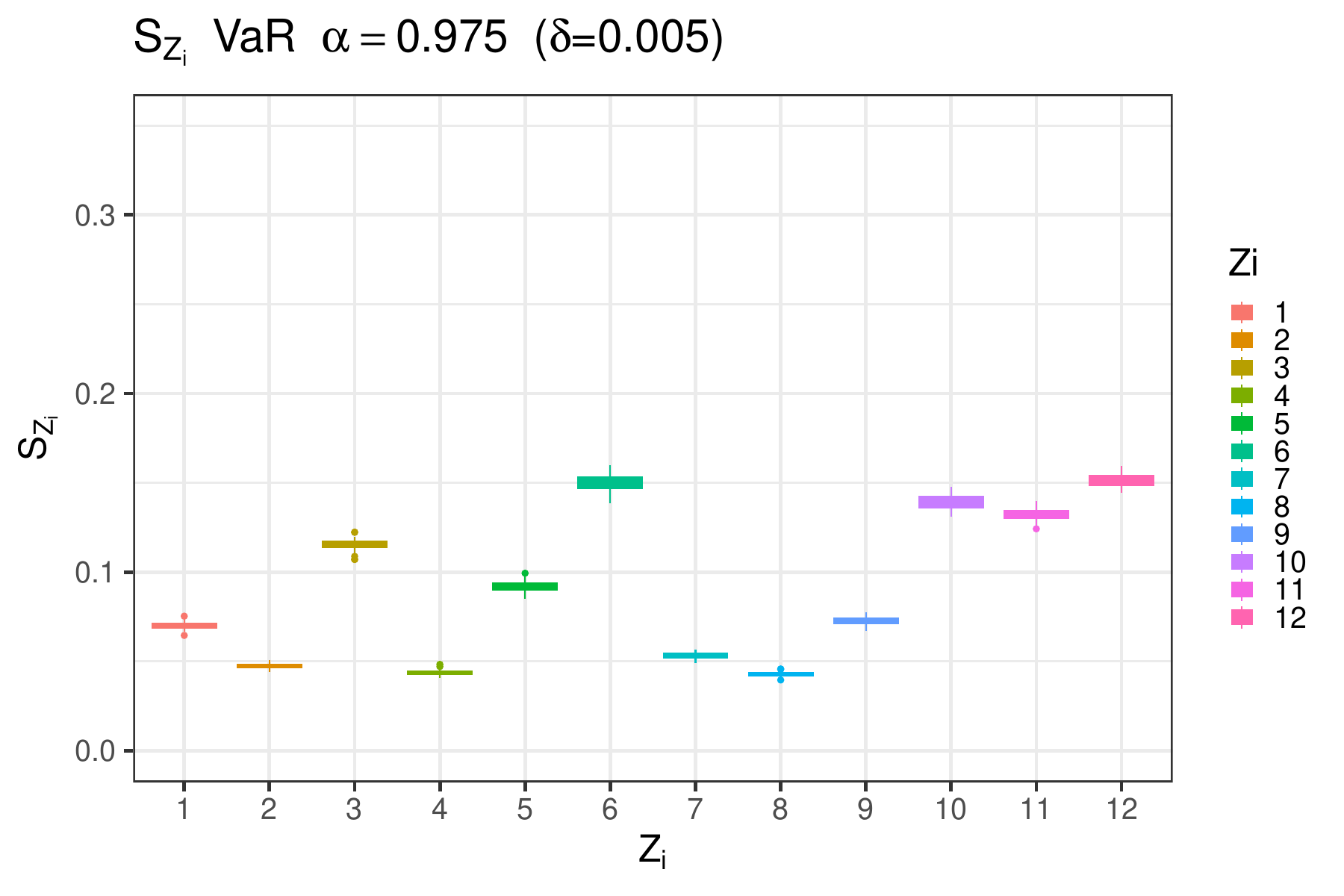}
    \includegraphics[width =0.45\textwidth]{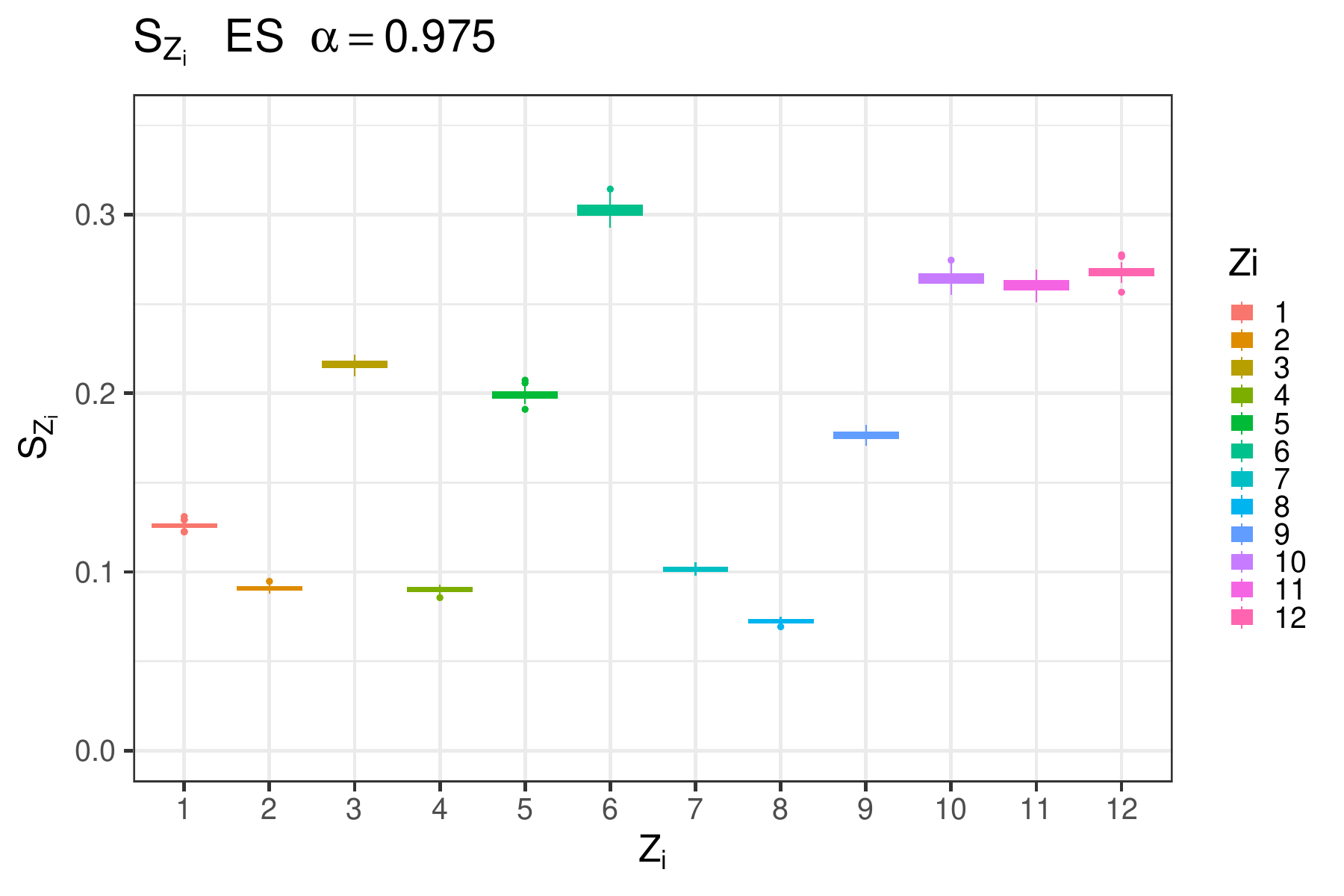}
    \caption{Marginal sensitivity to $Z_i$s of VaR (left; with  $\delta = 0.005$) and ES (right) with $\alpha = 0.975$.}
    \label{fig:sens-Zi}
\end{figure}

\begin{figure}[ht]
    \centering
    \includegraphics[width = 0.45\textwidth]{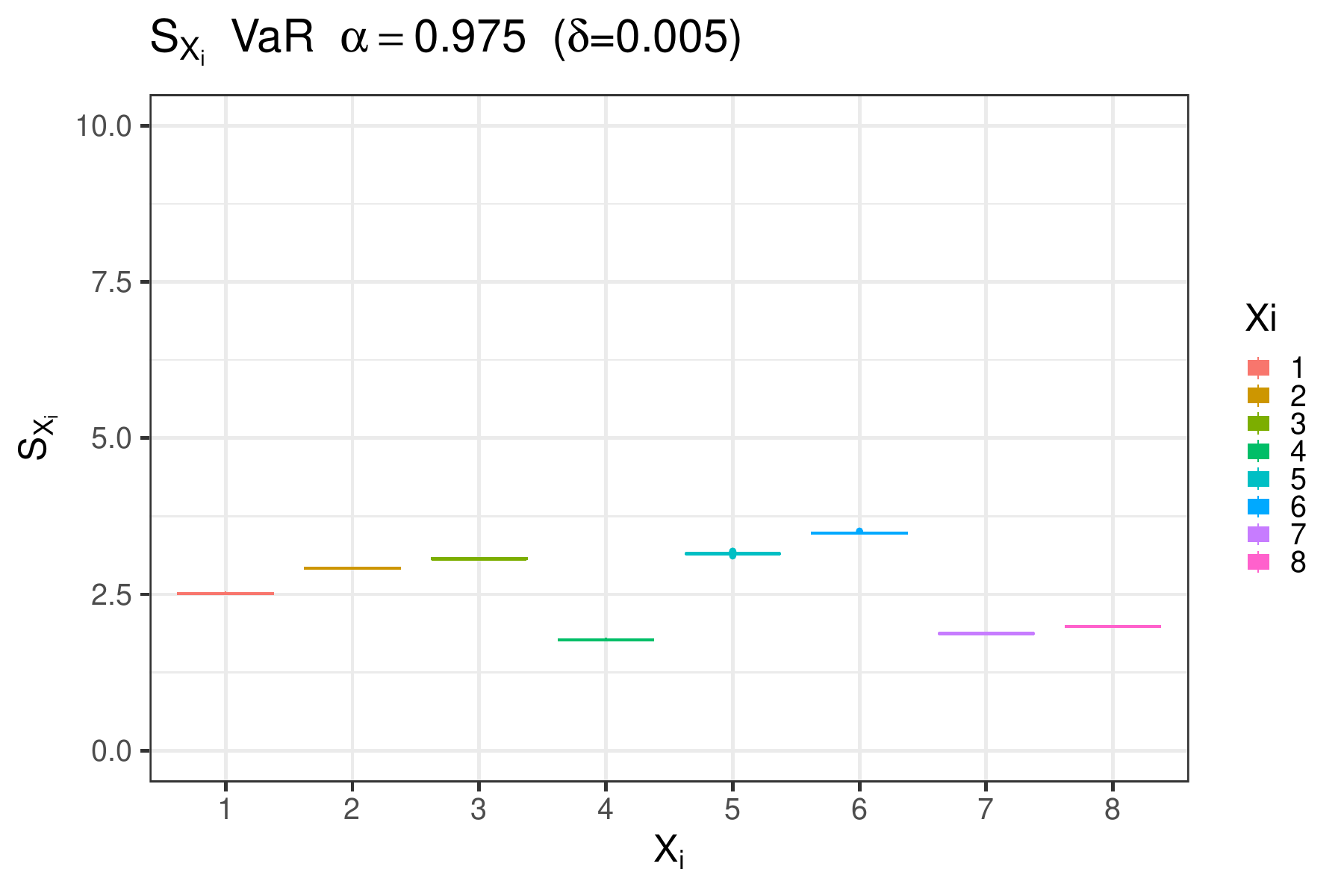}
    \includegraphics[width =0.45\textwidth]{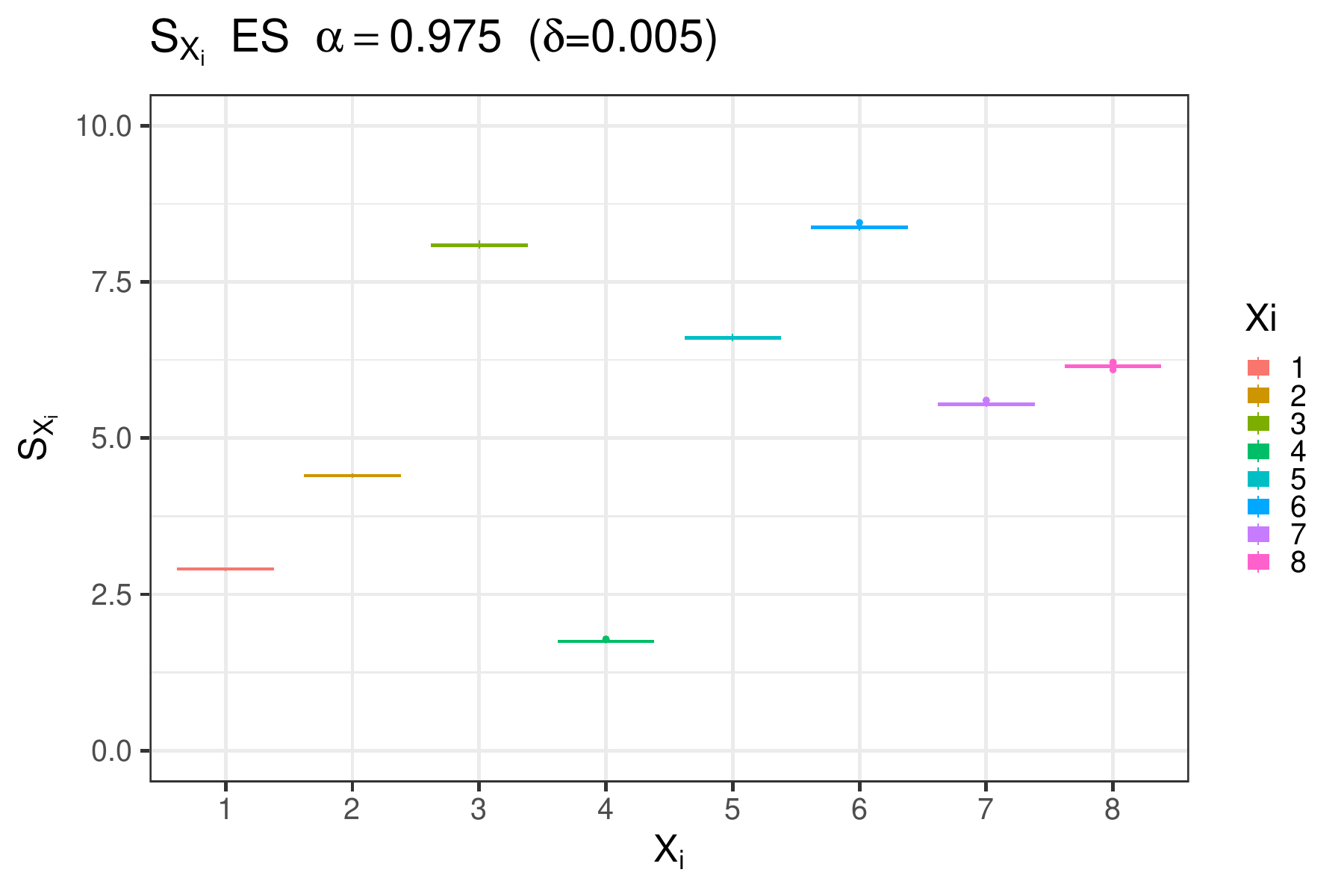}
    \caption{Marginal sensitivity to $X_i$s of VaR (left; with $\delta = 0.005$) and ES (right; with $\delta = 0.005$) with $\alpha = 0.975$.}
    \label{fig:sens-Xi}
\end{figure}

\begin{figure}[ht]
    \centering
    \includegraphics[width =0.45\textwidth]{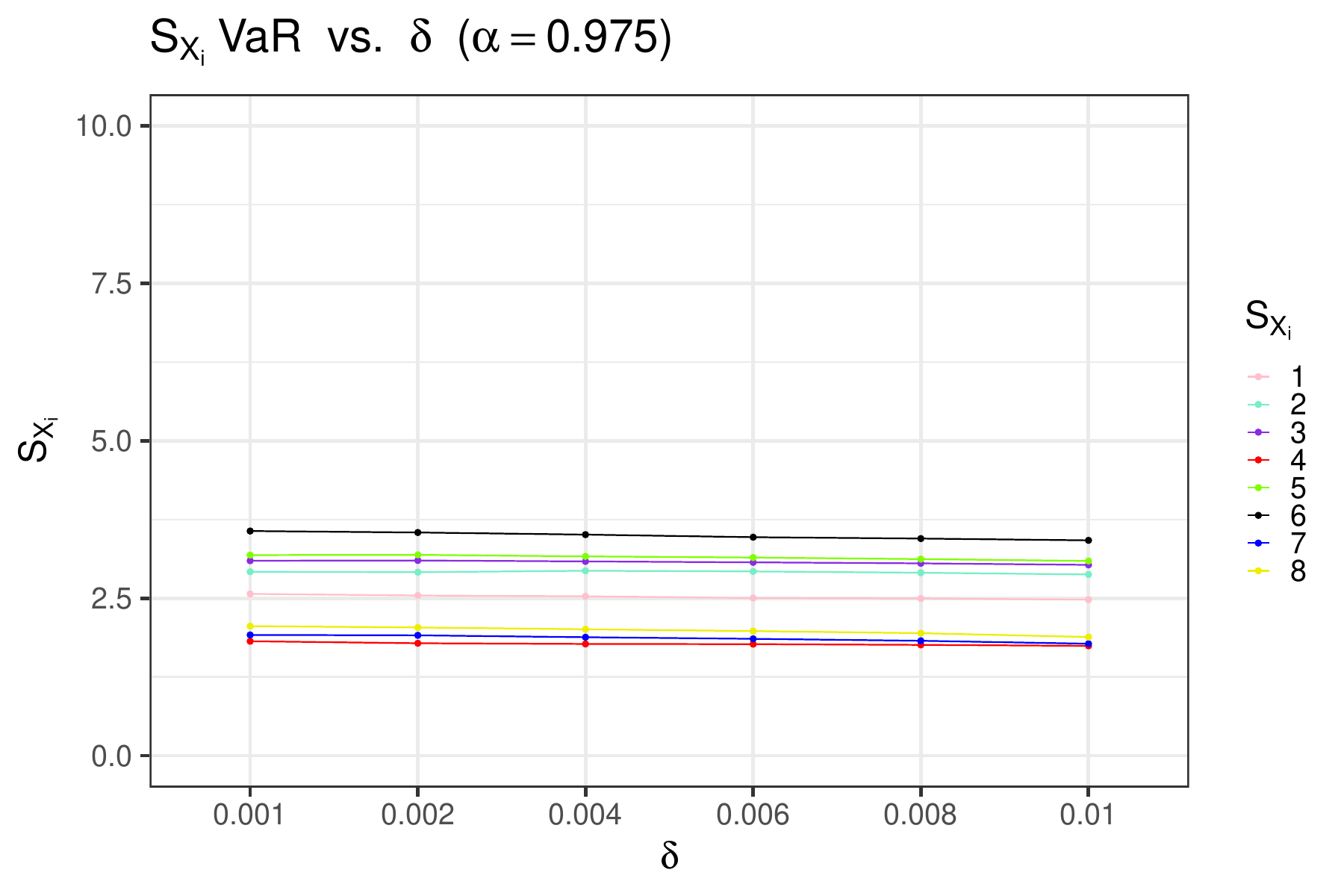}
    \includegraphics[width = 0.45\textwidth]{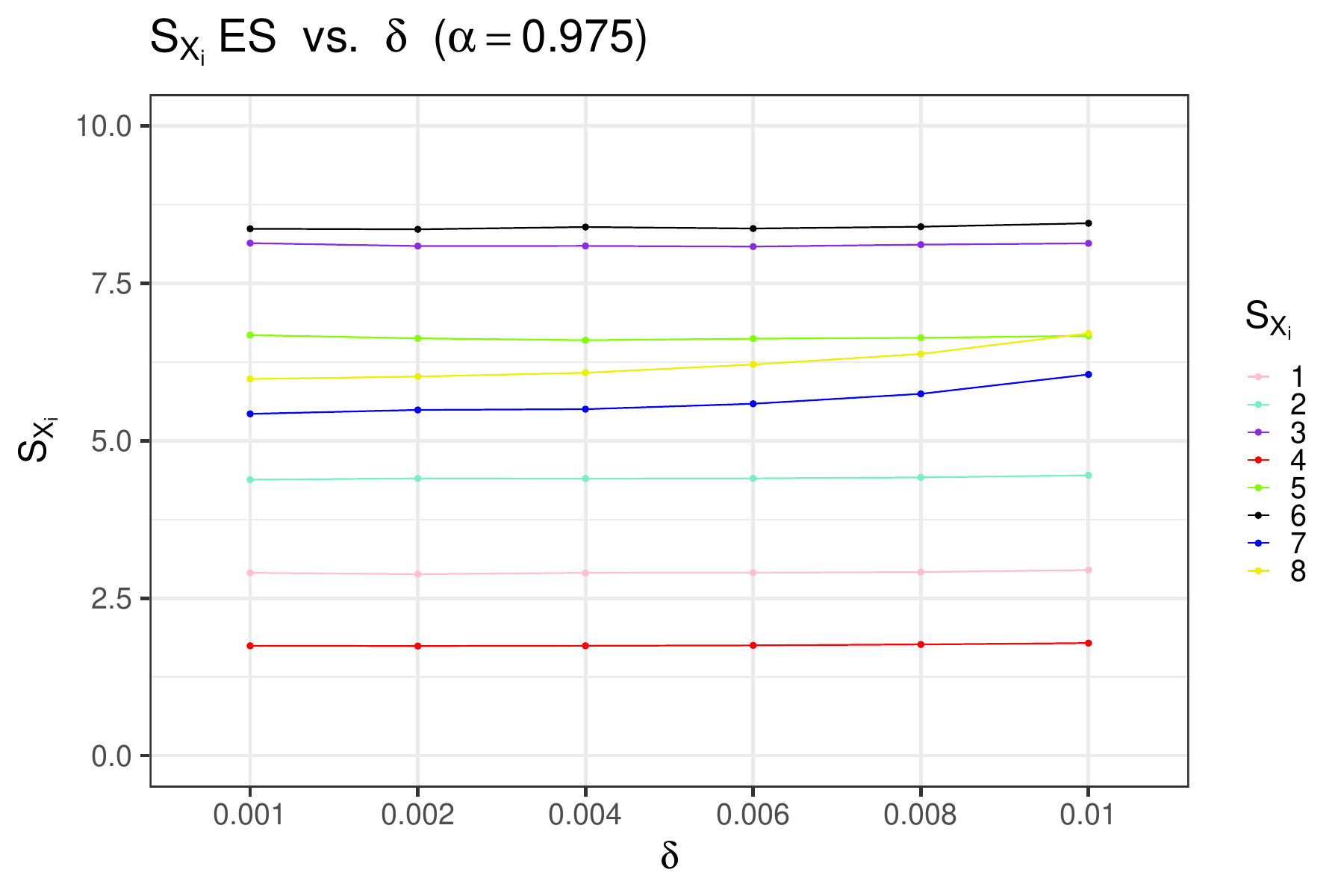}
    \\
    \includegraphics[width =0.45\textwidth]{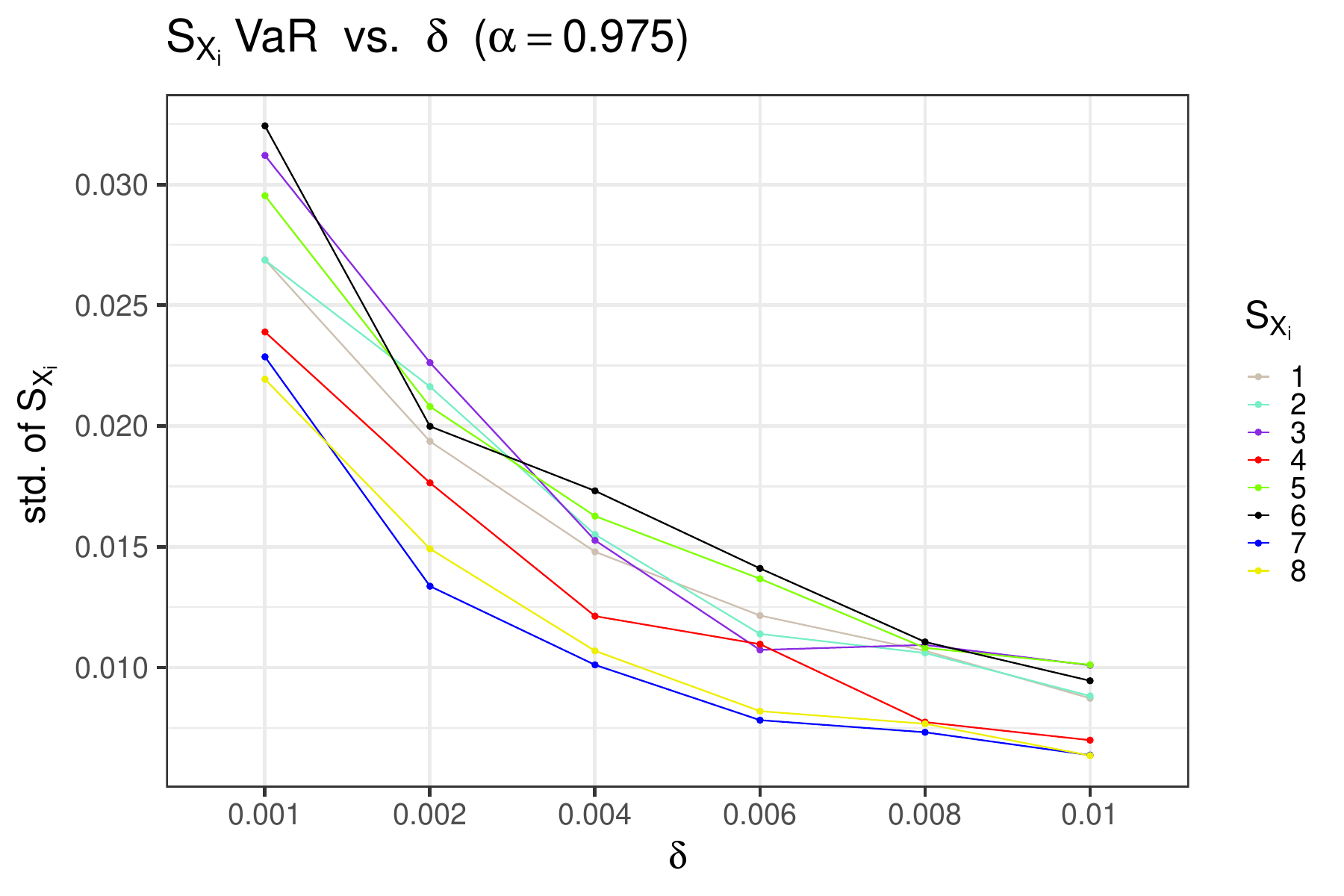}
    \includegraphics[width = 0.45\textwidth]{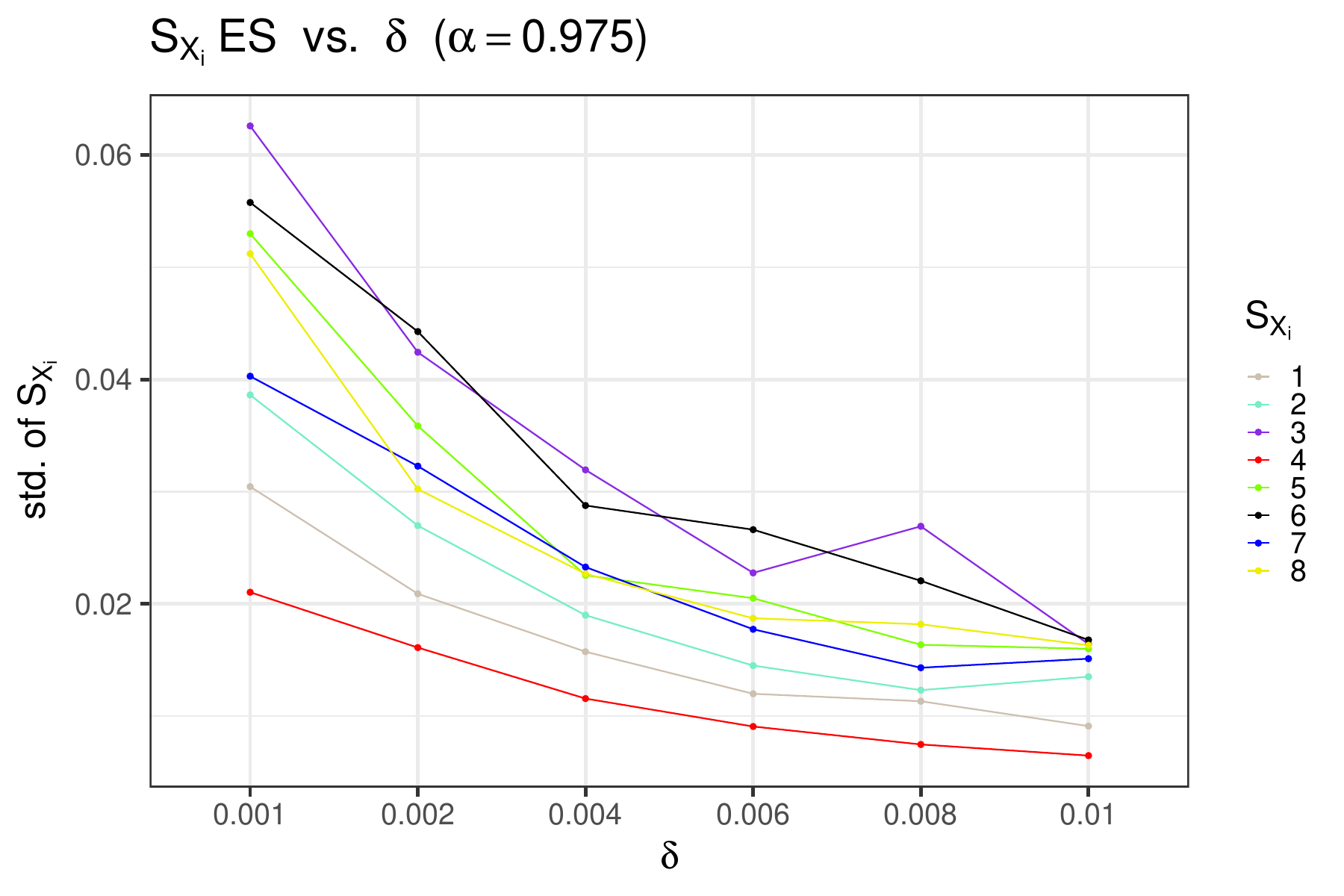}
    \caption{Marginal sensitivities to $X_i$ and their sample standard deviations for different choices of $\delta$. Top panels: Marginal sensitivity for VaR (left) and for ES (right) both with $\alpha = 0.975$. Bottom panels: Sample standard deviation of the sensitivity estimators for VaR (left) and ES (right).}
    \label{fig:sens-ES-delta}
\end{figure}

For estimating the sensitivities to each $X_i$, a different dataset is simulated. Specifically, for each $j = 1, \ldots,8$, we generate a dataset of size 500,000, in which all realisations of $(\X, \Z)$ satisfy $X_j \in \big(F_{X_j}^{-1}(d_j- \delta) , \, F_{X_j}^{-1}(d_j+ \delta)\big)$, for small $\delta>0$. Again, sensitivities were estimated by bootstrapping 100 times with replacement and bootstrap size 450,000. Figures \ref{fig:sens-Zi} and \ref{fig:sens-Xi} display box plots of the sensitivities to $Z_i$ and to $X_i$ for both VaR and ES.  Again a value of $\delta=0.005$ is used as a baseline; the effect of this choice on sensitivity estimates is discussed in the sequel (Figure \ref{fig:sens-ES-delta}). 

In Figure \ref{fig:sens-Zi}, where the  sensitivities to the $Z_i$s are plotted, we observe that business line 6 has a large sensitivity for VaR, the LoB with the largest CoV, see Table \ref{tab:LOB}. In the right panel we observe that for ES the sensitivities are ordered similarly to the CoV of the business lines, but with a larger spread compared to the case of VaR -- this could be attributed to the higher tail-sensitivity of the ES measure. Indeed, LoB 6 has the largest sensitivity, followed by 10, 11, and 12, which all have the same, second largest, CoV. Furthermore, LoB 2, 4, 7, and 8, which have the smallest CoVs, have small sensitivities for both VaR and ES.

In Figure \ref{fig:sens-Xi}, we depict the sensitivities to the $X_i$s. A similar picture emerges, with the sensitivities for VaR (left panel) being very close together and for ES (right panel) being more spread-out. For ES, we observe that
reinsurer 3, which has a layer on LoBs 5 and 6, and reinsurer 6, which has a layer on LoBs 11 ad 12, have the largest sensitivities. These LoBs have large sensitivities for ES, as seen in Figure \ref{fig:sens-Zi} (right panel). Thus, a default of these reinsurers would naturally lead to a large impact on the ES of the total loss. We also see that reinsurers 7 and 8 have large sensitivities for ES. 
This is in line with expectations, since reinsurer 7 and 8 take on the highest layers (between the 85\% and 95\% quantile) of 6 business lines each. Nonetheless, this concentration effect is not picked up by the VaR sensitivity, as in the left panel the sensitivities to $X_7,X_8$ are rather low. This points to the importance of selecting an appropriately tail-sensitive risk measure.

\begin{figure}[ht]
    \centering
    \includegraphics[width = 0.45\textwidth]{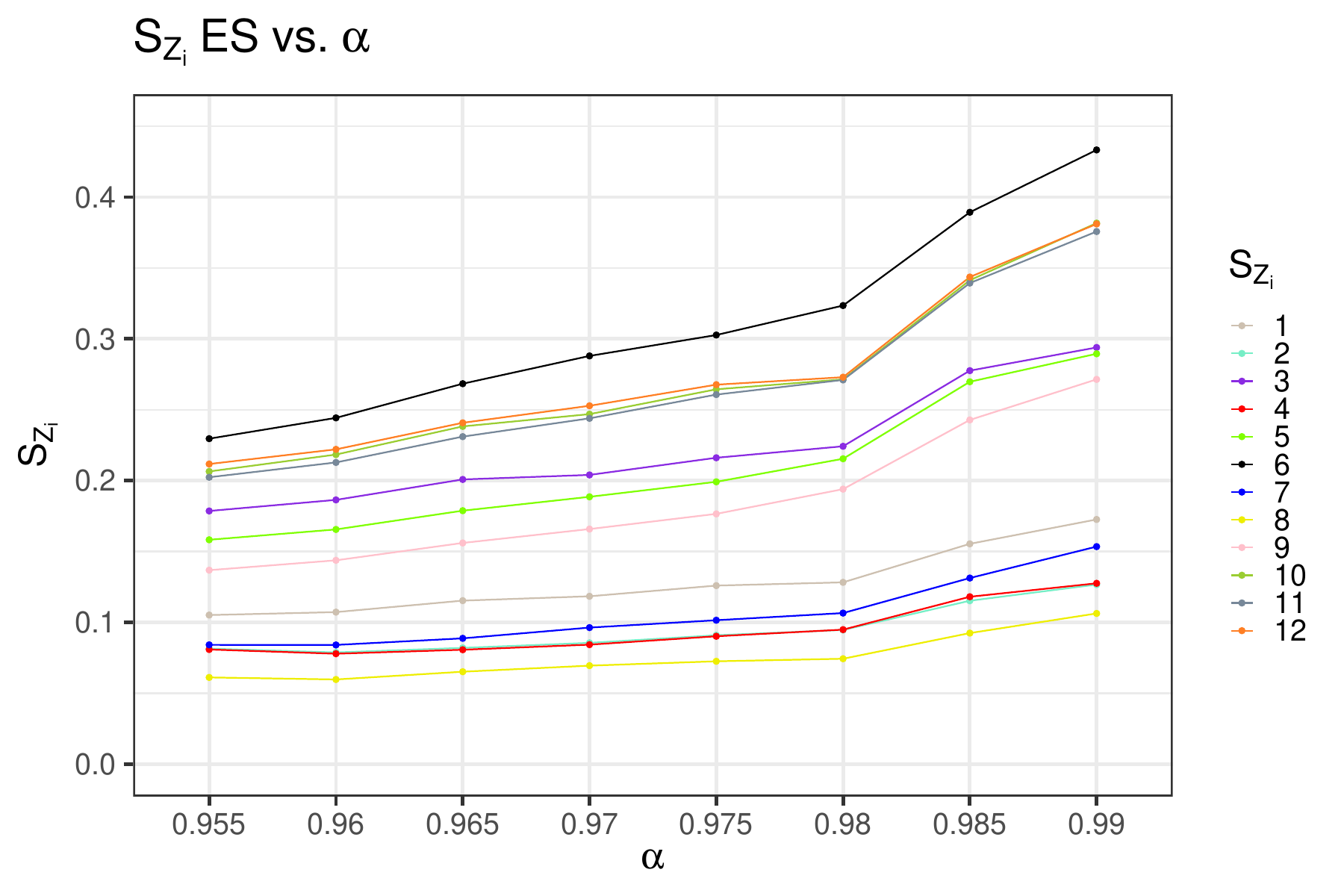}
    \includegraphics[width =0.45\textwidth]{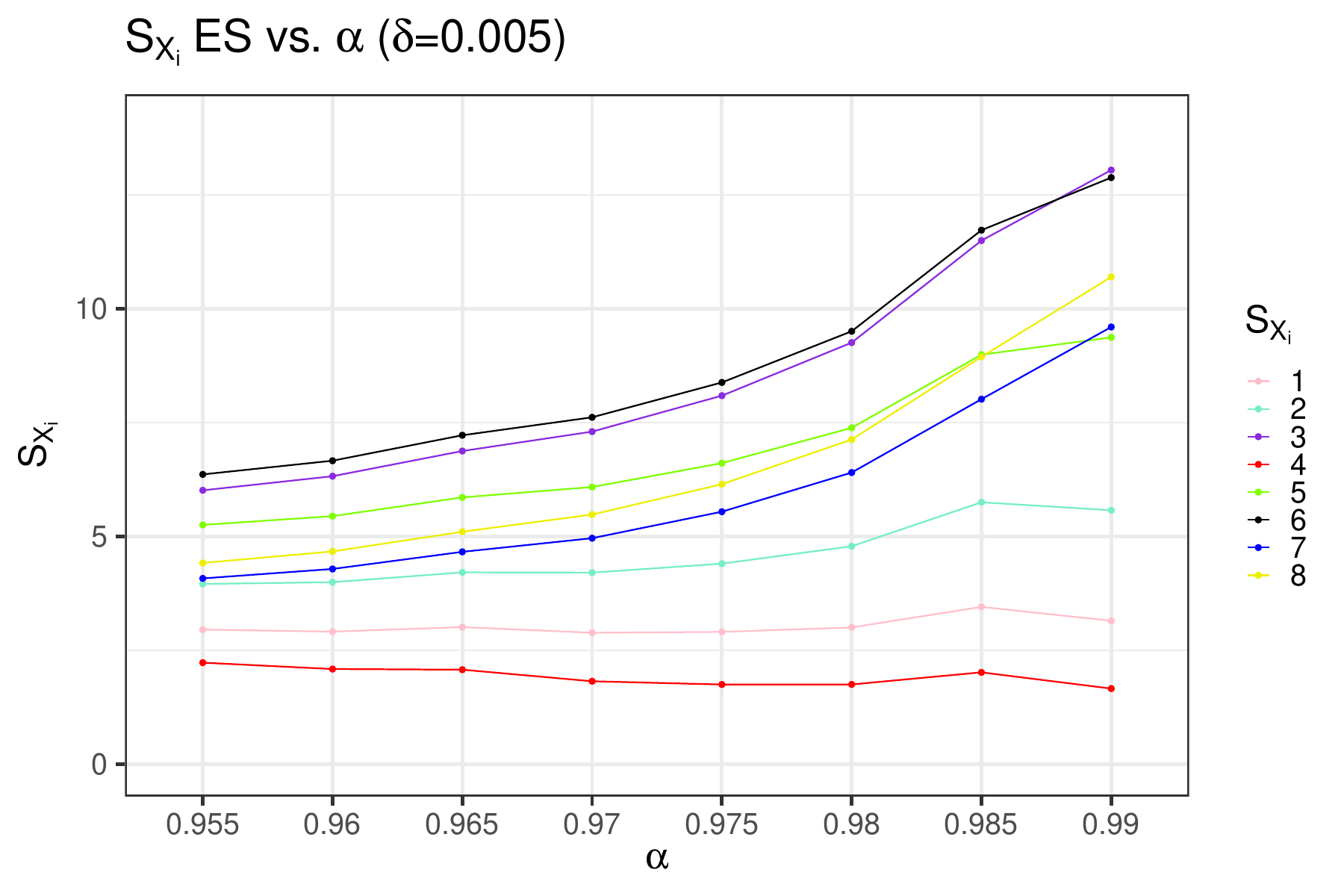}
    \caption{Marginal sensitivities of ES and different choices of $\alpha$ between 0.955 and 0.99. Left: sensitivity to $Z_i$s. Right: sensitivity to $X_i$s.}
    \label{fig:sens-ES-alphas}
\end{figure}
Figure \ref{fig:sens-ES-alphas} depicts the sensitivities of ES for different choices of $\alpha$, from 0.955 to 0.99. The left panel contains the sensitivities to the LoBs ($Z_i$) and the right panel the sensitivities to the reinsurers ($X_i$). We observe that the ordering of the risk factors is mostly consistent with respect to changes in confidence level. An exception to this are the sensitivities to $X_7$ and $X_8$ which increase faster (relative to others) with $\alpha$, as seen by the line crossings on the right panel. Once again this demonstrates the increased impact of default risk concentration at high loss quantiles.

Finally, in Figure \ref{fig:sens-ES-delta} top panels, we show the sensitivities to $X_i$s for VaR (left panel) and ES (right panel) with $\alpha = 0.975$, using different choices of $\delta$ for approximating the expectation conditional on $\{X_i = d_i\}$. We observe that the estimates are very stable for different choices of $\delta$. Furthermore, in the bottom panels of Figure \ref{fig:sens-ES-delta} we plot the standard deviation of the sensitivity estimators, thus choosing $\delta = 0.005$ provides a suitable bias and variance trade-off.

\section{Conclusion}\label{sec:conclusion}
Taking derivatives of model outputs in the direction of inputs is a foundational process for interpreting complex computational models. However, differential sensitivity measures typically require stringent assumptions on differentiability and Lipschitz continuity of the model function. This severely limits the scope of current methods of differential sensitivity analysis. We address the problem by noting that, when inputs are uncertain -- as is the case in settings ranging from Monte Carlo simulation to algorithmic prediction -- a global view can be more appropriate than a local one. For a global assessment, differentiation is required across the entire input space; but then, it is not the derivative of the model function as such that is of primary interest, but rather the derivative of a statistical functional of the output. Still, extant literature on sensitivity analysis of risk measures typically requires differentiability of the model aggregation function. 

In this paper, we overcome current limitations in the literature and derive expressions for derivatives of quantile-based risk measures of model outputs, in a general setting where aggregation functions contain step functions and thus are not Lipschitz continuous. The conditions we require are rather weak and the sensitivity measures obtained admit representations as conditional expected values, which allows their estimation by standard methods. There are multiple potential applications of our methodology. We demonstrate applications in the area of credit risk modelling, but also in assessing  sensitivity with respect to discrete random inputs. While our work is applicable in principle to discontinuous (e.g., tree-based) predictive models, addressing the idiosyncratic challenges of such exercises remains a topic for future work.

\ACKNOWLEDGMENT{
The authors would like to thank Wen Yuan (William) Tang for his help in implementing the numerical examples.
SP gratefully acknowledges the support of the Canadian Statistical Sciences Institute (CANSSI) and the Natural Sciences and Engineering Research Council of Canada (NSERC) with funding reference numbers DGECR-2020-00333 and RGPIN-2020-04289.
}

\begin{APPENDICES}

\section{Additional Sensitivity Formulas}\label{app:additional-sensitivies}

Here we provide additional results for marginal and cascade sensitivities, which are omitted from the main body of the text for reasons of concision. In Section \ref{app:cascade} we deal with cascade sensitivities of VaR, while in Section \ref{app::general-loss-model} of the electronic companion we present results and proofs for a more general model than \eqref{eq:loss-model}, that is for the loss $ L =  \sum_{j \in \mM} g_j(\X, \Z) \Id_{\{X_j \le d_j\}}$.

\subsection{Cascade Sensitivities to VaR}\label{app:cascade}
Here we report the cascade sensitivity formulas for VaR.

\begin{theorem}[Cascade Sensitivity VaR to $X_i$] \label{thm:VaR:cascade}
Let Assumptions \ref{asm: marginal VaR}, \ref{asm:diff-quantile} and \ref{asm:psi-combined} (for $Y=X_i)$  be fulfilled for the stressed model $L^{\bPsi}_\ep(X_i)$ and given $\alpha \in (0,1)$. Then, the cascade sensitivity for $\VaR_\alpha$ to input $X_i$ is given by, 
\begin{align*}
    \C_{X_i}\,[\,\VaR_\alpha\,]
    &= 
    \sum_{j \in\mM} \C_{X_i,X_j}
    + 
    \sum_{k \in\mN} \C_{X_i,Z_k},
 \end{align*}
where for all $k \in\mN$,
\begin{equation*}
    \C_{X_i,Z_k}\,
    = \sum_{j \in \mM} \left.\E\left[ \mfk(X_i) \, \partial_k g_j(\Z)\, \Psi_1^{(m+k)}(X_i, \V)  \Id_{\{X_j \le d_j\}} ~\right|~L = \q \right]\,,
\end{equation*}
and for $j \in\mM$,
\begin{equation*}
    \C_{X_i,X_j}\,
    = \frac{ c(\kappa;\, j) f_j(d_j)}{f\left(\q\right)}\;\;  \E\left.\left[\, \mfkinv(X_i)
    \Psi_1^{(j)}(X_i, \V)\left(\Id_{\{L \le \q + c(\kappa;\, j) g_j(\Z)\}} - \Id_{\{L \le\q\}}\right)~\right|~X_j = d_j\,\right] \,.
\end{equation*}
\end{theorem}

\begin{theorem}[Cascade Sensitivity VaR to $Z_i$] \label{thm:VaR:cascade-Zi}
Let Assumptions \ref{asm: marginal VaR}, \ref{asm:diff-quantile} and \ref{asm:psi-combined} (for $Y=Z_i)$  be fulfilled for the stressed model $L^{\bPsi}_\ep(Z_i)$ and given $\alpha \in (0,1)$. Then, the cascade sensitivity for $\VaR_\alpha$ to input $Z_i$ is given by, 
\begin{align*}
    \C_{Z_i}\,[\,\VaR_\alpha\,]
    &= 
    \sum_{j \in\mM} \C_{Z_i,X_j}
    + 
    \sum_{k \in\mN} \C_{Z_i,Z_k},
 \end{align*}
where for all $k \in\mN$,
\begin{equation*}
    \C_{Z_i,Z_k}\,
    = \sum_{j \in\mM} \left.\E\left[ \mfk(Z_i) \, \partial_k g_j(\Z)\, \Psi_1^{(m+k)}(Z_i, \V)  \Id_{\{X_j \le d_j\}} ~\right|~L = \q \right]\,,
\end{equation*}
and for $j  \in\mN$,
\begin{equation*}
    \C_{Z_i,X_j}\, 
    = \frac{ c(\kappa;\, j) f_j(d_j)}{f\left(\q\right)}\;\;  \E\left.\left[\, \mfkinv(Z_i)
    \Psi_1^{(j)}(Z_i, \V)\left(\Id_{\{L \le \q + c(\kappa;\, j) g_j(\Z)\}} - \Id_{\{L \le\q\}}\right)~\right|~X_j = d_j\,\right] \,.
\end{equation*}
\end{theorem}

\section{Proofs}\label{app:proofs}

This section contains proofs of Theorems \ref{thm:VaR:marginal} -   \ref{thm:ES:cascade-Zi},
\ref{thm:VaR:cascade}, and \ref{thm:VaR:cascade-Zi}.

\subsection{Proofs of Marginal Sensitivity: Theorems \ref{thm:VaR:marginal} and \ref{thm:ES:marginal}}
For the proof of the marginal sensitivities to $\VaR$ and $\ES$, we need the following lemma concerning sequences of functions that converge weakly to a Dirac delta function. For this, we first write the marginally stressed portfolios as
\begin{align*}
    L(Z_{i,\ep} )
    &:= L + \sum_{k = 1}^m \Delta_\ep g_k
    \quad \text{and} \quad
    L(X_{i,\ep} )
    :=L +  g_i(\Z)\,\left(\Id_{\{X_{i, \ep} \le d_i\}} -\Id_{\{X_i \le d_i\}}\right)
    \,,
\end{align*}
where we define $\Delta_\ep g_k := ( g_k(\Z_{-i}, \kappa_\ep(Z_i)) - g_k(\Z)) \Id_{\{X_k \le d_k\}}$. When the stress is clear from the context, we write $L_\ep = L(Z_{i,\ep} )$ and $L_\ep  =L(X_{i,\ep} )$.

\begin{lemma}
\label{lemma:dirac-marginal}
For fixed $d\in \R$, define the family of functions
\begin{equation*}
    \delta_\ep(x) = \tfrac{\left|\Id_{\{\kappa_\ep (x) \le d\}} - \Id_{\{x \le d\}}\right|}{\ep}\,,
\quad x\in \R\,, \quad \ep>0.
\end{equation*}
Then, $\delta_\ep$ converges weakly to a scaled Dirac delta function at $d$ for $\ep \searrow 0$. Moreover, for any family of measurable functions $h_\ep \colon \R^{m + n} \to \R$ such that $\lim_{\ep \searrow 0}\E\left[|h_\ep(\X, \Z)|\right] < \infty$,
the following holds:
\begin{align*}
   \lim_{\ep \searrow 0} \E\left[ \delta_\ep(X_i) \,h_\ep(\X, \Z)\right]
    &= -c(\kappa) \mfkinv(d)\, f_i (d)  \;\E\left[\, h_0(\X, \Z)~|~ X_i = d\,\right]\,,
\end{align*}
where $c(\kappa)$ is given in \eqref{eq:c-kappa}, and $h_0(\x, \z) = \lim_{\ep \searrow 0}h_\ep(\x, \z)$.
\end{lemma}

\proof{Proof of Lemma \ref{lemma:dirac-marginal}.}
First note that 
\begin{equation}\label{eq:indicator}
    \big|\Id_{\{\kappa_\ep (x) \le d\}} - \Id_{\{x \le d\}}\big|
    =
    - c(\kappa) \left(\Id_{\{\kappa_\ep (x) \le d\}} - \Id_{\{x \le d\}}\right)\,.
\end{equation}
Let $\xi$ be an infinitely often differentiable function. Using the change of variable $y = \kappa_\ep(x)$, we obtain
\begin{align*}
    \int_{- \infty}^{+\infty}\xi(x) \delta_\ep(x)\,dx  
    &= 
    - \frac{c(\kappa)}{\ep} \int_{- \infty}^{+\infty}\xi(x) \left(\Id_{\{\kappa_\ep (x) \le d\}} - \Id_{\{x \le   d\}}\right) \, dx
    \\
    &= 
    - \frac{c(\kappa)}{\ep} \int_{- \infty}^{+\infty}\frac{\xi(z)}{\frac{\partial}{\partial x}\kappa_\ep(z)}\Big|_{z = \kappa_\ep^{-1}(y)} \Id_{\{y \le d\}}  \, dy
    - \frac{1}{\ep} \int_{- \infty}^{d}\xi(x)\, dx\,.
\end{align*}
Letting $\Xi$ be a primitive of $\xi$ vanishing at $-\infty$, then 
\begin{align*}
    \int_{- \infty}^{+\infty}\xi(x) \delta_\ep(x)\,dx  
    &= 
    - \frac{c(\kappa)}{\ep} \left(\int_{- \infty}^{d} \frac{d}{dy}\Xi(\kappa_\ep^{-1}(y)) \, dy 
    -  \Xi(d)
    \right)
    = 
    - \frac{c(\kappa)}{\ep} \left(\Xi(\kappa_\ep^{-1}(d)) - \Xi(d)\right)\,.
\end{align*}
Taking the limit as $\ep\rightarrow 0$, we obtain that
\begin{equation*}
    \lim_{\ep \searrow 0}
    \int_{- \infty}^{+\infty}\xi(x) \delta_\ep(x)\,dx  
    =
    - c(\kappa)\,\xi(d) \mfkinv(d)\,.
\end{equation*}

For the second part of the statement, note that 
\begin{align*}
    \lim_{\ep \searrow 0} \,  \E\left[\delta_\ep(X_i) \, h_\ep(\X, \Z)\right]
    &= 
    \lim_{\ep \searrow 0} \, \int_{\R^{m+n}} \delta_\ep(x_i)  \,h_\ep(\x, \z)  \,f_{\X, \Z}(\x, \z) \,d\x\, d\z
    \\
    &= 
    - c(\kappa)\,\mfkinv(d)\int_{\R^{m +n-1}}  \,h_\ep(\x_{-i}, d, \z)  \,f_{\X, \Z}(\x_{-i}, d, \z)\frac{f_i(d)}{f_i(d)} \,d\x_{-i}\, d\z
    \\
    &= 
    - c(\kappa)\,\mfkinv(d)\,f_i(d)\E\left[  \,h_0(\X, \Z)  ~|~ X_i = d \,\right]\,.
\end{align*}
\vspace{-10pt}
\hfill\Halmos
\endproof

\begin{lemma}\label{lemma:dirac-marginal-Z}
For fixed $0<\alpha<1$ and $\z \in \R^n$, define the sequence of functions
\begin{equation*}
    \delta_\ep(l) = \tfrac{\Id_{\left\{l \le \q - \sum_{k = 1}^m\Delta_\ep g_k \right\}}
    - \Id_{\left\{l \le \q \right\}}
    }{\ep}\quad l\in \R\,, \quad \ep>0.
\end{equation*}
where $\Delta_\ep g_k = ( g_k(\z_{-i}, \kappa_\ep(z_i)) - g_k(\z)) \Id_{\{x_k \le d_k\}}$, $\z \in \R^n$, and $l \ge 0$. Then, $\delta_\ep$ converges weakly to a scaled Dirac delta function at $\q$ for $\ep \searrow 0$. Moreover, for any family of measurable functions $h_\ep \colon \R^{m + n} \to \R$ such that $\lim_{\ep \searrow 0}\E\left[|h_\ep(\X, \Z)|\right] < \infty$, the following holds:
\begin{equation}\label{eq:lemma-diac-marginal-Z}
    \lim_{\ep \searrow 0}\E\left[\delta_\ep(L) h_\ep(\X, L)\right] 
    = - f(\q)\sum_{k = 1}^m\left. \E\left[  \mfk(Z_i)  \partial_i \,g_k(\Z) \Id_{\{X_k \le d_k\}}\,h_0(\X, L)  \right|L = \q\right]\,.
\end{equation}
\end{lemma}

\proof{Proof of Lemma \ref{lemma:dirac-marginal-Z}.}
Let $\xi(\cdot)$ be an infinitely often differentiable function. Applying Taylor's Theorem  of $g_k$ around $z_i$, and using $\kappa_\ep(z_i) =z_i + \ep \mfk(z_i) + o(\ep)$, we obtain for all $k = 1, \ldots, n$, that 
\begin{equation}
    g_k(\z_{-i}, \kappa_\ep(z_i)) - g_k(\z)  
     = 
    \left( \kappa_\ep(z_i) - z_i \right)\partial_i \,g_k(\z)  + o\left( \kappa_\ep(z_i) - z_i\right)
    =
     \ep \mfk(z_i)\,\partial_i \,g_k(\z)  + o\left( \ep\right)
    \label{eq:taylor-margin-z}\,,
\end{equation}
where $\partial_i \, g_k(\z)= \tfrac{\partial}{\partial z_i}g_k(\z) $ is the derivative in the $i^\text{th}$ component. 
Thus, we have that for all $\z\in \R^{n}$, using the Mean Value Theorem for some $l^* \in (\q, \q - \Delta_\ep g\,] $ (or $l^* \in (\q - \Delta_\ep g\,, \q]$) in the second equation, and then \eqref{eq:taylor-margin-z} that
\begin{align*}
    \int_{-\infty}^{+\infty} \xi(l) \delta_\ep(l) \, d l
    & =
    \frac1\ep \int_{\q}^{\q - \sum_{k = 1}^m\Delta_\ep g_k}\xi(l) \, d l
     = - \frac{1}{\ep}\sum_{k = 1}^m\Delta_\ep g_k \;\xi(l^*)
    \\
    &= -  \Big( \sum_{k = 1}^m\, \mfk(z_i)\,\partial_i \,g_k(\z)\Id_{\{x_k \le d_k\}}  + o\left(1\right)\Big)\xi(l^*)\,.
\end{align*}
Taking the limit for $\ep \searrow 0$, we have
\begin{equation*}
    \lim_{\ep \searrow 0}\int_{-\infty}^{+\infty} \xi(l) \delta_\ep(l) \, d l
     = - \mfk(z_i)\sum_{k = 1}^m  \partial_i \,g_k(\z)\Id_{\{x_k \le d_k\}} \xi\left(\q\right)\,.
\end{equation*}
Equation \eqref{eq:lemma-diac-marginal-Z} follows using a similar argument as in Lemma \ref{lemma:dirac-marginal}.
\hfill\Halmos
\endproof


\proof{Proof of Theorem \ref{thm:VaR:marginal} (Marginal Sensitivity VaR)}
By Proposition 2.3 in \cite{Embrechts2013MMOR} it holds for all $\ep \ge 0$ that $F_{\ep}\left(\q(\ep)\right) = \alpha$. Setting $H(\ep, x):= F_\ep(x)$ the equation becomes $F_{\ep}\left(\q(\ep)\right) = H(\ep, \q(\ep))=\alpha$.
Taking derivative with respect to $\ep$ and evaluating at $\ep = 0$ (note that Assumptions \ref{asm: marginal VaR} and \ref{asm:diff-quantile} are fulfilled and that $\frac{\partial}{\partial x}H(0, x) = f(x)$ and  $\frac{\partial}{\partial \ep}H(\ep, x) = \tfrac{\partial}{\partial \ep} F_\ep (x )$), we obtain
\begin{equation}\label{eq:marginal:VaR:formula}
    f\left(\q\right) \; \left.\tfrac{\partial}{\partial \ep} \q(\ep)\right|_{\ep = 0} + \tfrac{\partial}{\partial \ep} F_\ep (\q )\Big|_{\ep = 0}= 0 
    \, \quad \text{and thus} \quad 
    \left.\tfrac{\partial}{\partial \ep} \q(\ep)\right|_{\ep = 0}
    = - \tfrac{1}{f\left(\q\right)} \frac{\partial}{\partial \ep} F_\ep (\q ) \,,
\end{equation}
whenever $\frac{\partial}{\partial \ep} F_\ep (\q )$ exists. Next, we show that the derivative of $F_\ep$ with respect to $\ep$ exists. \\
\underline{Part 1:}
We first consider the case of stressing $X_i$ and calculate
\begin{subequations}
\label{eq:maginal-VaR-deriv-ep}
\begin{align}
    F_\ep (\q)  - F(\q)
    &= 
    \;\P\left(L_{\ep} \le \q  \right) - \P\left(L \le \q\right)  
    \\
    &=  
    \E\big[\Id_{\left\{L \le \q  -  g_i(\Z)\,\left(\Id_{\{\kappa_\ep (X_i) \le d_i\}} - \Id_{\{X_i \le d_i\}}\right)\right\}}
      - \Id_{\{L \le \q\}}\big]
     \\
         &=  
    \E\left[\left|\Id_{\{\kappa_\ep (X_i) \le d_i\}} - \Id_{\{X_i \le d_i\}}\right|\,\left( \Id_{\left\{L \le \q  + c(\kappa)g_i(\Z)\,\right\}}
      - \Id_{\{L \le \q\}}\right)\right]
\,,
\end{align}
\end{subequations}
where the last equality follows from \eqref{eq:indicator}.
Invoking Lemma \ref{lemma:dirac-marginal} we obtain
\begin{align*}
    \tfrac{\partial}{\partial \ep} F_\ep (\q ) 
    &=-c(\kappa)\mfkinv(d_i)\, f_i(d_i) \;\;\left.\E\left[\, \left( \Id_{\left\{L \le \q + c(\kappa)g_i(\Z)\,\right\}}
      - \Id_{\{L \le \q\}}\right)~\right|~ X_i = d_i \right]\,.
\end{align*}
Combining with Equation \eqref{eq:marginal:VaR:formula} concludes the first part.

\underline{Part 2:}
Next, we consider the case of stressing $Z_i$. For this, it holds that
\begin{equation*}
    F_\ep (\q)  - F(\q)
    = \E\big[\big(\Id_{\left\{L \le  \q - \sum_{k = 1}^m\Delta_\ep g_k\, \right\}}
    - \Id_{\{L \le \q\}}\big)
    \big]\,.
\end{equation*}
Applying Lemma \ref{lemma:dirac-marginal-Z} and Equation \eqref{eq:marginal:VaR:formula} conclude the proof.
\hfill\Halmos
\endproof


\proof{Proof of Theorem \ref{thm:ES:marginal} (Marginal Sensitivity ES).}
We first calculate the sensitivity to $X_i$, and in a second step to $Z_i$.

\underline{Part 1:}
To calculate the sensitivity to $X_i$, we observe that
\begin{align}
    \frac{\ES_\alpha(L_\ep) - \ES_\alpha(L)}{\ep}
    &= \frac{1}{\ep(1 - \alpha)} \E\big[ \big(L_\ep -  \q(\ep)\big)_+ -  \left(L -  \q\right)_+ \big] + \tfrac{\q(\ep) - \q}{\ep}
    \nonumber
    \\
    &= \underbrace{\frac{1}{\ep(1 - \alpha)} \E\big[ \big(L_\ep -  \q(\ep)\big)_+ - \left(L_\ep -  \q\right)_+ \big]}_{:=A(\ep)}
    + 
    \underbrace{\E\left[\left(L_\ep -  \q\right)_+ - \left(L -  \q\right)_+ \right]}_{:=B(\ep)}
    + \underbrace{\tfrac{\q(\ep) - \q}{\ep}}_{:=C(\ep)}\,.
    \label{eq:pf-marginal-es-decomp}
\end{align}
To calculate the expectation in $A(\ep)$, we use  integration by parts in the third equation, and interpret $\int_b^a h(x) \, dx =- \int_{a}^b h(x) \, dx $, if $a < b$.
\begin{align*}
     A(\ep)\, \ep (1-\alpha)
    &= \int_{\q(\ep)}^{+ \infty}\big(y - \q(\ep)\big)\, d F_\ep(y) - \int_{\q}^{+ \infty}(y - \q)\, d F_\ep(y)\\
    &= \int_{\q(\ep)}^{\q}y \, d F_\ep(y) - 
    \q(\ep) \left(1 - \alpha\right) + \q \left(1 - F_\ep(\q)\right)\\
    &= \q F_\ep(\q)- \q(\ep) \,\alpha 
    - \int_{\q(\ep)}^{\q}F_\ep(y)\, dy  
    - \q(\ep) \left(1 - \alpha\right) + \q \left(1 - F_\ep(\q)\right)\\
    &= \big(\q - \q(\ep)\big)- \int_{\q(\ep)}^{\q}F_\ep(y)\, dy \,.
\end{align*}
Next, we collect parts $A(\ep)$ and $C(\ep)$, and use the Mean Value Theorem, that is there exists a $q^* \in (\q(\ep) , \q]$ (or $q^* \in (\q , \q(\ep)]$, if $\q < \q(\ep)$) such that $\int_{\q(\ep)}^{\q}F_\ep(y)\, dy   = (\q - \q(\ep)) \, F_\ep(q^*)$. Thus,  
\begin{align*}
    A(\ep) + C(\ep) &= \frac{1}{\ep(1 - \alpha)}\big( (\q - \q(\ep)) \, (1 - F_\ep(q^*)\big)+ \frac{\q(\ep) - \q}{\ep}\\[0.25em]
     &= \frac{(\q(\ep) - \q)}{\ep} \, \big(1 - \frac{ 1 - F_\ep(q^*)}{1 - \alpha}  \big).
\end{align*}
Talking the limit for $\ep \searrow0$, and noting that the derivative of the quantile function with respect to $\ep$ exists by Theorem \ref{thm:VaR:marginal}, we obtain
$\lim_{\ep \searrow 0} A(\ep) + C(\ep) = 0$.
For part $B(\ep)$ we obtain using \eqref{eq:indicator}
\begin{align*}
    B(\ep) 
    &= 
    \tfrac{1}{\ep(1 - \alpha)}\E\left[\left(L + g_i(\Z)\,\left(\Id_{\{X_{i, \ep} \le d_i\}} -\Id_{\{X_i \le d_i\}}\right) -  \q\right)_+ -   \left(L -  \q\right)_+ \right]
    \\
    & =
    \tfrac{1}{\ep(1 - \alpha)}\E\left[\Big|\Id_{\{X_{i, \ep} \le d_i\}} -\Id_{\{X_i \le d_i\}} \Big|\Big(\left(L - c(\kappa) g_i(\Z)\,-  \q\right)_+ -   \left(L -  \q\right)_+\Big) \right]\,.
\end{align*}
Applying Lemma \ref{lemma:dirac-marginal}, we obtain 
\begin{equation*}
    \lim_{\ep \searrow 0}B(\ep) 
    =
    \frac{-c(\kappa) \mfkinv(d_i)\, f_i(d_i)}{1 - \alpha}\E\left[\left(L - c(\kappa) g_i(\Z)\,-  \q\right)_+ -   \left(L -  \q\right)_+~|~X_i = d_i \right]\,.
\end{equation*}

\underline{Part 2:}
For the sensitivity to $Z_i$, we write similarly to part 1, $  \tfrac{1}{\ep}\big(\ES_\alpha(L_\ep) - \ES_\alpha(L) \big)     = A(\ep) + B(\ep) + C(\ep)$,
where $A(\ep)$ and $C(\ep)$ are the same as in \eqref{eq:pf-marginal-es-decomp}, while $B(\ep)$ is
\begin{subequations}\label{eq:rewrite-B-ep-indicator}
\begin{align}
    B(\ep)
    &= 
    \tfrac{1}{\ep (1 - \alpha)}\, \E\big[\big(L + \sum_{k = 1}^m\Delta_\ep g_k\, -  \q\big)_+ - \left(L - \q\right)_+ \big]
    \\
    &= 
    \tfrac{1}{\ep (1 - \alpha)}\, \E\big[\left(L   -  \q\right)\left(\Id_{\{L\le \q\}} - \Id_{\{L \le \q - \sum_{k = 1}^m\Delta_\ep g_k\}}\right)  + \sum_{k = 1}^m\Delta_\ep g_k \Id_{\{L \ge \q - \sum_{k = 1}^m\Delta_\ep g_k\}}\big]\,,
\end{align}
\end{subequations}
where in the last equality we used that $\Id_{\{L > \q\}} =1 - \Id_{\{L \le \q\}} $.
Note that the argument that $A(\ep) + C(\ep)$ converges to 0 for $\ep \searrow 0$ only depends on the fact that $F_\ep$ converges to $F$ for $\ep \searrow 0$. Thus, also here, it holds that $\lim_{\ep \searrow 0} A(\ep) + C(\ep) = 0$. To calculate the limit of $B(\ep)$, we apply Lemma \ref{lemma:dirac-marginal-Z} to the first term, which turns out to be equal to zero. For the second term, note that $\frac1\ep \Delta_\ep g_k$ converges to $\mfk(Z_i) \partial_i\, g_k(\Z)\Id_{\{X_k \le d_k\}}$ $\P$-a.s. for $\ep \searrow 0$, see also Equation \eqref{eq:taylor-margin-z}. Thus, 
\begin{align*}
    \lim_{\ep \searrow 0} B(\ep)
    &= 
    \frac{1}{1 - \alpha}\sum_{k = 1}^m
    \E\left[\mfk(Z_i)\partial_i\, g_k(\Z)\Id_{\{X_k \le d_k\}}\Id_{\{L \ge \q \}}\right]
    = 
    \sum_{k = 1}^m
    \E\left[\mfk(Z_i)\partial_i\, g_k(\Z)\Id_{\{X_k \le d_k\}}~|~ L \ge \q \right]\,.
\end{align*}
\hfill\Halmos
\endproof

\subsection{Proof of Cascade Sensitivity: Theorems \ref{thm:ES:cascade}, \ref{thm:ES:cascade-Zi},
\ref{thm:VaR:cascade}, and \ref{thm:VaR:cascade-Zi}.
}

For the proofs of the cascade sensitivities to $\VaR$ and $\ES$, we need the following lemmas concerning sequences of functions that converge weakly to Dirac delta functions. For this, we first provide a representation of the stressed loss, when stressing $X_i$. 
For a stress function $\kappa_\ep$ and a Rosenblatt transform $\Psi$, we define for all $j\in \mM$ and fixed $\mathbf{v}$,
\[a_{\ep,j}(x):=|\Id_{\{\eta_{\ep,j}(x) \le d_j\}} -\Id_{\{x \le d_j\}}|\,,\]
where $\eta_{\ep,j}(x):=\Psi^{(j)}\left(\kappa_\ep\left(\Psi^{(j), -1}(x,\mathbf{v})\right), \mathbf{v}\right)$ and $\Psi^{(j), -1}$ denotes the inverse in the first component of $\Psi^{(j)}$. Further, we let
$A_{\ep,j}:=a_{\ep,j}(X_j)$,
where it is implicit that $\mathbf{v}$ is replaced by $\V$. Note that $X_j = \Psi^{(j)}(X_i, \V)$ $\P$-a.s., and therefore
\begin{equation*}
    \Psi^{(j)}\left(X_{i, \ep}, \V\right)
    = \Psi^{(j)}\left(\kappa_\ep(X_i), \V\right)
    = \Psi^{(j)}\left(\kappa_\ep\left(\Psi^{(j), -1}(X_j, \V)\right), \V\right)
    = \eta_{\ep,j}(X_j)
    \quad \P\text{-a.s}.\,.
\end{equation*}

\begin{lemma}[Stressed Portfolio Loss]\label{lemma:stress-pl}
For a stress $X_{i, \ep}$, the stressed portfolio admits representation
\begin{equation*}
    L^\bPsi(X_{i,\ep} )
    =L +  \sum_{k = 1}^m\tilde{\Delta}_\ep g_k
    -  \sum_{j = 1}^mc(\kappa;\,j) g_j(\Z)\,A_{\ep, j}
    \,,
\end{equation*}
where $\tilde{\Delta}_\ep \, g_k = \left(g_k\left(\Psi^{(\Z)}(X_{i, \ep}, \V)\right) - g_k(\Z)\right) \Id_{\{\Psi^{(k)}(X_{i, \ep}, \V) \le d_k\}}$.
\end{lemma}
\proof{Proof of Lemma \ref{lemma:stress-pl}.}
We obtain
\begin{align*}
    L^\bPsi(X_{i,\ep} )
    &=
    L +  \sum_{k = 1}^m\tilde{\Delta}_\ep g_k
    +  \sum_{j = 1}^m g_j(\Z)\,\left(\Id_{\{\eta_{\ep,j}(X_j) \le d_j\}} -\Id_{\{X_j \le d_j\}}\right)
    \\
    &= 
    L +  \sum_{k = 1}^m\tilde{\Delta}_\ep g_k
    -  \sum_{j = 1}^m c(\kappa;\, j)g_j(\Z)\,A_{\ep, j}
    \,,
\end{align*}
since by Assumption \ref{asm:psi-combined} it holds that $\Id_{\{\eta_{\ep,j}(x) \le d_j\}} -\Id_{\{x \le d_j\}}  = -c(\kappa;\, j) a_{\ep, j}(x)$ for all $j  \in\mM$.
\hfill\Halmos
\endproof

\begin{lemma}\label{lemma:dirac-cascade-A}
Let $\K\subset  \mM$ and its complement  $\K^\complement = \mM / \K$ and define the sequence of functions
\begin{equation*}
    \delta^\K_\ep(\x) = \tfrac{1}{\ep}\prod_{k \in \K}a_{\ep, k}(x_k)\prod_{l \in \K^\complement}a_{\ep, l}^\complement(x_l),\,\qquad \ep>0,
\end{equation*}
where $a_{\ep, k}^\complement(x) = 1 - a_{\ep, k}(x)$.

Then, for all functions $h_\ep \colon \R^{m + n} \to \R$ such that $\lim_{\ep \searrow 0}\E\left[|h_\ep(\X, \Z)|\right] < \infty$,
the following holds:
\begin{enumerate}[label = \roman*)]
    \item if $\K$ contains one element, $\K = \{k\}$, then
    \begin{equation*}
        \lim_{\ep \searrow 0}\,\E\left[ \delta_\ep^\K( \X) \, h_\ep(\X, \Z)\right] 
        = - c(\kappa;\, k)\, f_k(d_k)\;\E\left.\left[\,\mfkinv\left(X_i\right) \, \Psi_1^{(k)}(X_i, \V)\,h_0(\X, \Z)~\right|~ X_k = d_k\right] \,.
    \end{equation*}

    \item if $\K$ contains two or more elements, then 
    \begin{equation*}
        \lim_{\ep \searrow 0}\,\E\left[ \delta_\ep^\K( \X) \, h_\ep(\X, \Z)\right] = 0\,.
    \end{equation*}
\end{enumerate}
\end{lemma}

\proof{Proof of Lemma \ref{lemma:dirac-cascade-A}.}
First, let $\K = \{k\}$ and note that $\lim_{\ep \searrow 0} a_{\ep, j}^\complement(x)
    = \lim_{\ep \searrow 0} 1 - a_{\ep, j}(x)
    = 1$, for all $j \in\mM$ and $x\in\R$.
Next, we calculate the inverse of $\eta_{\ep,k}(x)$ in $x$, which is given by
\begin{equation*}
    \eta_{\ep,k}^{-1}(x)
    =
    \Psi^{(k) }\left(\kappa_\ep^{-1}\left(\Psi^{(k), -1}(x, \mathbf{v})\right), \mathbf{v}\right)
    =
    \Psi^{(k)}\left(\kappa_\ep^{-1}(x), \mathbf{v}\right).
\end{equation*}
Its derivative is, noting that $\eta_{0,k}(x)= \eta^{-1}_{0,k}(x)= x$,
\begin{equation*}
    \lim_{\ep \searrow 0}
    \tfrac{1}{\ep}\left(\eta_{\ep,k}^{-1}(x) - x\right)
    =
    \Psi^{(k)}_1\left(x, \mathbf{v}\right) \mfkinv(x)\,.
\end{equation*}
Using similar arguments as in the proof of Lemma \ref{lemma:dirac-marginal}, replacing $\kappa_\ep^{-1}$ with $\eta_{\ep,k}^{-1}$, we obtain that
\begin{equation*}
    \lim_{\ep \searrow 0} 
    \E[\delta_\ep^k(\X)h_\ep(\X, \Z)]
    =
    -c(\kappa;\, k)  f_i (d_k)  \;\E\left[\,\mfkinv(X_i)\,\Psi^{(k)}_1\left(X_i, \V\right) h_0(\X, \Z)~|~ X_i = d_k\,\right]\,.
\end{equation*}

Next, assume that $\K = \{k, j\}$ contains two indices and let $\xi\colon \R^2 \to \R$ be an infinitely often differentiable function. Then, using \eqref{eq:indicator} and the following change of variable $y_j = \eta_{\ep,j}(x_j)$ in the first equation \begin{align*}
    &\int_{- \infty}^{+\infty}\int_{- \infty}^{+\infty}\xi(x_j, x_k) \delta_\ep^\K(x_j, x_k)\,dx_j dx_k  \\
    &= 
    -c(\kappa)\frac{1}{\ep} \int_{- \infty}^{+\infty}\int_{- \infty}^{+\infty}\xi(x_j, x_k) \left(\Id_{\{\eta_{\ep,j}} (x_j) \le d_j\} - \Id_{\{x_j \le   d_j\}}\right)  \,dx_j\, a_{\ep, k}(x_k) \prod_{l \neq j, k}a_{\ep, l}^\complement(x_l) \,dx_k 
    \\
    &= 
     -c(\kappa)\int_{- \infty}^{+\infty}\left( \frac{1}{\ep}\left(
     \int_{- \infty}^{+\infty}\frac{\xi(\eta_{\ep,j}^{-1}(y_j), x_k)}{\eta_{\ep,j}^\prime(\eta_{\ep,j}^{-1}(y_j))} \Id_{\{y_j \le d\}}  \, dy_j
    -  \int_{- \infty}^{d_j}\xi(x_j, x_k)\, dx_j\right)
    \right)
     a_{\ep, k}(x_k) \prod_{l \neq j, k}a_{\ep, l}^\complement(x_l)\, dx_k 
    \,.
\end{align*}
Define the function $\Xi(x,y)$, such that $\frac{d}{dx}\Xi(x,y) = \xi(x,y)$, so that 
\begin{equation}\label{proof:eq:lemma-multi-dirac}
    \frac{1}{\ep}\int_{- \infty}^{+\infty}\frac{\xi(\eta_{\ep,j}^{-1}(y_j), x_k)}{\eta_{\ep,j}^\prime(\eta_{\ep,j}^{-1}(y_j))} \Id_{\{y_j \le d\}}  \, dy_j
    -  \int_{- \infty}^{d}\xi(x_j, x_k)\, dx_j
    = 
    \frac{1}{\ep} \left(\Xi(\eta_{\ep,j}^{-1}(d, x_k)) - \Xi(d, x_k)\right)\,.
\end{equation}
The limit of \eqref{proof:eq:lemma-multi-dirac} for $\ep \searrow 0$ exists, moreover $a_{\ep, k}(x)$ converges to 1, for $\ep \searrow 0$, while $a_{\ep, l}^\complement(x)$, $l \neq \{j,k\}$, converge to 0 for $\ep \searrow 0$. Thus, we  obtain that $\delta^\K_\ep(\cdot)$ converges weakly to 0, for $\ep \searrow 0$. 

The cases when $\K$ contains more than two indices follow analogous.
\hfill\Halmos
\endproof

\begin{lemma}\label{lemma:dirac-cascade-g}
Define the sequence of functions
\begin{equation*}
    \delta_\ep(l) = \tfrac{\Id_{\left\{l \le \q - \sum_{k = 1}^m\tilde{\Delta}_\ep g_k \right\}}
    - \Id_{\left\{l \le \q \right\}}
    }{\ep}\,,
\end{equation*}
where $\tilde{\Delta}_\ep g_k = \left(g_k\left(\Psi^{(\Z)}(\kappa_\ep(x_i), \bv)\right) - g_k(\z)\right) \Id_{\{\Psi^{(k)}(\kappa_\ep(x_i), \bv) \le d_k\}}$, $\z \in \R^n$, $x_i \in \R$, and $l \ge 0$. Then, $\delta_\ep$ converges weakly to a scaled Dirac delta function at $\q$ for $\ep \searrow 0$. Moreover, for any function  $h_\ep \colon \R^{m + n} \to \R$ such that $\lim_{\ep \searrow 0}\E\left[|h_\ep(\X, \Z)|\right] < \infty$, the following holds:
\begin{equation*}
    \lim_{\ep \searrow 0}\E\left[\delta_\ep(L) h_\ep(\X, L)\right] 
    = - f(\q)\sum_{k = 1}^m \sum_{l = 1}^n\left. \E\left[  \mfk(X_i)  \partial_l \,g_k(\Z) \Psi_1^{(m+l)}(X_i, \V) \Id_{\{X_k \le d_k\}}\,h_0(\X, L)  \right|L = \q\right]\,.
\end{equation*}
\end{lemma}

\proof{Proof of Lemma \ref{lemma:dirac-cascade-g}.}
This proof follows along the lines of the proof of Lemma \ref{lemma:dirac-marginal-Z}.  
Note that $z_l = \Psi^{(m+l)}(x_i, \bv)$, and that the Taylor approximation of $g_k\left(\Psi^{(\z)}(\kappa_\ep(x_i), \bv)\right)$ around $\ep  = 0$, becomes, using first an approximation of $g_k$ around $\z$, then of $\bPsi^{(m+l)}$ around $x_i$, for all $l = 1, \ldots, n$, and finally for $\kappa_\ep$ around $\ep = 0$
\begin{align*}
    g_k\left(\Psi^{(\Z)}(\kappa_\ep(x_i), \bv)\right) - g_k(\z)
    &= 
    \sum_{l = 1}^n \partial_l \, g_k(\z) \left(\Psi^{(m+l)}(\kappa_\ep(x_i), \bv) - z_l \right) + o\left(\Psi^{(m+l)}(\kappa_\ep(x_i), \bv) - z_l\right)
    \\
    &= 
    \sum_{l = 1}^n \partial_l \, g_k(\z) \,\Psi_1^{(m+l)}(x_i, \bv) \, (\kappa_\ep(x_i) - x_i)+ o\left(\kappa_\ep(x_i)\right)
    \\
    &= 
    \ep \sum_{l = 1}^n \partial_l \, g_k(\z) \,\Psi_1^{(m+l)}(x_i, \bv) \, \mfk(x_i) 
    + o\left(\ep\right)
    \,.
\end{align*}
The reminder of the proof follows analogous steps to those in the proof of Lemma \ref{lemma:dirac-marginal-Z}.
\hfill\Halmos
\endproof

\proof{Proof of Theorem \ref{thm:VaR:cascade} (Cascade Sensitivity VaR to $X_i$).}
Analogous to the proof of Theorem \ref{thm:VaR:marginal}, we use Equation \eqref{eq:marginal:VaR:formula} and, thus, we only need to calculate $\frac{\partial}{\partial \ep}F_\ep (\q)|_{\ep = 0}$. Using Lemma \ref{lemma:stress-pl}, we obtain 
\begin{align*}
    F_\ep (\q )  - F(\q)
    &= 
    \E\big[\Id_{\left\{L \le \q 
    -\sum_{k = 1}^m \tilde{\Delta}_\ep g_k
    +  \sum_{j = 1}^m c(\kappa; \, j) g_j(\Z)
    A_{\ep, j}
    \right\}}
    -\Id_{\left\{L \le \q  \right\}}\big]\,,
\end{align*}
where we recall that $A_{\ep, j} = |\Id_{\{\eta_{\ep,j}(X_j) \le d_j\}} -\Id_{\{X_j \le d_j\}}|$ and denote its complement by $A_{\ep, j}^\complement$, i.e., $A_{\ep, j}^\complement = 1 - A_{\ep, j}$. 
Next, as $A_{\ep, j}$ are indicators, we can rewrite the expectation and split it into multiple sums, as follow: The first expectation corresponding to all $A_{\ep, j}^\complement$ \eqref{eq:cascade-var-1}, and then we sum over all possible combinations of $A_{\ep, j}$ and $A_{\ep, k}^\complement$. 
\begin{subequations}
\begin{align}
    & F_\ep (\q )  - F(\q)
    =
    \E\Big[\prod_{i = 1}^m A_{\ep, i}^\complement\left(\Id_{\left\{L \le \q 
    -\sum_{k = 1}^m \tilde{\Delta}_\ep g_k\right\}}
    -\Id_{\left\{L \le \q  \right\}}\right)\Big]
    \label{eq:cascade-var-1}
    \\
     & \qquad\qquad +
    \sum_{k = 1}^m \;\sum_{\stackrel{i_1, \ldots, i_k = 1}{i_1 < \cdots < i_k}}^m
    \E\Bigg[
    \prod_{j = 1}^k A_{\ep, i_j} \prod_{\stackrel{l  = 1}{l \not\in \{i_1, \ldots, i_k\}}}^m A_{\ep, l}^\complement
   \left(\Id_{\left\{L \le \q 
    -\sum_{r = 1}^m \tilde{\Delta}_\ep g_r
    +  \sum_{j = 1}^k c(\kappa;\,j) g_{i_j}(\Z)
    \right\}}
    -\Id_{\left\{L \le \q  \right\}}\right) \Bigg]\,.
    \nonumber
\end{align}
\end{subequations}
For the first expectation above (Equation \eqref{eq:cascade-var-1}), we apply Lemma \ref{lemma:dirac-cascade-g} and that $\lim_{\ep \searrow 0}A_{\ep, k}^\complement = 1$ for all $k = 1, \ldots, m$. For the other terms, we apply Lemma \ref{lemma:dirac-cascade-A}. Specifically, we observe that only the summands that contains exactly one $A_{\ep, k}$ do not converge to 0. Thus, we obtain the limit, noting that for all $k = 1, \ldots, m$, $\tilde{\Delta}_\ep \, g_k$ converges to 0, for $\ep \searrow 0$,
\begin{align*}
    \lim_{\ep \searrow 0} & \frac{ F_\ep (\q )  - F(\q)}{\ep}
    = 
    - \sum_{j = 1}^m \sum_{l = 1}^n  f(\q)\E\big[  \mfk(X_i)  \partial_l \,g_j(\Z) \Psi_1^{(m+l)}(X_i, \V) \Id_{\{X_j \le d_j\}}  \big|L = \q\big]
    \\
    & \quad -\sum_{j = 1}^m c(\kappa;\,j) f_j(d_j) 
    \E\big[\mfkinv(X_i) \Psi_1^{(j)}(X_i, \V) \big(\Id_{\left\{L \le \q
    +  c(\kappa;\,j) g_j(\Z)
    \right\}}
    -\Id_{\left\{L \le \q  \right\}}\big)
    ~|~ X_j = d_j\big]\,.
\end{align*}
Combining with Equation \eqref{eq:marginal:VaR:formula} concludes the proof.
\hfill\Halmos
\endproof

\proof{Proof of Theorem \ref{thm:ES:cascade} (Cascade Sensitivity ES to $X_i$).}
We write analogous to the proof of Theorem \ref{thm:ES:marginal}
\begin{equation*}
    \lim_{\ep \searrow 0}
    \tfrac{1}{\ep}\left(\ES_\alpha(L_\ep) - \ES_\alpha(L)\right) 
    =
    \lim_{\ep \searrow 0}
    A(\ep) + B(\ep)+ C(\ep)
    =
    \lim_{\ep \searrow 0}
    B(\ep)\,.
\end{equation*}
For part $B(\ep)$, we proceed similar to the proof of Theorem \ref{thm:VaR:cascade} and write, using the notation from the proof of Theorem  \ref{thm:VaR:cascade} and Lemma \ref{lemma:stress-pl}
\begin{subequations}
\begin{align}
 B(\ep)(1-\alpha) \ep
    &= 
    \E\Big[\Big(L +  \sum_{r = 1}^m\tilde{\Delta}_\ep g_r
    -  \sum_{k = 1}^m c(\kappa;\,k) g_k(\Z)\,A_{\ep , k} -  \q\Big)_+ - \left(L -  \q\right)_+\Big]
    \nonumber
    \\
    &=
    \E\Big[\prod_{i = 1}^m A_{\ep, i}^\complement\Big(\big(L +  \sum_{r = 1}^m\tilde{\Delta}_\ep g_r
     -  \q\big)_+ - \left(L -  \q\right)_+\Big)\Big]
     \label{eq:B-ep-1}
    \\
    & \quad +
    \sum_{k = 1}^m \sum_{\stackrel{i_1, \ldots, i_k = 1}{i_1 < \cdots < i_k}}^m\E\Big[ \prod_{j = 1}^k A_{\ep, i_j} \prod_{\stackrel{l = 1}{l \not\in \{i_1, \ldots, i_k\}}}^m A_{\ep, l}^\complement
    \nonumber
    \\
    & \qquad \times
    \Big(\big(L +  \sum_{r = 1}^m\tilde{\Delta}_\ep g_r
    -  \sum_{j = 1}^k c(\kappa;\,j) g_{i_j}(\Z)\,A_{\ep , i_j} -  \q\big)_+ - \big(L -  \q\big)_+\Big)\Big]\,.
    \label{eq:B-ep-2}
\end{align}
\end{subequations}
To calculate the limit of the expectation in Equation \eqref{eq:B-ep-1}, we rewrite similar to \eqref{eq:rewrite-B-ep-indicator}
\begin{subequations}
\begin{align}
    \big(L +  \sum_{r = 1}^m\tilde{\Delta}_\ep g_r
     -  \q\big)_+ - \left(L -  \q\right)_+
     &= 
     \big(L -  \q\big)_+ \big(\Id_{\{L \le \q\}} - \Id_{\{L \le \q  - \sum_{k = 1}^m\tilde{\Delta}_\ep g_k\}}\big)
     \label{eq:l-q-alpha-1}
     \\
     & \quad +  \sum_{r = 1}^m\tilde{\Delta}_\ep g_r
     \Id_{\{L \ge \q  - \sum_{k = 1}^m\tilde{\Delta}_\ep g_k\}}\,.
     \label{eq:l-q-alpha-2}
\end{align}
\end{subequations}
For \eqref{eq:l-q-alpha-1} we apply Lemma \ref{lemma:dirac-cascade-g}, noting that $A_{\ep, k}^\complement$ converges to 1, for all $k = 1, \ldots, m$, as $\ep \searrow 0$. For \eqref{eq:l-q-alpha-2}, we note that for all $k = 1, \ldots, m$, it holds $\P$-a.s. (see the Proof of Lemma \ref{lemma:dirac-cascade-g}) that
\begin{equation*}
\lim_{\ep \searrow 0} \tfrac{\tilde{\Delta}_\ep g_k}{\ep} = 
\sum_{l = 1}^n \partial_l\, g_k(\Z) \Psi_1^{(m + l)}(X_i, \V) \mfk(X_i) \Id_{\{X_k \le d_k\}}\,.
\end{equation*}
For all the other summands in Equation \eqref{eq:B-ep-2} we apply Lemma \ref{lemma:dirac-cascade-A}. Collecting, we obtain that 
\begin{subequations}
\begin{align}
    & (1 - \alpha)\; \lim_{\ep \searrow 0} B(\ep)
    = 
    \nonumber
    \\
    &  \sum_{k = 1}^m \sum_{l = 1}^n\left. f(\q) \E\left[  \mfk(X_i)  \partial_l \,g_k(\Z) \Psi_1^{(m+l)}(X_i, \V) \Id_{\{X_k \le d_k\}}\,\big(L -  \q\big)_+ \right|L = \q\right]
    \label{eq:cascade-ES-term-1}
    \\
    & + \sum_{j = 1}^m \sum_{l = 1}^n \E\left[\partial_l\, g_j(\Z) \Psi_1^{(m + l)}(X_i, \V) \mfk(X_i) \Id_{\{X_j \le d_j\}} \Id_{\{L \ge \q\}}\right]
    \nonumber
    \\
    & - \sum_{j = 1}^m c(\kappa;\,j) f_j(d_j)\E\left[\mfkinv(X_i) \Psi_1^{(j)}(X_i, \V)
    \left(\big(L 
    -  c(\kappa;\,j) g_j(\Z)\,-  \q\big)_+ - \big(L -  \q\big)_+\right)
    ~|~ X_j = d_j\right]\,.
    \nonumber
\end{align}
\end{subequations}
Due to the conditioning event, \eqref{eq:cascade-ES-term-1} is equal to 0. 
\hfill\Halmos
\endproof

\proof{Proof of Theorems \ref{thm:VaR:cascade-Zi} and \ref{thm:ES:cascade-Zi} (Cascade Sensitivities to $Z_i$).}
The proofs follow by as the stressed portfolio for a stress on $Z_i$, admits an analogous representation as when stressing $X_i$, with the difference that the inverse Rosenblatt transform starts at $Z_i$ instead of $X_i$, see Eqs. \eqref{eq:stressed-cascade-loss-X} and \eqref{eq:stressed-cascade-loss-Z}\,.

\hfill\Halmos
\endproof

\section{Sensitivity to General Loss Models}\label{app::general-loss-model}
In this section, we generalise the loss model \eqref{eq:loss-model} to include cases where the functions $g_j$ depends on both $\Z$ and $\X$. Specifically, we let 
\begin{equation}\label{eq:loss-model-como}
    L :=  \sum_{j \in \mM} g_j(\X, \Z) \Id_{\{X_j \le d_j\}},
\end{equation}
where a stress on $X_i$ results in the stressed loss model
\begin{equation}\label{eq:loss-model-como-stressed}
    L_\ep(X_i) 
        :=
    \sum_{j \neq i, \, j \in \mM} g_j(X_{i, \ep},\X_{- i}, \Z) \Id_{\{X_j \le d_j\}}
    + 
    g_i(X_{i, \ep}\X_{-i}, \Z) \Id_{\{X_{i, \ep} \le d_j\}}\,.
\end{equation}
We only present the sensitivities to $X_i$, as the sensitivities to $Z_i$ are not impacted by the model generalisation. We observe in the next result that the marginal sensitivity to $X_i$ of the loss model \eqref{eq:loss-model-como} accounts for both the stress in the indicator and the stress via the functions $g_j$, $j \in\mM$.

\begin{theorem}[Marginal Sensitivity -- General Loss Model]\label{thm:VaR:marginal-como}
Let Assumptions \ref{asm: marginal VaR} and \ref{asm:diff-quantile} be fulfilled for the loss model \eqref{eq:loss-model-como} and for fixed $\alpha \in (0,1)$.
Then, the marginal sensitivity for $\VaR$ to $X_i$ for loss model \eqref{eq:loss-model-como} is 
\begin{align*}
   \S_{X_i}\,[\,\VaR_\alpha\,]
   &=
   \sum_{j\in\mM} \, \big( 
   \E\big[\mfk(X_i) \partial_i g_j(\X, \Z) \Id_{\{X_j \le d_j\}} ~\big|~L =  \q\big]\big)
   \\
   & \quad 
   +c(\kappa) \mfkinv(d_i)
   \frac{f_i(d_i)}{f\left(\q\right)} \,
   \E\big[ \big( \Id_{\{L   \le \q + c(\kappa)g_i(\X, \Z)\}} - \Id_{\{L \le \q\}}\big) ~\big|~X_i = d_i\big]\,.
\end{align*}
The marginal sensitivity for ES to $X_i$ for the loss model \eqref{eq:loss-model-como} is 
\begin{align*}
   \S_{X_i}\,[\,\ES_\alpha\,]
   &=
   \sum_{j \in\mM} \,  \big(
   \E\big[\mfk(X_i) \,\partial_i g_j(\X, \Z)\, \Id_{\{X_j \le d_j\}} ~\Big|~L \ge \q\big]\big)
   \\
   & \quad 
   - \frac{c(\kappa) \mfkinv(d_i)f_i(d_i) }{1-\alpha}
   \,
   \E\big[\left(L - c(\kappa)g_i(\X, \Z) - \q\right)_+ - (L - \q)_+ ~\Big|~X_i = d_i\big]\,.
\end{align*}
\end{theorem}

\begin{theorem}[Cascade Sensitivity VaR -- General Loss Model]\label{thm:VaR:cascade-general}
Let Assumptions \ref{asm: marginal VaR}, \ref{asm:diff-quantile}, and \ref{asm:psi-combined} (for $Y=X_i)$ be fulfilled for the stressed model \eqref{eq:loss-model-como-stressed} and given $\alpha \in (0,1)$. Then, the cascade sensitivity for $\VaR_\alpha$ to input $X_i$ is given by
\begin{align*}
    \C_{X_i}\,[\,\VaR_\alpha\,]
    &= 
    \sum_{j \in\mM} \C_{X_i,X_j}
    + 
    \sum_{k \in\mN} \C_{X_i,Z_k},
 \end{align*}
where for all $k \in\mN$,
\begin{equation*}
    \C_{X_i,Z_k}\,
    = \sum_{j \in\mM} \left.\E\left[ \mfk(X_i) \, \partial_{m+k} g_j(\X, \Z)\, \Psi_1^{(m+k)}(X_i, \V)  \Id_{\{X_j \le d_j\}} ~\right|~L = \q \right]\,,
\end{equation*}
and for $j\in\mM$,
\begin{small}
\begin{align*}
    \C_{X_i,X_j}\,
    &=
   \left( \sum_{r = 1}^m \,  
   \E\left[\mfk(X_i) \partial_j g_r(\X, \Z) \Psi_1^{(j)}(X_i, \V)\Id_{\{X_r \le d_r\}} ~\Big|~L =  \q\right]\right)
   \\
   & \quad 
   +c(\kappa;\, j) 
   \frac{f_j(d_j)}{f\left(\q\right)} \,
   \E\left[ \mfkinv(X_i)\Psi_1^{(j)}(X_i, \V)\left( \Id_{\{L   \le \q + c(\kappa;\, j)g_i(\X, \Z)\}} - \Id_{\{L \le \q\}}\right) ~\Big|~X_j = d_j\right]\,.
\end{align*}
\end{small}
\end{theorem}

\begin{theorem}[Cascade Sensitivity ES -- General Loss model]\label{thm:ES:cascade-general}
Let Assumptions \ref{asm: marginal VaR}, \ref{asm:diff-quantile}, and \ref{asm:psi-combined} (for $Y=X_i)$ be fulfilled for the stressed model \eqref{eq:loss-model-como-stressed} and given $\alpha \in (0,1)$. Then, the cascade sensitivity for $\ES_\alpha$ to input $X_i$ has decomposition \eqref{eq:cascade-xi-es-decomp}, where for all $k\in\mN$,
\begin{equation*}
    \C_{X_i,Z_k}\,
    = \sum_{j \in\mM} \left.\E\left[ \mfk(X_i) \, \partial_{m+k} g_j(\X, \Z)\, \Psi_1^{(m+k)}(X_i, \V)  \Id_{\{X_j \le d_j\}} ~\right|~L \ge \q \right]\,,
\end{equation*}
and for $j  \in\mM$,
\begin{small}
\begin{align*}
    \C_{X_i,X_j}\,
    &=
    \sum_{r = 1}^m \,  \left(
   \E\left[\mfk(X_i) \partial_j g_r(\X, \Z) \Psi_1^{(j)}(X_i, \V)\Id_{\{X_r \le d_r\}} ~\Big|~L \ge  \q\right]\right)
   \\
   & \quad 
   -c(\kappa;\,j) 
   \frac{f_j(d_j)}{1-\alpha} \,
   \E\left[ \mfkinv(X_i)\Psi_1^{(j)}(X_i, \V)\Big( (L   - c(\kappa;\,j)g_j(\X, \Z) - \q)_+ - (L - \q)_+\Big) ~\Big|~X_j = d_j\right]\,.
\end{align*}
\end{small}
\end{theorem}

\subsection{Proof of Sensitivity to General Loss Model: Theorems \ref{thm:VaR:marginal-como}, \ref{thm:VaR:cascade-general}, and \ref{thm:ES:cascade-general}}

\proof{Proof of Theorem \ref{thm:VaR:marginal-como} (Marginal Sensitivity - General Loss Model).}
The stressed loss model has representation (using Equation \eqref{eq:indicator})
\begin{equation*}
    L_\ep(X_i)
    =
    L + \sum_{j = 1}^m \Delta_\ep g_j 
    -
    c(\kappa) g_i\left(X_{i, \ep}, \X_{-i}, \Z\right) A_\ep 
    \,,
\end{equation*}
where $\Delta_\ep g_j := \left(g_j(X_{i, \ep}, \X_{-i}, \Z) - g_j(\X, \Z)\right)\Id_{\{X_j \le d_j\}}$ and $A_\ep = |\Id_{\{X_{i, \ep} \le d_i\}} - \Id_{\{X_i \le d_i\}}|$.
To prove the case for VaR, note that
\begin{subequations}
\begin{align}
    F_\ep(\q) - F(\q)
    &= 
    \E\left[
    A_\ep^\complement \left( \Id_{\{L + \sum_{j = 1}^m \Delta_\ep g_j \le \q\}} - \Id_{\{L \le \q\}}\right)\right]
    \label{proof:eq-VaR-marginal-a}
    \\
    & \quad + 
    \E\left[
    A_\ep \left( \Id_{\{L + \sum_{j = 1}^m \Delta_\ep g_j - c(\kappa) g_i(X_{i, \ep}, \X_{-i}, \Z) \le \q\}} 
    - \Id_{\{L \le \q\}}\right)
    \right]
    \label{proof:eq-VaR-marginal-b}
    \,,
\end{align}
\end{subequations}
Applying Lemma \ref{lemma:dirac-marginal-Z} to \eqref{proof:eq-VaR-marginal-a}, noting that $\lim_{\ep \searrow 0} A_\ep^\complement = 1$, and Lemma \ref{lemma:dirac-marginal} to \eqref{proof:eq-VaR-marginal-b}, noting that $\lim_{\ep \searrow 0}\sum_{j = 1}^m \Delta_\ep g_j = 0$ concludes the proof for VaR.

To prove the case of ES, we have from the proof of Theorem \ref{thm:ES:marginal} that
\begin{subequations}
\begin{align} 
    \lim_{\ep \searrow 0} \frac {\ES_\alpha(L_\ep) - \ES_\alpha(L)}{\ep}
    &=
    \lim_{\ep \searrow 0}\frac{1}{\ep (1-\alpha)}\, \E[(L_\ep - \q)_+ - (L-\q)_+]
    \notag
    \\
    &=
    \lim_{\ep \searrow 0}\frac{1}{\ep (1-\alpha)}\, \Bigg\{\E\Big[
    A_\ep^\complement \underbrace{\left((L+\sum_{j = 1}^m \Delta_\ep g_j-\q)_+ - (L-\q)_+\right)}_{=B_\ep}
    \Big]
    \notag
    \\
    &\quad + 
    \E\left[
    A_\ep \left(\Big(L + \sum_{j = 1}^m \Delta_\ep g_j -c(\kappa) g_i(X_{i,\ep}, \X_{-i}, \Z) -\q\Big)_+ - (L-\q)_+\right)
    \right]\Bigg\}\,.
    \label{proof:eq:ES-marginal-como}
\end{align}
\end{subequations}
Next, we see that  
\begin{equation*}
    B_\ep=
    (L-\q)_+ \left(\Id_{\{L \le \q \}} - \Id_{\{L \le \q -\sum_{j= 1}^m \Delta_\ep g_j\}}\right)
    +
    \sum_{j = 1}^m \Delta_\ep g_j \Id_{\{L \ge \q -  \sum_{j = 1}^m \Delta_\ep g_i\}}\,.
\end{equation*}
Using similar arguments as in the proof of Theorem \ref{thm:ES:cascade} (in particular  applying Lemma \ref{lemma:dirac-cascade-g}), we observe that 
\begin{equation*}
    \lim_{\ep \searrow 0} \frac{1}{\ep (1-\alpha)}\E[A_\ep^\complement\; B_\ep ]
    =
      \sum_{j = 1}^m \E \left[\partial_i g_j(\X, \Z) \mfk(X_i) \Id_{\{X_j \le d_j\}} ~|~L \ge \q \right]\,.
\end{equation*}
Applying Lemma \ref{lemma:dirac-marginal} to \eqref{proof:eq:ES-marginal-como} concludes the proof.
\hfill\Halmos
\endproof

The proofs of Theorems \ref{thm:VaR:cascade-general}, and \ref{thm:ES:cascade-general} follow along the lines of the proofs of Theorems \ref{thm:VaR:cascade} and \ref{thm:ES:cascade}.

\section{Additional proofs}\label{elec-app-additional-proofs}

\subsection{Proof of Mixture Stress}

\proof{Proof of mixture stress stress properties.}\label{app:proofs mixture}
We prove that the mixture stress in Table \ref{tab:stresses}  is well-defined. First, the stress function and its inverse are given by 
\begin{equation*}
    \kappa_\ep(x) = F_{i, \ep}^{-1}\left(F_i(x)\right)
    \quad \text{and} \quad
    \kappa^{-1}_\ep(x) = F_i^{-1}\left(F_{i, \ep}(x)\right)\,,
\end{equation*}
where $F_{i, \ep} = (1 -\ep) F_i + \ep\, G$. By construction, it holds that $F_{i, \ep}(x)$ is continuous and strictly increasing in $x$ for all $\ep \ge 0$. Furthermore, $F_{i,\ep}$ and $F_{i,\ep}^{-1}$ converge pointwise, as $\ep \searrow 0$, to $F_i$ and $F_i^{-1}$ respectively, thus the stress function fulfils \ref{asm:kappa-lim} and  \ref{asm:kappa-lim-invert}. Next, if $G(x) \le F_i(x)$ for all $x \in \R$, then $F_{i, \ep}(x)  = F_i(x) + \ep\,(G(x) - F_i(x)) \le F_i(x)$, and therefore $F_{i, \ep}^{-1}(u) \ge F_i^{-1}(u)$ for all $u \in [0,1]$. Thus $\kappa_\ep(x) = F_{i, \ep}^{-1}\left(F_i(x)\right) \ge x$ and the stress function fulfils \ref{asm:kappa-ineq} \ref{asm:kappa-ineq-ge}. The case when the stress function fulfils \ref{asm:kappa-ineq} \ref{asm:kappa-ineq-le} if $G(x) \ge F_i(x)$ for all $x \in \R$, follows similarly.

Now, to check property \ref{asm:kappa-mfk}, note that
\begin{equation*}
    \mfk(x)
    = \lim_{\ep \searrow 0}\tfrac{1}{\ep}\left(\kappa_\ep(x) - x\right)
    = \lim_{\ep \searrow 0} \tfrac{1}{\ep} \left(F_{i,\ep}^{-1}\big(F_i(x)\big) - F_{i}^{-1}\big(F_{i}(x)\big)\right)
    = \frac{\partial}{\partial \ep}F_{i,\ep}^{-1}(F_i(x))\Big|_{\ep = 0}\,.
\end{equation*}
Further, from the relation $\frac{\partial}{\partial \ep}F_{i,\ep}^{-1}(x)|_{\ep = 0} = - \frac{\frac{\partial}{\partial \ep} F_{i, \ep}(y)}{f_i(y)}|_{\ep = 0, y = F_i^{-1}(x)}=
\frac{F_i(y)-G(y)}{f_i(y)}|_{y = F_i^{-1}(x)}$, it follows that 
\begin{equation*}
    \mfk(x)
    = \frac{F_i(x) - G(x)}{f_i(x)}\,.
\end{equation*}
To obtain the expression for \ref{asm:kappa-mfk-inv}, note that the stress function $\kappa_\ep(x)$ is differentiable in $x$, for all $\ep$ in a neighbourhood of 0, and $\frac{\partial}{\partial x}\kappa_\ep(x)|_{\ep = 0}\not=0$, then property \ref{asm:kappa-mfk-inv} is satisfied with $\mfkinv(x) = \frac{-\mfk(x)}{\frac{\partial}{\partial x}\kappa_\ep(x)|_{\ep = 0}}$, which follows immediately from an application of the chain rule applied to the identity $\kappa_\ep^{-1}(\kappa_\ep(x))=x$. Moreover,
\begin{align*}
    \tfrac{\partial}{\partial x}\kappa_\ep(x) 
    &= -\frac{f_i(x)}{\frac{\partial }{\partial x}F_{i, \ep}(F_{i,\ep}^{-1}(F_i(x)))|_{\ep=0}}
    = -1\,,
\end{align*}
which concludes the proof.
\hfill\Halmos
\endproof

\subsection{Proof of Bivariate Inverse Rosenblatt Transform: Proposition \ref{prop: inv-rosenblatt}}

\proof{Proof of Proposition \ref{prop: inv-rosenblatt}.}
The first two cases follow from Proposition 4.2 in \cite{Pesenti2021RA}. Assume $(X_i, X_j)$ follow a Archimedean copula. By independence of the cascade sensitivity to the choice of Rosenblatt transform, we have can choose \citep{Ruschendorf1993}
\begin{equation*}
    \Psi^{(j)}_1(X_i, \V)
    =
    F_{j|i}^{-1}(V_j | X_i)
    \quad \P\text{-a.s.}\,,
\end{equation*}
where a.s. $V_j = F_{X_j \, |\, X_i}(X_j\, |\, X_i)$ and $F_{X_j \, |\, X_i}(x\,|\,y)= \P\left( X_j \le x \, |\, X_i = y\right)$ denotes the conditional distribution of $X_j$ given $X_i = y$. We observe that $\Psi^{(j)}_1(X_i, \V)$ only depends on $V_j$ and we may write $\Psi^{(j)}_1(X_i, V_j)$ instead of $\Psi^{(j)}_1(X_i, \V)$.

We use Sklar's theorem to write the conditional distribution and quantile functions as
\begin{equation}
\label{pf: eqn: copula rep}
    F_{j|i}(x_j\, |\,x_i) = \Cop_{j|i}(F_j(x_j)\, |\,F_i(x_i))\quad \text{and } \quad
    F^{-1}_{j|i}(v\, |\,x_i) = F_j^{-1} \left(\Cop_{j|i}^{-1}(v\, |\,F_i(x_i))\right)\,.
\end{equation}
Taking derivative of the conditional quantile function with respect to the conditioning variable
\begin{equation*}
     \Psi_1^{(j)}(y, v)
     =\frac{\partial}{\partial y} F^{-1}_{j|i}(v\, |\,y)
     = \frac{f_i(y)}{f_j\left( F^{-1}_{j|i}(v\, |\,y)\right)}
     \frac{\partial}{\partial z} \Cop_{j|i}^{-1} (v \, | \, z )\Big|_{z = F_i(y)} \,.
\end{equation*}
By definition of $V_j$, it holds $\P$-a.s. that $ F_{j|i}^{-1}(V_j \, |X_i) = X_j$, thus  
\begin{equation}\label{proof:eq:archimedean-psi}
     \Psi_1^{(j)}(X_i, V_j)
    =\frac{f_i(X_i)}{f_j\left( X_j\right)}\frac{\partial}{\partial z} \Cop_{j|i}^{-1} (V_j \, | \, z )\Big|_{z = F_i(X_i)}\,.
\end{equation}
Next, we calculate the derivative (with respect to the conditioning argument) of the inverse of an conditional Archimedean copula with generator $\psi$. The conditional Archimedean copula and its inverse are given by \citep{cambou2017SC}
\begin{subequations}
\begin{align}
    \Cop_{j|i} (x \, | \, y )
    &= \frac{\dot{\psi} \left(\psi^{-1}(y) + \psi^{-1}(x)\right)}{\dot{\psi}\left(\psi^{-1}(y)\right)}  \quad  \text{and}\label{pf: eqn: cond-arch}\\[0.25em]
    \Cop_{j|i}^{-1} (v \, | \, y )
    &= \psi \left[ (\dot{ \psi})^{-1}\left\{v\, \dot{\psi}\left(\psi^{-1}(y)\right) \right\}- \psi^{-1}(y)\right] \,, 
    \nonumber
\end{align}
\end{subequations}
where $\dot{\psi}(x) = \frac{d}{dx}\psi(x)$. 
Taking derivative 
\begin{small}
\begin{align*}
    \frac{\partial}{\partial y}\, \Cop_{j|i}^{-1} (v \, | \, y )
    &= \dot{\psi} \left[ (\dot{ \psi})^{-1}\left\{v\, \dot{\psi}\left(\psi^{-1}(y)\right) \right\}- \psi^{-1}(y)\right] 
    \frac{1}{\dot{\psi}\left(\psi^{-1}(y)\right)}\, \left\{\frac{v \,\ddot{\psi}\left(\psi^{-1}(y)\right)}{\overset{..}{\psi}\left((\dot{\psi})^{-1}\{v\, \dot{\psi}\left( \psi^{-1}(y) \right)\} \right)} 
     - 1\right\}\,.
\end{align*}
\end{small}
Next, we use the definition of $V_j$ and Equations \eqref{pf: eqn: copula rep} and \eqref{pf: eqn: cond-arch} to write
\begin{align*}
    V_j = \Cop_{j|i}(F_j(X_j)\, |\,F_i(X_i))
     = \frac{\dot{\psi} \left(\psi^{-1}(U_i) + \psi^{-1}(U_j)\right)}{\dot{\psi}\left(\psi^{-1}(U_i)\right)}\,,
\end{align*}
where $U_j = F_j(X_j)$ and $U_i = F_i(X_i)$. 
Using the above, we obtain
\begin{equation*}
    \left.\frac{\partial}{\partial y}\, \Cop_{j|i}^{-1} (V_j\, | \, y )\right|_{y = U_i}
    =
    \frac{\dot{\psi}\left(\psi^{-1} (U_j) \right)}{\dot{\psi}\left(\psi^{-1} (U_i) \right)}
    \left\{ \frac{\ddot{\psi}\left(\psi^{-1}(U_i)\right)}{\ddot{\psi}\left(\psi^{-1}(U_i) + \psi^{-1}(U_j) \right)} \frac{\dot{\psi} \left(\psi^{-1}(U_i) + \psi^{-1}(U_j)\right)}{\dot{\psi}\left(\psi^{-1}(U_i)\right)}- 1\right\}\,.
\end{equation*}
Combining with Equation \eqref{proof:eq:archimedean-psi} concludes the proof.
\hfill\Halmos
\endproof

\subsection{Proof of Sensitivity to Discrete Random Variable: Theorem \ref{thm:VaR:marginal-discrete}.}

\proof{Proof of Theorem \ref{thm:VaR:marginal-discrete} (Marginal Sensitivity - Discrete).}
The stressed loss can be written as
\begin{equation*}
    T_{W,\ep}
    = T -c(\kappa) \sum_{k = 1}^r \Delta_k \, h(W, \Y)A_{\ep, k}\,,
\end{equation*}
where $A_{\ep, k} = |\Id_{\{\kappa_\ep(U) \le p_k\}} - \Id_{\{U \le p_k\}}|$ for $k \in\{ 1, \ldots, r\}$.
To prove the formula for VaR, we calculate similarly to the proof of Theorem \ref{thm:VaR:cascade}
\begin{align*}
    \P\left(T_{W,\ep} \le \q \right)&- \P\left(T \le \q \right)
    \\
    &=
    \E\left[
    \Id_{\{T \le \q + c(\kappa) \sum_{k = 1}^r \Delta_k \, h(W, \Y) A_{\ep, k}\}}
    -\Id_{\{T \le \q\}}
    \right]
    \\
    &=
    \sum_{k = 1}^r \sum_{\stackrel{i_1, \ldots, i_k = 1}{i_1 < \cdots < i_k}}^r 
    \E\left[
    \prod_{j = 1}^k A_{\ep, i_j} 
    \prod_{\stackrel{l =1}{l \not\in \{i_1, \ldots i_k\} }}^r A_{\ep, l}^\complement
    \left(
       \Id_{\{T \le \q + c(\kappa) \sum_{j = 1}^k \Delta_j \, h(W,\Y)\}}
    -\Id_{\{T \le \q\}}
    \right)\right]\,.
\end{align*}
Applying Lemma \ref{lemma:dirac-cascade-A} we obtain
\begin{align*}
    \frac{\partial}{\partial \ep} F_\ep(\q)
    &=
    -\sum_{k = 1}^r c(\kappa)\E\left[ \mfk^{-1}(U)\left(
       \Id_{\{T \le \q + c(\kappa) \Delta_k \, h(W, \Y) \}}
    -\Id_{\{T \le \q\}}
    \right) ~|~ U= p_k\right]
    \\
    &= 
    -c(\kappa) \sum_{k = 1}^r \mfk^{-1}(p_k) \E\left[ \left(
       \Id_{\{T \le \q + c(\kappa)  \Delta_k \, h(W, \Y) \}}
    -\Id_{\{T \le \q\}}
    \right)~|~ W = w_k \right]\,.
\end{align*}
Combining with Equation \eqref{eq:marginal:VaR:formula} concludes the proof for VaR.

Second, we prove the case ES. From the proof of Theorem \ref{thm:ES:marginal}, we have
\begin{align*}
    \ES_\alpha(T_{W,\ep}) - \ES_\alpha(T)
    &=
    \tfrac{1}{ (1-\alpha)}\, \E[(T_{W,\ep} - \q)_+ - (T-\q)_+]
    \\
    &=
    \tfrac{1}{ (1-\alpha)}\, \E\left[\Big(T -c(\kappa) \sum_{k = 1}^r \Delta_k \, h(W,\Y)A_{\ep, k}  - \q\Big)_+ - (T-\q)_+\right]
    \\
    &=
    \tfrac{1}{ (1-\alpha)}\,     \sum_{k = 1}^r \sum_{\stackrel{i_1, \ldots, i_k = 1}{i_1 < \cdots < i_k}}^m 
    \E\left[
    \prod_{j = 1}^k A_{\ep, i_j} 
    \prod_{\stackrel{l =1}{l \not\in \{i_1, \ldots i_k\} }}^m A_{\ep, l}^\complement
    \right.
    \\
    &\quad \times 
    \left.
    \left(
     \Big( T -c(\kappa) \sum_{j = 1}^k \Delta_{i_j}  h(W,\Y)  - \q\Big)_+ - (T-\q)_+
    \right)\right]\,.
\end{align*}
Next, we apply Lemma \ref{lemma:dirac-cascade-A} and obtain
\begin{small}
\begin{align*}
    \lim_{\ep \searrow 0}\frac1\ep \left(\ES_\alpha(L_\ep) - \ES_\alpha(L)\right)
    &=
    -\frac{c(\kappa)}{1-\alpha}\sum_{k =1 }^r
    \mfk^{-1}(p_k) 
    \E\left[
     \Big( T -c(\kappa)  \Delta_kh(W,\Y)  - \q\Big)_+ - (T-\q)_+
    ~\Big|~ U =p_k
    \right]\,,
\end{align*}
\end{small}
which concludes the proof.
\hfill\Halmos
\endproof

\section{Model specification for the numerical example of Section \ref{sec: numerical example RI}}\label{app:reinsurance-model}

The insurance company has $n = 12$ lines of business (LoB) each following a LogNormal distribution,  $Z_k\sim \mathcal{LN}(\mu_k, \sigma_k^2)$, $k \in\mN$. The parameters $\mu_k$ and $\sigma_k$ are such that $\E[Z_i] = e^{\mu_i + \frac12 \sigma_i^2} = 100$ (reflecting the business volume) and $\mathrm{CoV}_i = \sqrt{\text{var}(Z_i)}/ \E[Z_i] = \sqrt{e^{\sigma_i^2}-1}$, where $\mathrm{CoV}$ denotes the coefficients of variation. The considered CoV and lines of business are reported in Table \ref{tab:LOB} and the correlation matrix $\mathbf R$ in Table \ref{tab:cor-Z}. These figures are taken from the Solvency II Standard Formula parameters \citep{lloydsSCR}.

\begin{table}[tbp]
  \centering
  \caption{Correlation matrix $\boldsymbol{R}$ of $Z$ (source: \cite{lloydsSCR}).}
  \begin{scriptsize}
    \begin{tabular}{l@{\hskip 1.25em} S[table-format=3.2] S[table-format=3.2] S[table-format=3.2] S[table-format=3.2] S[table-format=3.2] S[table-format=3.2] S[table-format=3.2] S[table-format=3.2] S[table-format=3.2] S[table-format=3.2] S[table-format=3.2] S[table-format=3.2] }
    \toprule \toprule
      & 
      \multicolumn{1}{c}{$Z_1$} & \multicolumn{1}{c}{$Z_2$} & \multicolumn{1}{c}{$Z_3$} & \multicolumn{1}{c}{$Z_4$} & \multicolumn{1}{c}{$Z_5$} & \multicolumn{1}{c}{$Z_6$} & \multicolumn{1}{c}{$Z_7$} & \multicolumn{1}{c}{$Z_8$} & \multicolumn{1}{c}{$Z_9$} & \multicolumn{1}{c}{$Z_{10}$} & \multicolumn{1}{c}{$Z_{11}$} &
      \multicolumn{1}{c}{$Z_{12}$}  \\[0.5em]
\cmidrule{2-13}    
    $Z_1$ & 1 & 0.5 & 0.5 & 0.25 & 0.5 & 0.25 & 0.5 & 0.25 & 0.5 & 0.25 & 0.25 & 0.25 \\
    $Z_2$ & 0.5 & 1 & 0.25 & 0.25 & 0.25 & 0.25 & 0.5 & 0.5 & 0.5 & 0.25 & 0.25 & 0.25 \\
    $Z_3$ & 0.5 & 0.25 & 1 & 0.25 & 0.25 & 0.25 & 0.25 & 0.5 & 0.5 & 0.25 & 0.5 & 0.25 \\
    $Z_4$ & 0.25 & 0.25 & 0.25 & 1 & 0.25 & 0.25 & 0.25 & 0.5 & 0.5 & 0.25 & 0.5 & 0.5 \\
    $Z_5$ & 0.5 & 0.25 & 0.25 & 0.25 & 1 & 0.5 & 0.5 & 0.25 & 0.5 & 0.5 & 0.25 & 0.25 \\
    $Z_6$ & 0.25 & 0.25 & 0.25 & 0.25 & 0.5 & 1 & 0.5 & 0.25 & 0.5 & 0.5 & 0.25 & 0.25 \\
    $Z_7$ & 0.5 & 0.5 & 0.25 & 0.25 & 0.5 & 0.5 & 1 & 0.25 & 0.5 & 0.5 & 0.25 & 0.25 \\
    $Z_8$ & 0.25 & 0.5 & 0.5 & 0.5 & 0.25 & 0.25 & 0.25 & 1 & 0.5 & 0.25 & 0.25 & 0.5 \\
    $Z_9$ & 0.5 & 0.5 & 0.5 & 0.5 & 0.5 & 0.5 & 0.5 & 0.5 & 1 & 0.25 & 0.5 & 0.25 \\
    $Z_{10}$ & 0.25 & 0.25 & 0.25 & 0.25 & 0.5 & 0.5 & 0.5 & 0.25 & 0.25 & 1 & 0.25 & 0.25 \\
    $Z_{11}$ & 0.25 & 0.25 & 0.5 & 0.5 & 0.25 & 0.25 & 0.25 & 0.25 & 0.5 & 0.25 & 1 & 0.25 \\
    $Z_{12}$ & 0.25 & 0.25 & 0.25 & 0.5 & 0.25 & 0.25 & 0.25 & 0.5 & 0.25 & 0.25 & 0.25 & 1 \\
    \end{tabular}
      \end{scriptsize}
  \label{tab:cor-Z}%
\end{table}%

We assume that the reinsurers' critical variables follow a standardised student $t$ distributions with $\nu = 4$ degrees of freedom, i.e. $X_i \sim t(4)$, for all $i = 1, \ldots, 8$. (Note that the choice of marginal distribution for  $X_i$ is irrelevant since we consider only the event $\{X_i \le d_i\}$.) The default probabilities $\P(X_i \le d_i) = q_i$ are set to $q_i = 0.015 $, $ i = 1, \ldots, 6$ and $q_i = 0.01$, $ i = 7,8$.
 
We assume that $(\X,\Z)$ has a multivariate $t$ copula with $\nu=4$ degrees of freedom and correlation parameter matrix $\mathbf\Sigma=\{\sigma_{i,j}\}_{i,j=1,\dots,m+n}$ \citep[Sec.~7.3]{mcneil2015quantitative}. The elements of $\mathbf \Sigma$ comprise the pairwise Pearson correlations of multivariate $t_\nu$-distributed random vector arising from monotone transformations of elements of $(\X,\Z)$, so that each has $t_\nu$ marginals. 
As the (multivariate) margins of multivariate $t_\nu$ distributions are again multivariate $t_\nu$,  we start by specifying the correlation parameters of the vectors $\X$ and $\Z$ separately and then consider the dependence across the two vectors' elements. First, $\X$ has standardised $t_\nu$ margins and hence follows a multivariate $t_\nu$ distribution. Furthermore, we assume that the dependence structure is homogeneous, such that $\sigma_{i,j}={\rm Corr}(X_i,X_j) =\lambda>0$, for all $i\neq j, ~i,j\in\mM$. Second, $\Z$ has a multivariate $t_\nu$ copula with correlation parameter matrix $\mathbf R=\{r_{k,l}\}_{k,l\in\mN}$, given in Table \ref{tab:LOB}, thus $\sigma_{m+k,m+l}:=r_{k,l},~k,l \in \mN$.
Third, to specify the elements of $\mathbf\Sigma$ characterising the dependence of $(X_j,Z_k)$, i.e. $\sigma_{j,m+k},~j \in \mM,~k\in\mN$, we build a dependence model that links gross losses to reinsurance defaults using a single factor model, reflecting the homogeneity in the dependence of $\X$. The common factor is a function of the gross losses and acts as a proxy for industry effects. 

By the representation of multivariate $t$ distributions as normal mixtures \citep[Sec.~6.2]{mcneil2015quantitative}, we can represent each $Z_k$ as 
$$
Z_k =F_{Z_k}^{-1}\left(t_\nu\big(\sqrt{W}\tilde Z_k\big)\right),\quad k\in\mN,
$$
where $W\sim {\rm InvGamma}(\nu/2,\nu/2-1)$, such that $\E[W]=1$, and $(\tilde Z_1,\dots,\tilde Z_n)$ are multivariate standard normal, with correlation matrix $\mathbf R$. Then the random variables $(\sqrt{W}\tilde Z_1,\dots,\sqrt{W}\tilde Z_n)$ are multivariate $t_\nu$ distributed, with correlation matrix $\mathbf R$ and margins standardised to have unit variance. (Note that this is slightly different to the standard $t_\nu$ construction, which has margins with variance $\nu/(\nu-2)$. This choice, which does not affect the dependence model, is made to simplify moment calculations.) Define: 
\begin{equation*}
    \beta:= var\Big(\sum_{k=1}^n\tilde Z_k\Big)=\sum_{k, l \in\mN} r_{k,l}\,,
    \quad \text{and} \quad 
    \Psi:=\frac {1}{\sqrt{\beta}} \sum_{k\in\mN}\tilde Z_k\sim {\rm N}(0,1)\,.
\end{equation*}
Then the factor model becomes:
\begin{align*}
    X_j & = \sqrt{W}\left(\sqrt{\lambda} \Psi  + \sqrt{1-\lambda}\Theta_j\right),\quad j \in\mM,
\end{align*}
where $\Theta_1,\dots,\Theta_m$ are i.i.d. standard normal variables,  independent of $(\tilde Z_1,\dots,\tilde Z_n,W)$. It follows easily that $\E[X_i]=0,~var(X_i)=1$ and ${\rm Corr}(X_i,X_j)=\lambda$ is fulfilled. Furthermore, the cross-correlation values are:
\begin{align*}
    \sigma_{i,m+k}&={\rm Corr}(X_i,t_\nu^{-1}\left( F_{Z_k}(Z_k)\right)
    = 
    {\rm Corr}\big(\sqrt{W}\left(\sqrt{\lambda} \Psi  + \sqrt{1-\lambda}\Theta_i\right)~,~\sqrt{W}\tilde Z_k\big)
    \\
    &=
    \sqrt{\lambda}~ {\rm Corr}\big(\Psi ~,~\tilde Z_k\big)
    =
    \sqrt{\tfrac{\lambda}{\beta}}~ {\rm Corr}\Big( \sum_{l \in \mN}\tilde Z_l~,~\tilde Z_k\Big)
    =\sqrt{\tfrac{\lambda}{\beta}} \sum_{l\in\mN }r_{k,l}\,.
\end{align*}
This completes the dependence model specification.

\end{APPENDICES}


\bibliographystyle{informs2014}
\bibliography{references.bib}

\end{document}

%% file: preamble.tex
\usepackage{dsfont}
\usepackage{amssymb}
\usepackage{enumitem}
\usepackage[colorinlistoftodos]{todonotes}

\usepackage{booktabs}
\usepackage{siunitx}
\usepackage{color}
\usepackage[normalem]{ulem}
\usepackage{xr-hyper}
\makeatletter
\newcommand*{\addFileDependency}[1]{
  \typeout{(#1)}
  \@addtofilelist{#1}
  \IfFileExists{#1}{}{\typeout{No file #1.}}
}
\makeatother

\newcommand*{\myexternaldocument}[1]{%
    \externaldocument{#1}%
    \addFileDependency{#1.tex}%
    \addFileDependency{#1.aux}%
}
\myexternaldocument{main}
\myexternaldocument{e-companion}

\newcommand{\ep}{{\varepsilon}}
\newcommand{\Id}{{\mathds{1}}}

\newcommand{\Cop}{{\textrm{C}}}

\newcommand{\mfkinv}{{\mathfrak{K}^{-1}}}
\newcommand{\mfk}{{\mathfrak{K}}}

\newcommand{\q}{{q_\alpha}}

\newcommand{\E}{{\mathbb{E}}}
\newcommand{\N}{{\mathbb{N}}}
\renewcommand{\P}{{\mathbb{P}}}
\newcommand{\R}{{\mathbb{R}}}

\newcommand{\x}{{\boldsymbol{x}}}
\newcommand{\z}{{\boldsymbol{z}}}
\newcommand{\V}{{\boldsymbol{V}}}
\newcommand{\bv}{{\boldsymbol{v}}}

\newcommand{\X}{{\boldsymbol{X}}}
\newcommand{\Y}{{\boldsymbol{Y}}}
\newcommand{\Z}{{\boldsymbol{Z}}}

\newcommand{\bPsi}{{\boldsymbol{\Psi}}}

\newcommand{\C}{{\mathcal{C}}}
\renewcommand{\S}{{\mathcal{S}}}
\newcommand{\K}{{\mathcal{K}}}
\newcommand{\mM}{{\mathcal{M}}}
\newcommand{\mN}{{\mathcal{N}}}

\newcommand{\VaR}{{\mathrm{VaR}}}
\newcommand{\ES}{{\mathrm{ES}}}

\usepackage[plainpages=false,hyperfootnotes=false]{hyperref}
\hypersetup{
  colorlinks   = true, 
  urlcolor     = blue, 
  linkcolor    = red, 
  citecolor   = blue 
}